\documentclass[twocolumn]{aastex63} 
\usepackage{amsmath}
\usepackage[T1]{fontenc}
\usepackage[utf8]{inputenc}
\usepackage{wrapfig}
\usepackage{longtable}
\usepackage{dirtree}
\usepackage{multirow}

\received{June 18, 2021}
\revised{September 13, 2021}
\submitjournal{PSJ}

\shorttitle{A Snowball in Hell}

\shortauthors{Harman et al.}

\graphicspath{{./}{figures/}}

\newcommand{\tableref}[1]{\hyperref[#1]{Table \ref{#1}}}
\newcommand{\figureref}[1]{\hyperref[#1]{Fig. \ref{#1}}}

\begin{document}

\correspondingauthor{C. E. Harman}
\email{sonny.harman@nasa.gov}

\author[0000-0003-2281-1990]{C. E. Harman}
\affiliation{Planetary Systems Branch, 
            Space Science and Astrobiology Division, 
            NASA Ames Research Center, 
            Moffett Field, CA 94035, USA}

\author[0000-0002-5893-2471]{Ravi Kumar Kopparapu}
\affiliation{Planetary Environments Laboratory, 
            NASA Goddard Space Flight Center, 
            Greenbelt, MD 20771, USA}

\author[0000-0001-7409-5688]{Gu\dh{}mundur Stef\'ansson}
\affiliation{Princeton University, 
            Princeton, NJ 08540, USA}
\affiliation{Henry Norris Russell Fellow}

\author[0000-0002-9082-6337]{Andrea S.J. Lin}
\affiliation{Department of Astronomy \& Astrophysics, 
            The Pennsylvania State University, 
            University Park, PA 16802, USA}
\affiliation{Center for Exoplanets and Habitable Worlds, 
            The Pennsylvania State University, 
            University Park, PA 16802, USA}

\author[0000-0001-9596-7983]{Suvrath Mahadevan}
\affiliation{Department of Astronomy \& Astrophysics, 
            The Pennsylvania State University, 
            University Park, PA 16802, USA}
\affiliation{Center for Exoplanets and Habitable Worlds, 
            The Pennsylvania State University, 
            University Park, PA 16802, USA}

\author[0000-0002-3385-8391]{Christina Hedges}
\affiliation{Bay Area Environmental Research Institute,
            Moffett Field, CA 94035, USA}
\affiliation{Astrophysics Branch, 
            Space Science and Astrobiology Division, 
            NASA Ames Research Center, 
            Moffett Field, CA 94035, USA}

\author[0000-0003-1240-6844]{Natasha E. Batalha}
\affiliation{Planetary Systems Branch, 
            Space Science and Astrobiology Division, 
            NASA Ames Research Center, 
            Moffett Field, CA 94035, USA}


\title{A Snowball in Hell: The Potential Steam Atmosphere of TOI-1266c} 

\begin{abstract}

TOI-1266c is a recently discovered super-Venus in the radius valley orbiting an early M dwarf. However, its notional bulk density ($\sim$2.2 g cm$^{-3}$) is consistent with a large volatile fraction, suggesting that it might have volatile reservoirs that have survived billions of years at more than twice the Earth's insolation. On the other hand, the upper mass limit paints a picture of a cool super Mercury dominated by >50\% iron core ($\sim$9.2 g cm$^{-3}$) that has tiptoed up to the collisional stripping limit and into the radius gap. Here, we examine several hypothetical states for TOI-1266c using a combination of new and updated open-source atmospheric escape, radiative-convective, and photochemical models. We find that water-rich atmospheres with trace amounts of H$_{2}$ and CO$_{2}$ are potentially detectable (SNR $>\sim 5$) in less than 20 hours of JWST observing time. \replaced{By prescribing high altitude water and ice clouds, we}{We also} find that water vapor spectral features are not substantially impacted \added{by the presence of high-altitude water or ice clouds} due the presence of \added{a} significant amount of water above the cloud-deck, although further work with self-consistent cloud \replaced{modeling}{models} is needed. Regardless of its mass, however, TOI-1266c represents a unique proving ground for several hypotheses related to the evolution of sub-Neptunes and Venus-like worlds, particularly those near the radius valley.

\end{abstract}


\section{Introduction \label{sec:intro}}
The list of exoplanet detections from space-based observatories like the Kepler Space Telescope \citep[e.g.,][]{borucki2010kepler, twicken2016detection} and the Transiting Exoplanet Survey Satellite \citep[TESS;][]{ricker2014transiting, barclay2018revised}, as well as ground-based endeavors such as WASP \citep[e.g.][]{pollacco2006wasp}, HATNet \citep[e.g.][]{bakos2004wide,hellier2012seven}, TRAPPIST \citep{jehin2011trappist, gillon2017seven, delrez2018speculoos}, and the Habitable-zone Planet Finder (HPF) Spectrograph \citep{mahadevan2012habitable, mahadevan2014habitable}, is rapidly growing. These detections enhance our understanding of planetary occurrence rates \citep{batalha2014exploring} as well as enable robust statistical insights into planet populations \citep[e.g.,][]{dressing2013occurrence, burke2015terrestrial, fulton2017california, hardegree2019kepler}. In particular, the presence of a gap in the radius distribution of planets \citep{rogers2015most, fulton2017california, fulton2018california} highlights the cumulative effects of a planet's host star \citep[e.g.,][]{owen2012planetary, owen2017evaporation}, formation \citep[e.g.,][]{lee2014make, lee2016breeding, ginzburg2018core, gupta2020signatures}, and/or evolution \citep[e.g.,][]{luger2015habitable}, although disentangling these effects will likely require more sensitive observations \citep{loyd2020current}.

The recent discovery of two planets orbiting TOI-1266 \citep{stefansson2020, demory2020} offers a rare opportunity to begin connecting some of these planetary processes through observations. The outer planet, c (1.673\deleted{$^{+0.087}_{-0.11}$} R$_{\oplus}$\added{)}\deleted{; 1.56$^{+0.15}_{-0.13}$R$_{\oplus}$)}\deleted{ (using published radius parameters and 1-$\sigma$ uncertainties from}\deleted{ \citeauthor{stefansson2020} and \citeauthor{demory2020}, respectively),} is smaller than the inner planet, b (2.458\deleted{$^{+0.083}_{-0.073}$} R$_{\oplus}$\added{) (see Table \ref{tab:planet_params} for the reported uncertainties and a comparison between the \citeauthor{stefansson2020} and \citeauthor{demory2020} values)}\deleted{; 2.37$^{+0.16}_{-0.12}$ R$_{\oplus}$) \citep{stefansson2020, demory2020}}. This puts TOI-1266c in the `radius valley' \citep{fulton2017california, fulton2018california}. This type of `straddler' planetary system \citep{owen2020testing} can be leveraged to constrain the temporal evolution of host star's EUV flux\replaced{, but}{. However,} the \replaced{flipped orientation (with}{fact that} the smaller planet \replaced{to the}{is} outside of the larger one \citep{weiss2018california} hints at significant migration that may \replaced{obfuscate}{defeat} first-order attempts to reproduce their present-day bulk compositions through atmospheric escape alone \citep{bean2021nature}. 

The \replaced{mass of planet c has}{bulk composition of TOI-1266c is poorly constrained because the mass measurement has} large 1-$\sigma$ uncertainties\replaced{ (1.9$^{+2.3}_{-1.3}$ and 2.2$^{+2.0}_{-1.7}$ M$_{\oplus}$, again from \citeauthor{stefansson2020} and \citeauthor{demory2020}, respectively)}{: 1.9$^{+2.3}_{-1.3}$ M$_{\oplus}$ from \citeauthor{stefansson2020} using radial velocity constraints, and 2.2$^{+2.0}_{-1.7}$ M$_{\oplus}$ from \citeauthor{demory2020} based on an analysis of the transit timing variations}. This broad range of possible planet masses covers several different planet types including both rocky terrestrials and gas-dominated sub-Neptunes. Currently, compositional constraints are insufficient to rule out significant volatile inventories for small exoplanets under even higher instellation \added{than what planet c receives} \citep{dai2019homogeneous}. Formation models suggest that sub-Neptunes/super-Earths like TOI-1266b and c can end up as part of distinct water- or silicate-rich populations if planet embryos aggregate material from a more well-sampled protoplanetary disk \citep[e.g.][]{liu2019super}. It may also be easier to form volatile-rich mini-Neptunes if additional gas sources, such as envelope enrichment sourced from various accreted ices, are considered \citep[e.g.][]{venturini2017formation}. Taken together, TOI-1266c \replaced{may be}{could be} a rare example of a volatile-rich super-Earth, contrasting with the more well-populated family of volatile-poor, rocky super-Earths such as LHS 3844b \citep{kane2020volatilepoor} and TOI-849b \citep{armstrong2020remnant}. 

\begin{table}[htb]
    \centering
    \begin{tabular}{ll|l}
    Property & \citeauthor{stefansson2020} & \citeauthor{demory2020}  \\ \hline \hline
    TOI-1266: & & \\ \hline
     Spectral type & M2 & M3 \\
     Mass [M$_{\odot}$]   & 0.437 $\pm$ 0.021 & 0.45 $\pm$ 0.03 \\
     Radius [R$_{\odot}$] & 0.4232$^{+0.0077}_{-0.0079}$ & 0.42 $\pm$ 0.02 \\
     Temperature [K]      & 3563 $\pm$ 77 & 3600 $\pm$ 150  \\
     Age [Gyr]            & 7.9$^{+4.2}_{-5.2}$ & $\sim$5 \\
     Luminosity [L$_{\odot}$]  & 0.02629$^{+0.00071}_{-0.00075}$ & --- \\
      & & \\
     Planet b: & & \\ \hline
     Mass [M$_{\oplus}$]         & 6.9$^{+5.5}_{-4.0}$ & 13.5$^{+11.0}_{-9.0}$ \\
     Radius [R$_{\oplus}$]       & 2.458$^{+0.083}_{-0.073}$ & 2.37$^{+0.16}_{-0.12}$ \\
     Semi-major axis [au]        & 0.0745$^{+0.0046}_{-0.0069}$ & 0.0736$^{+0.0016}_{-0.0017}$ \\
     Instellation [S$_{\oplus}$] & 4.72$^{+1.0}_{-0.66}$ & 4.9$^{+1.0}_{-0.8}$ \\
     Equilibrium temp.$^{*}$ [K] & 410.0$^{+21.0}_{-15.0}$ & 413 $\pm$ 20 \\
     & & \\
     Planet c: & & \\ \hline
     Mass [M$_{\oplus}$]         & 1.9$^{+2.3}_{-1.3}$ & 2.2$^{+2.0}_{-1.5}$ \\
     Radius [R$_{\oplus}$]       & 1.673$^{+0.087}_{-0.110}$ & 1.56$^{+0.15}_{-0.13}$ \\
     Semi-major axis [au]        & 0.1037$^{+0.0026}_{-0.0025}$ & 0.1058$^{+0.0023}_{-0.0024}$ \\
     Instellation [S$_{\oplus}$] & 2.42$^{+0.23}_{-0.22}$ & 2.3$^{+0.5}_{-0.4}$ \\
     Equilibrium temp.$^{*}$ [K] & 347.1$^{+7.9}_{-8.0}$ & 344 $\pm$ 16 
    \end{tabular}
    \caption{We use the \citeauthor{stefansson2020} stellar and planetary parameters as the default in this study, and include values for planet b. The planet mass is reported with the 1-$\sigma$ error. Equilibrium temperature is calculated assuming an albedo of 0. We also include the values reported by \citet{demory2020} for reference, which agree within error. \label{tab:planet_params}}
\end{table}

\added{In this paper, we explore the potential states of the planet by focusing on the H-C-O chemistry of three families of scenarios, all dominated by water: H$_{2}$+H$_{2}$O, H$_{2}$+CO$_{2}$+H$_{2}$O, and O$_{2}$+H$_{2}$O. This encompasses two potential intermediate states (as we describe below) as well as a hypothetical super-Venus transitional state, in which CO$_{2}$ begins to represent a significant portion of the envelope mass. We omit two other classes of atmospheres (Venus-like and sub-Neptune) for the sake of brevity. Simulations of exo-Venus atmospheres \citep[][]{schaefer2011atmospheric, lincowski2018evolved, lustig2019detectability} largely resemble Venus' CO$_{2}$-dominated atmosphere at present or with smaller CO$_{2}$ inventories earlier in its history \citep[e.g.][]{way2020venusian}. For exo-Venuses in particular, more exploration is warranted to cover the expected diversity of planetary conditions and composition \citep[see the review by][]{madhusudhan2016exoplanetary}. A number of studies on the diversity of sub-Neptune atmospheres \citep[e.g.,][]{lavvas2019photochemical, chouqar2020properties} and retrievals \citep[e.g.,][]{benneke2019sub, mikal2020transmission} can be found in the literature; see also the review by \citet{bean2021nature}. 

}

\deleted{But estimating the composition and any potential observables for TOI-1266c remains difficult. The degeneracies in relating bulk composition, atmosphere-to-solid planet fraction, and mean density \citep[e.g.][]{rogers2010framework, dorn2015can, welbanks2019degeneracies} pose significant challenges in predicting the atmospheres of exoplanets even if the mass and radius are well constrained, and even spectroscopic observations of the planet may not necessarily break this degeneracy \citep{batalha2017challenges}. Additional radial velocity observations--such as precision RVs with HPF \citep{mahadevan2012habitable, mahadevan2014habitable}, NEID or CARMENES--are required to constrain the planet's mass, which will narrow the range of possible compositions \citep[e.g.][]{valencia2013bulk}, but in the meantime we can begin exploring the connections between composition, evolution, and observations.
}
\begin{figure*}[ht]
    \centering
    \includegraphics[width=\textwidth]{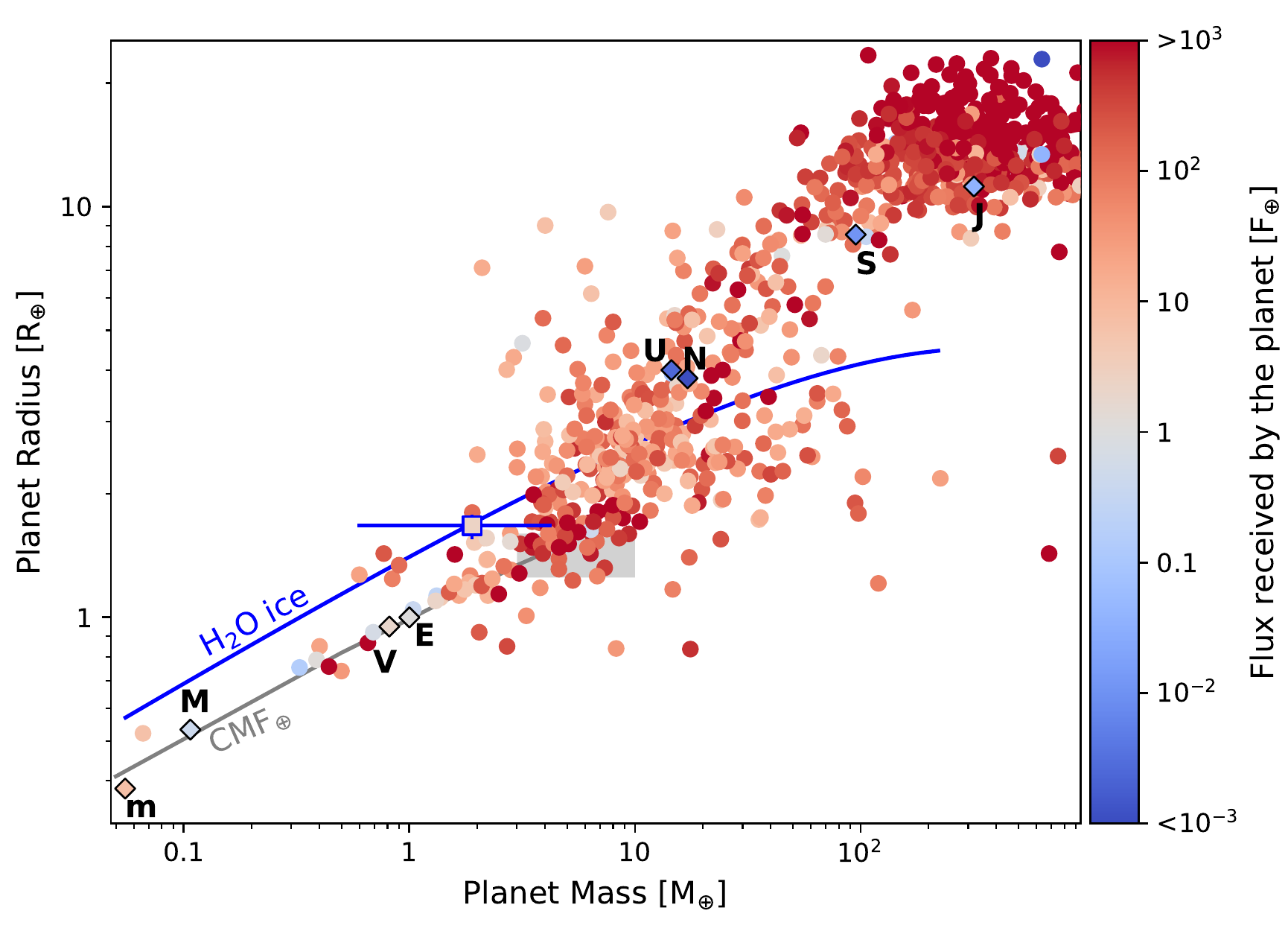}
    \caption{A mass-radius diagram for currently known exoplanets (small circles), the solar system planets (diamonds with letter designations), and TOI-1266c (square). Error bars for the mass and radius of TOI-1266c are also shown. The symbols are colored by instellation normalized to what the Earth receives ($\sim$1367 W/m$^{2}$). The gray shaded region denotes the commonly accepted mass-radius regime for super-Earths. Two compositional trends are identified on the plot: the mass-radius relationship for a pure water ice planet \citep[`H$_{2}$O ice';][]{seager2007mass}, and for a planet with the same core mass fraction as the Earth \citep[`CMF$_{\oplus}$';][]{unterborn2016scaling}.}
    \label{fig:mass-radius}
\end{figure*}

\subsection{Motivating the `snowball' scenario}\label{subsec:snowball}

The current mass of TOI-1266c affords a variety of different possible compositions that largely fork into two main solutions: a water-rich steam world, and a dense Mercury-like planet. As shown by \citet{stefansson2020}, TOI-1266c is broadly consistent with a water-dominated planet \citep[\figureref{fig:mass-radius}; see also][]{aguichine2021massradius}. This can be seen when comparing to modeled mass-radius relationships for more complex compositions. For example, \deleted{the radius estimates for} an Earth-like planet surrounded by up to 50 wt\% (percent water by mass) \replaced{yield}{can have} radii of up to 1.47 R$_{\oplus}$ \citep{fu2009interior}, which is 12\% smaller than the observed radius for TOI-1266c (though this is still within 2$\sigma$). \replaced{As an intermediate step, t}{T}he model grid from \citet{zeng2019growth} includes \replaced{a}{several} 50\% water/Earth-like rocky core planet\added{s} with \deleted{a 1,000 K }isothermal pure-H$_{2}$O atmosphere\added{s} that roughly match\deleted{es} TOI-1266c's mass and radius. \replaced{Moving away from an `Earth-like' composition, the generic}{More generalized} formulation\added{s} for ice/rock/iron fractional compositions \replaced{of \citet{fortney2007planetary} suggests that TOI-1266c is $\sim60$\% water ice by mass, while the model from \citet{noack2016water} suggests a lower limit of $\sim$77 wt-\% H$_{2}$O}{suggest that TOI-1266c is $\sim$60\% water ice \citep{fortney2007planetary} or $\gtrsim$77\% liquid water \citep{noack2016water} by mass.}\deleted{, with only a $\sim$50 km radius difference if the remaining fraction is iron or silicates.} These estimates are comparable to the maximum ice content expected for planets formed beyond the ice line \citep[e.g.][]{mordasini2009extrasolar}, suggesting that in this scenario, both planet b and c migrated inwards to their present locations. \added{However, none of these mass-radius relationships have self-consistent atmospheres, which can lead to underestimates in the estimated radius \citep[e.g.,][]{turbet2020revised}.} \deleted{Given TOI-1266c's current equilibrium temperature of $\sim$250 K, assuming an albedo$\sim$0.3, it is unlikely that there is a substantial supercritical water component, and instead, TOI-1266c is `liquid' water-dominated \citep{brugger2017constraints, mousis2020irradiated}. }

\replaced{Such s}{S}team atmospheres are often discussed within the context of the loss of water either during the magma ocean phase of a planet, immediately following accretion \citep[e.g.,][]{zahnle1988evolution, schaefer2010chemistry, hamano2013emergence, katyal2019evolution} or as a result of stellar brightening inducing a runaway greenhouse state \citep[e.g.][]{kasting1988runaway}. However, abundant water (and other heavy gases) \replaced{can}{could} also be accreted \citep{kral2020formation} or outgassed later in a planet's life \citep{kite2020atmosphere, kite2020exoplanet, kite2021water}, potentially setting up a scenario in which the runaway state is entered much later in the planet's life. 

One unavoidable consequence of a potentially water-dominated planet receiving slightly more irradiation than Venus is atmospheric escape. Venus is thought to have lost its water at some point in its history \citep[e.g.,][]{kasting1988runaway, way2020venusian} due to high temperatures allowing significant amounts of water in the upper atmosphere, where it could be photolyzed and create hydrogen atoms that then escape the atmosphere \citep[e.g.,][]{kasting1983loss}. A high hydrogen escape flux would also drag along oxygen \citep[e.g.,][]{zahnle1986mass, schaefer2016predictions, luger2015extreme, tian2015history}. Of particular note for M dwarfs is their prolonged pre-main sequence lifetimes, during which they are 10-100 times brighter than their main sequence luminosities \citep{luger2015extreme, luger2015habitable}. As a result, planets like TOI-1266c have multiple avenues by which their atmospheric composition can evolve in time, even to the extent that the planet loses its atmosphere entirely \citep[e.g.][]{kreidberg2019absence, poppenhaeger2020x}. \added{A second issue is that planets orbiting M dwarfs may be volatile-poor due to the high impact velocities during formation \citep[e.g.,][]{lissauer2007planets}.} However, if the planet began with larger initial volatile inventories \citep[e.g.][]{luger2015habitable}, replenished its atmospheric volatiles through regassing from the interior \citep[e.g.][]{moore2020keeping}, or experienced slower than expected atmospheric escape, TOI-1266c may still have a substantial envelope today.

On the other extreme, the 2-$\sigma$ upper mass limit of 6.4 M$_{\oplus}$ for TOI-1266c would indicate that the planet is >50\% iron core \citep[following the mass-radius relationship given by][]{noack2016water}, near the size limit driven by collisional stripping from impacts during accretion \citep{marcus2010minimum}. TOI-1266c would then represent a super Mercury at less than half of Mercury's instellation, which would have implications for the composition of its atmosphere. At lower planetary masses, TOI-1266c would likely still have an iron core but could also have a modest H$_{2}$-He envelope, on the order of 0.2-0.5\% of the total planet mass \citep{lopez2014understanding, zeng2019growth}. 

However, \replaced{two}{several theoretical} arguments \replaced{work to rule out}{make it difficult to form} an iron core with a substantial H$_{2}$-He fraction. First, a gas-rich initial composition is inconsistent with rocky planet formation models, which suggest that substantial accumulation of H$_{2}$ and He from the protoplanetary disk requires a minimum core mass $\gtrsim$5-20 M$_{\oplus}$ \citep[e.g.,][]{rafikov2006atmospheres, rafikov2011constraint}. \replaced{Additionally, p}{P}lanets with small rocky cores are \added{also }subject to atmospheric `boil-off' supported by the planet's inability to cool rapidly enough \citep{owen2016atmospheres}, potentially followed by loss driven by the cooling of the core \citep[e.g.][]{misener2021cool} and/or photo-evaporation \citep[e.g.][]{owen2017evaporation}. Boil-off would prevent the accumulation of more than a few tenths of a percent H$_{2}$, even if the planet began with $>$10\% by mass H$_{2}$ \citep{owen2016atmospheres}\deleted{, given the high equilibrium temperature for TOI-1266c during the super-luminous pre-main sequence phase of the host star \citep[e.g.,][]{luger2015extreme}, assuming it was in its present location}. Core-powered mass loss and photoevaporation would further reduce the amount of H$_{2}$, ultimately leaving behind an evaporated core \citep{luger2015habitable} that would be inconsistent with current mass and radius constraints for TOI-1266c. These arguments effectively rule out a H$_{2}$-dominated state, although it is important to note that more complex models of the mass and radius evolution of volatile-rich planets still suggest a peak for planets with $\sim$1\% H$_{2}$/He atmospheres \citep{chen2016evolutionary}. Additionally, secondary loss processes like ion pickup could further modify the atmosphere \citep[see][for an overview]{gronoff2020atmospheric}, although the stellar wind for TOI-1266 is only modest at present ($\lesssim 4 \times 10^{-14}$ M$_{\odot}$/yr, based on rotation constraints) \citep{johnstone2015stellar}.

We can combine this with the limited information we have about the solar system giants. Uranus and Neptune have interior `high-metallicity' (elements heavier than He) components ranging from 75-90\% of the their total mass, depending on whether silicates and/or ices are assumed to be the 'metal' \citep[e.g.,][]{hubbard1981interiors, helled2010interior}. This is consistent with other estimates regarding the internal composition of sub-Neptune-sized exoplanets \citep[e.g.,][]{wolfgang2015rocky}. This translates to roughly 11-13 and 13-15 M$_{\oplus}$ for Uranus and Neptune, respectively \citep{helled2010interior, dodson2010formation}, larger than the 2-$\sigma$ upper bound on TOI-1266c's mass estimate. The ice component \citep[thought to be the majority of the core;][]{hubbard1981interiors} is expected to be of supersolar metallicity \citep{lodders1994origin}, with Neptune having a higher oxygen-to-hydrogen fraction (corresponding to a higher water ice fraction). This \replaced{w}{c}ould also push the C/O ratio to higher values, approaching $\sim$1 \citep[][see also the review by \citeauthor{mousis2020key}, \citeyear{mousis2020key}]{ali2014measured}\replaced{, although s}{S}everal studies have motivated C/O ratios closer to 0.5 and a modest amount of ammonia ice \citep{nettelmann2016uranus}, which would be physically consistent with protoplanetary material that experienced full clathration \citep{mousis2020key}. This is further complicated by the fact that assumptions about atmospheric structure and what type of adiabat the temperature profile follows (ranging from dry to wet) shift the retrieved atmospheric metallicity by a factor of a few \citep{cavalie2017thermochemistry}.

Taken together, there is a compelling case to assume that TOI-1266c is volatile-rich, and may even be an eroded sub-Neptune core. \added{We take CO$_{2}$ as the major carbon-bearing species, as the CO/CH$_{4}$ transition and the CO$_{2}$ fraction peak at intermediate temperatures and pressures for higher-metallicity atmospheres \citep{lodders2002atmospheric, venot2014atmospheric}; likewise, N$_{2}$ dominates over NH$_{3}$ \citep{burrows1999chemical}.} \deleted{We can begin exploring the potential states of the planet by focusing on the H-C-O chemistry of three families of scenarios, all dominated by water: H$_{2}$+H$_{2}$O, H$_{2}$+CO$_{2}$+H$_{2}$O, and O$_{2}$+H$_{2}$O. This encompasses two of the potential intermediate states (where H$_{2}$ is still present, or when water has been lost and O$_{2}$ is accumulating) as well as a hypothetical super-Venus transitional state, in which CO$_{2}$ begins to represent a significant portion of the envelope mass. We can further motivate the choice of CO$_{2}$ as the major carbon-bearing species with work suggesting that the CO/CH$_{4}$ transition and the CO$_{2}$ fraction peak at intermediate temperatures and pressures for higher-metallicity atmospheres \citep{lodders2002atmospheric, venot2014atmospheric}; likewise, N$_{2}$ dominates over NH$_{3}$ \citep{burrows1999chemical}. In this paper, we focus on steam atmospheres, omitting two classes of atmospheres (Venus-like and sub-Neptune) for the sake of brevity. Simulations of exo-Venus atmospheres \citep[][]{schaefer2011atmospheric, lincowski2018evolved, lustig2019detectability} largely resemble Venus' CO$_{2}$-dominated atmosphere at present or with smaller CO$_{2}$ inventories earlier in its history \citep[e.g.][]{way2020venusian}. For exo-Venuses in particular, more exploration is warranted to cover the expected diversity of planetary conditions and composition \citep[see the review by][]{madhusudhan2016exoplanetary}. A number of studies on the diversity of sub-Neptune atmospheres \citep[e.g.,][]{lavvas2019photochemical, chouqar2020properties} and retrievals \citep[e.g.,][]{benneke2019sub, mikal2020transmission} can be found in the literature; see also the review by \citet{bean2021nature}.}

\section{Methods \label{sec:methods}}

We use a one-dimensional radiative-convective cloud-free model from \cite{Kopp2013, Kopp2014}, which was updated from the original version \citep{kasting1988runaway,kasting1993} with new H$_{2}$O and CO$_{2}$ absorption coefficients. We employ inverse climate calculations in which the vertical temperature profile is specified, and radiative fluxes from the planet are back-calculated to determine the equivalent incident stellar flux. The atmosphere is divided into 101 layers. The model uses a moist pseudoadiabat extending from the ``surface'' (assumed to be at 100 bar) up to an isothermal stratosphere of 200 K. The surface temperature is varied until the effective solar flux ($S_{\text{eff}}$) matches the observed incident flux on the planet. S$_{\text{eff}}$ is calculated from the ratio between the net outgoing IR flux (F$_{\text{IR}}$) and the net incident solar flux (F$_{\text{sol}}$), both evaluated at the top of the atmosphere. Essentially, by changing the surface temperature to match the incident stellar flux in our model, we are making sure that energy balance is maintained. The model top pressure is set to 10\replaced{$\mu$ bar}{ $\mu$bar}. Short-wave and long-wave fluxes are calculated using a $\delta$-2-stream approximation \citep{Toon1989} using separate eight-term, correlated-k coefficients for H$_{2}$O. 

We also use a one-dimensional photochemical model that is a fork of \texttt{Atmos}\footnote{\href{ https://github.com/VirtualPlanetaryLaboratory/atmos}{\texttt{Atmos} on GitHub}} \citep{arney2017pale} with a modified version of the C-H-O photochemical network from \texttt{VULCAN}\footnote{\href{ https://github.com/github.com/shami-EEG/VULCAN2}{\texttt{VULCAN} on GitHub}} \citep[][see Appendix A]{tsai2017vulcan}. We deliberately set aside nitrogen chemistry for this study because of \deleted{both} the additional complexity necessary to include it in our model, \replaced{as well as}{and because of} the uncertainties associated with speciation (we will return to this briefly in Section \ref{subsec:disc-chemical}). Initial tests with a 0-dimensional chemical equilibrium model suggest that in the relatively oxidizing water-dominated scenarios we explore here, nitrogen is largely present as N$_{2}$, which would contribute to a higher mean molecular weight for the atmosphere but have few other practical impacts. \added{The atmosphere is assumed to be well-mixed below the `surface' at 100 bars. }The model also includes newly-measured water vapor photolysis cross sections \citep{ranjan2020photochemistry}. We use the ultraviolet through near-infrared spectra for GJ 581 \citep{france2016muscles, youngblood2016muscles, loyd2016muscles} as a proxy for TOI-1266, as they have comparable effective temperatures, luminosities, and ages, within uncertainties \citep{selsis2007habitable, dragomir2012search, gaia2018gaia}. GJ 581 is technically a variable star, but its brightness variations are \replaced{relatively small}{$<$1\%} \citep{dragomir2012search}.

\replaced{We}{In our photochemical simulations, we} ensure that the total mixing ratio is unity by using He as the remainder of the atmosphere. \added{The amount of He added is generally $\lesssim$1\% by volume.} As an aside, He abundances could be reduced by drag-off if escape fluxes are high, but we find using another gas (such as Ar) for this purpose has no qualitative impact on our photochemical results. We have also chosen a vertical eddy diffusion parameter K$_{\text{zz}} = 10^{10}$ cm$^{2}$/s consistent with other preliminary studies of hot Jupiters \citep[e.g.,][]{venot2014atmospheric}, noting that we have no constraints on the internal heat flux, rotation rate, and magnetic field strengths in order to constrain this value \citep[e.g.,][and references therein]{visscher2010deep}. We explore the effect of differing K$_{\text{zz}}$ later, but to first order, lower values of K$_{\text{zz}}$ decrease the vertical extent of the well-mixed region of the atmosphere, but do not significantly impact the results described below.

\added{Given TOI-1266c's current equilibrium temperature (see \tableref{tab:planet_params}), it is unlikely that there is a substantial supercritical water component, and instead, TOI-1266c is `liquid' water-dominated \citep{brugger2017constraints, mousis2020irradiated}. } While the lower atmospheres in all of the cases we present here are above the critical temperature of water, the upper atmosphere passes through the temperature range where water would normally condense. We include in our calculations an updated H$_{2}$O saturation vapor pressure over water and ice \citep{meyer1983asme, haar1984steam}, and moderate the condensation loss frequency to ensure that the atmosphere is not substantially supersaturated where liquid water can condense ($\geq$233 K), and below this the condensation over ice is allowed to decrease, reflecting higher possible supersaturations \citep{wallace2006atmospheric, korolev2003supersaturation}, consistent with observations of cirrus clouds on Earth \citep[e.g.,][]{kramer2009ice}. This is in keeping with other studies of temperate water-dominated atmospheres \citep[e.g.,][]{piette2020temperature}. We do not, however, include either aerosols to serve as cloud condensation nuclei nor the necessary microphysical models to capture cloud formation processes, nor the radiative effects of clouds, and caution that the estimated cloud properties are solely illustrative. We return to this later in the Section \ref{sec:discussion}.

\begin{figure*}[ht]
    \centering
    \includegraphics[width=\textwidth]{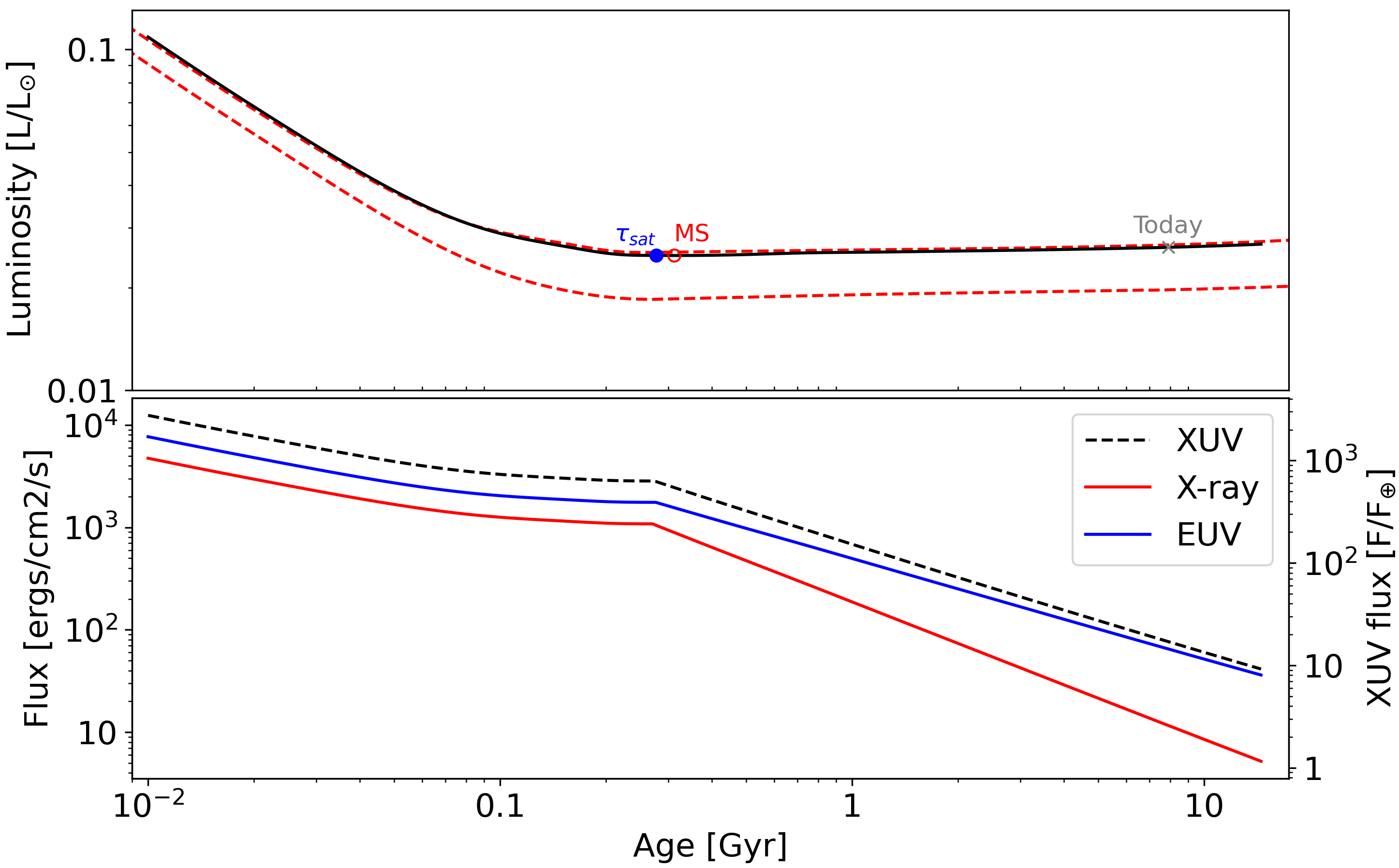}
    \caption{Top panel: The notional luminosity evolution of TOI-1266\added{ (black curve)}. The two \added{red dashed }curves denote the 0.4 and 0.45 M$_{\odot}$ stellar evolution tracks of \citet{hidalgo2018updated}, which have been interpolated and normalized to match the observed luminosity at the current age estimate (denoted with the gray X and labeled `Today') for TOI-1266. The XUV saturation time is denoted as $\tau_{\text{sat}}$, and the start of the main sequence with `MS'. Bottom panel: The X-ray (red), EUV (blue), and total XUV (black dashed) reflect the combination of both the evolving luminosity (top panel) and the saturation parameterizations from \citet{peacock2020hazmat}. For reference, the XUV flux is shown on the right-hand axis as normalized to what the Earth receives ($\sim$4.5 erg/cm$^{2}$/s).}
    \label{fig:luminosity-uv-flux}
\end{figure*}

One additional constraint on the composition of the atmosphere is the cumulative impact of \added{historic }X-ray ($X$) and extreme ultraviolet ($EUV$) radiation-driven mass loss (the sum is represented as $XUV$). To do this, we have developed a simple model of atmospheric loss\footnote{\href{https://github.com/sonny-harman/snowball}{GitHub repository for \textsc{Snowball}}}\added{ separate from the other two models}. The \added{escape }model interpolates the BaSTI luminosity evolution grid\footnote{\href{http://basti-iac.oa-abruzzo.inaf.it/}{http://basti-iac.oa-abruzzo.inaf.it/}} of \citet{hidalgo2018updated} to the observed mass and luminosity of the host star (see \figureref{fig:luminosity-uv-flux}, top panel), similar to \citet{barnes2020vplanet}. The stellar luminosity evolution of \citeauthor{hidalgo2018updated} agrees with other stellar evolution models \citep{baraffe2015new} from $\sim$0.01-10 Gyr (the interval for the \citeauthor{baraffe2015new} grid), and include time points back to 0.01 Myr and out to the end of the main sequence, even if this is beyond the age of the universe. Given the large uncertainty in the age of TOI-1266, this larger stellar age range allows for a more complete uncertainty analysis. We then use the X-ray and EUV scaling relationships from \citet{peacock2020hazmat} (see Fig. \ref{fig:luminosity-uv-flux}, bottom panel), rather than the empirical scaling with respect to the bolometric luminosity (L$_{\text{bol}}$) from \citet{sanz2011estimation}, for two reasons. First, the EUV luminosity (L$_{\text{EUV}}$) from \citeauthor{sanz2011estimation} is above 1\% of L$_{\text{bol}}$ for $\sim$0.2 Gyr, and above 0.1\% for over 1 Gyr, due to the lack of an EUV saturation threshold for younger stars. The second issue with using the parameterizations of \citeauthor{sanz2011estimation} is that for a dimmer star like TOI-1266, there is a discontinuity in the calculated X-ray luminosity (L$_{\text{X}}$) saturation timescale ($\tau_{i}\sim$0.33 Gyr) using their Equation 5, such that the post-saturation L$_{\text{X}}$ is briefly higher than it is in the saturated regime. During our initial tests, we chose to empirically set the saturation timescale to when the time-dependent X-ray luminosity falls below the saturated value, resulting in $\tau_{i}\sim$0.47 Gyr, which produced a negligible change in the total mass lost. We instead chose to use the X-ray and EUV scaling relationships from \citet{peacock2020hazmat} in order to address these two discrepancies. \citeauthor{peacock2020hazmat} note a saturation of $\sim$10$^{-2}$ F$_{\text{EUV}}$/F$_{\text{bol}}$ in their simulations of $\sim$0.4 M$_{\odot}$ stars, slightly higher in magnitude but qualitatively consistent with work for larger stars \citep[e.g.,][]{king2020euv}. Additionally, the log-linear dependence on age after the saturated period is comparable with other early M dwarf studies \citep[e.g.,][]{stelzer2013uv}. These \added{specific values from \citet{peacock2020hazmat} }are converted from flux ratios to luminosity ratios using the luminosity, distance, and 2MASS J-band magnitude listed in \citet{stefansson2020}, assuming that the J-band fluxes are proportional to the bolometric luminosity\footnote{Changes in stellar effective temperature are $\lesssim$10\% for 0.4-0.5 M$_{\odot}$ stars over their lifetimes \citep{baraffe2015new, hidalgo2018updated}, which would shift the wavelength peak by <50 nm.}.

Our escape model also takes advantage of the parameterizations available to distinguish between the radiation/recombination-, energy-, photon-, and diffusion-limited escape regimes \citep{murray2009atmospheric,owen2016uv, lopez2017born}. This is not strictly necessary, \replaced{given the intermediate XUV fluxes experienced by TOI-1266c,}{as the XUV fluxes at TOI-1266c are much less than those experienced by hot Jupiters thought to be undergoing radiation/recombination-limited escape. \citep{murray2009atmospheric}.} \replaced{but}{However,} depending on our choices for the mass loss efficiency (a.k.a., the heating efficiency), atmospheric composition, and XUV luminosity saturation, the atmosphere transitions \replaced{into these different escape regimes}{between the energy-, photon-, and diffusion-limited escape regimes} at different times. 

We also assume that our escape calculations are largely insensitive to exospheric temperature, except for across the critical XUV flux identified for hot Jupiters \citep{koskinen2007stability}. This is motivated by the interesting coincidence of the critical XUV flux necessary to drag off atomic oxygen from a terrestrial planet's atmosphere \citep[$\sim$40 times the XUV flux received by Earth today;][]{luger2015extreme} and the XUV flux at which H$_{3}^{+}$ cooling becomes ineffective at moderating thermospheric temperatures for gas giants \citep{koskinen2007stability}. For gas giants, this transition produces an order of magnitude increase in the thermospheric temperature and atmospheric scale heights \citep[][their Fig. 1a]{koskinen2007stability}. Temperature changes are accurately assessed as a secondary effect in the critical XUV flux relationship for planets in their host star's main sequence habitable zone shown by \citet{luger2015extreme}, since the critical XUV flux goes as T$^{-1/4}$. However, including a 10$\times$ change in temperature would cause the critical XUV flux to decrease by more than a factor of two (that is, the onset of oxygen drag-off would occur at lower fluxes). For habitable zone planets like those studied by \citeauthor{luger2015extreme} and \citet{ramirez2014habitable}, this assumption does not introduce significant errors over the $\lesssim$1 Gyr that these planets spend enduring the superluminous phase of their host stars. As we focus on water-dominated scenarios, the prevalence of molecular hydrogen and oxygen in the thermosphere suggests a closer resemblance to the upper atmosphere of Earth or Jupiter ($\sim$1,000-2,000 K) than the CO$_{2}$-dominated atmospheres of Venus and Mars \citep[$\sim$200-300 K;][]{mueller2008neutral}, but under high instellation, even CO$_{2}$-dominated thermospheres are $\gtrsim$ 10,000 K \citep{tian2009thermal}. This, combined with the potential for Lyman-$\alpha$ cooling at high XUV fluxes \citep[e.g.,][]{murray2009atmospheric}, suggests that above $\sim$180 erg cm$^{-2}$ s$^{-1}$ the exospheric temperature is $\sim$10,000 K, and $\sim$1,000 K below this flux threshold. Since TOI-1266c receives >180 erg cm$^{-2}$ s$^{-1}$ for nearly 3.5 Gyr, we find that escape of atomic oxygen continues for 2 Gyr longer than if we were to adopt the critical XUV flux suggested by \citeauthor{luger2015extreme} of $\sim$400 erg cm$^{-2}$ s$^{-1}$ for a planet with TOI-1266c's current mass and radius. This results in lower potential for accumulated oxygen abundances in nearly every scenario for TOI-1266c\added{, consistent with prior work that demonstrated oxygen accumulation and escape self-consistently for XUV fluxes 10-100 times larger than what the Earth receives today \citep[\figureref{fig:luminosity-uv-flux}; e.g.,][]{zahnle1986mass}}.

We assume mass loss efficiencies ($\eta_{XUV}$) in line with other authors, including $\eta_{XUV}\sim$0.1-0.15 for X-ray-dominated H$_{2}$ escape for a planet of comparable size to TOI-1266c \citep{owen2012planetary, bolmont2017water}, and $\eta_{XUV}\sim$0.01 for H$_{2}$O following \citet{lopez2017born}, based on protoplanetary disk photoevaporation studies \citep{ercolano2010metallicity}. These are meant only as order-of-magnitude approximations, since the efficiency is dependent on planetary mass, radius, and envelope composition and its radiative properties, as well as the flux of high-energy radiation from its host star, and as such will evolve \citep[e.g.,][]{murray2009atmospheric, owen2013kepler}. A planned next step is to use the flux-dependent efficiencies of \citet{bolmont2017water}, noting that there is still some uncertainty when comparing these to efficiencies for close-in giant planets \citep[e.g.,][]{koskinen2014thermal}.

Because of the inherent uncertainties associated with almost every aspect of the atmospheric escape as well as the planet's mass and composition, we employ a Monte Carlo approach and perform a suite of escape simulations over the range of parameter uncertainties set out in \tableref{tab:mcranges}. \added{The atmospheric composition values are drawn from a log-uniform distribution, whereas all other values are drawn from linear uniform ranges. }Values are \replaced{selected}{generated} using the Latin Hypercube sampling (LHS) method in the \href{https://smt.readthedocs.io/}{Surrogate Modeling Toolbox} \citep{bouhlel2019python}, which leverages the Enhanced Stochastic Evolutionary algorithm \citep{jin2003efficient} to optimize the \href{http://forge.scilab.org/index.php/p/scidoe/}{Design of Experiments Toolbox} (pyDOE) implementation. One important caveat is that this set of simulations assumes the maximum amount of water available for a given mass and radius, following the relationships \replaced{given}{derived} by \citet{noack2016water}. \added{We note that the mass-radius relationships from \citet{noack2016water} do not include an atmosphere, but in our simulations the vertical extent of the region between 1 bar and $\sim$10 mbars is $\sim$400-600 km, which would change the apparent radius by less than the reported uncertainty. }\replaced{This in turn highlights}{The \citeauthor{noack2016water} mass-radius relationship places} a physically-motivated lower limit for the planet's mass of $\sim$1.6 M$_{\oplus}$ from the lower bound on the planet's radius, where the planet would be 100\% water. If future observational constraints on the planet's mass are below this threshold, the planet must have a non-negligible amount of H$_{2}$ at present. More complex compositional mixes are beyond the scope of the present work, but abundant H$_{2}$ in TOI-1266c's atmosphere would most likely eliminate the possibility of oxygen accumulation from hydrogen loss, as well as posing an interesting conundrum for the formation and evolution mechanisms highlighted above that would remove an H$_{2}$-dominated atmosphere.

\begin{table}[htb]
    \centering
    \begin{tabular}{l|c|c}
    Property [units] & Default Value & Tested Range \\ \hline \hline
    Stellar age [Gyr] & 7.9 Gyr & 2.7--12.1 \\
    Current luminosity [L$_{\odot}$] & 0.02629 & 0.02554--0.027 \\ \hline
    Planet mass [M$_{\oplus}$] & 1.9 & 1.6--6.4 \\
    Planet radius [R$_{\oplus}$] & 1.673 & 1.563--1.76 \\
    Atm. mass [M$_{\oplus}$] & --- & \emph{see note [1]} \\ \hline
    Atm. composition [vmr] & & \\
    \multicolumn{1}{c|}{H$_{2}$} & --- & 10$^{-6}$--1 \\
    \multicolumn{1}{c|}{ He} & --- & 10$^{-6}$--1 \emph{[2]} \\
    \multicolumn{1}{c|}{ H$_{2}$O} & --- & 10$^{-6}$--1 \\
    \multicolumn{1}{c|}{ CO$_{2}$} & --- & 10$^{-6}$--1 \\
    Escape efficiency & & \\
    \multicolumn{1}{c|}{ H$_{2}$} & 0.1 & 0.01--0.4 \\
    \multicolumn{1}{c|}{ H$_{2}$O} & 0.01 & 0.01--0.4 \\
    \multicolumn{1}{c|}{ CO$_{2}$, O$_{2}$} & 0.01 & 0.01--0.1 \\
    \end{tabular}
    \caption{Monte Carlo test ranges for \replaced{variables}{for the variables in the atmospheric escape simulations}. \emph{[1]} The planetary volatile abundance was set to the maximum water abundance allowed for the selected mass and radius given by \citet{noack2016water}. \emph{[2]} The helium abundance \added{in the escape simulations} was set by scaling the solar He:H$_{2}$ ratio (0.3367) by the value drawn from the stated range, given the selected H$_{2}$ abundance. The atmospheric composition was normalized as a final step. \label{tab:mcranges}}
\end{table}

Lastly, we use the Planetary Spectrum Generator\footnote{https://psg.gsfc.nasa.gov/index.php} \citep[PSG;][]{psg:2018} to produce synthetic transmission spectra for the scenarios outlined here. PSG is an online radiative transfer suite that integrates the latest radiative transfer methods and spectroscopic parameterizations, and includes a realistic treatment of multiple scattering in layer-by-layer spherical geometry. It can synthesize planetary spectra (atmospheres and surfaces) for a broad range of wavelengths for any given observatory. We validate these results with {\fontfamily{pcr}\selectfont PandExo} \citep{batalha2017pandexo}.

\section{Results \label{sec:results}}

\subsection{Atmospheric Escape \label{results:escape}} 

\begin{figure*}[ht]
    \centering
    \includegraphics[width=\textwidth]{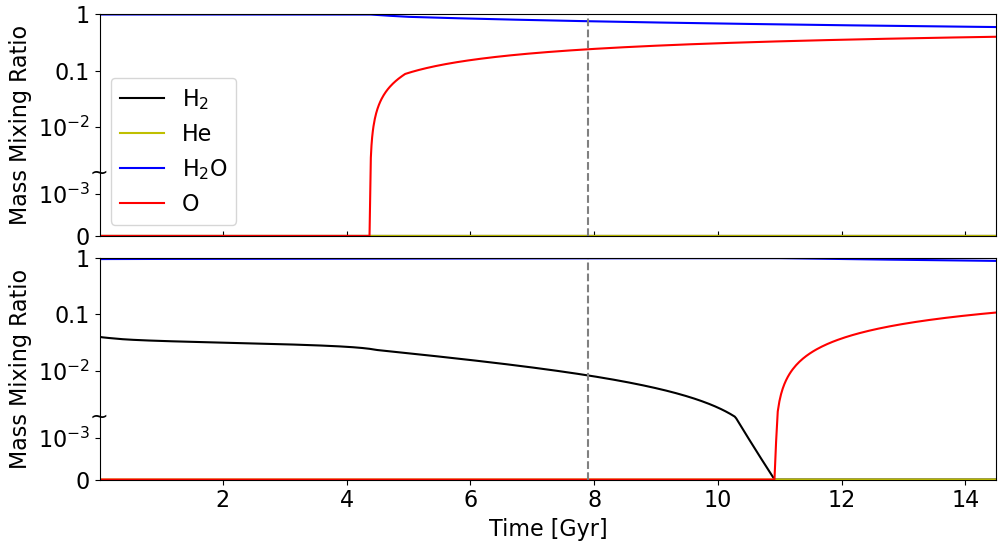}
    \caption{An example of one planetary evolution simulation, starting at $\sim$2 M$_{\oplus}$ and $\sim$1.7 R$_{\oplus}$, which ultimately evolves to match TOI-1266c's observed mass and radius at TOI-1266's estimated age (vertical dashed line). Top panel: 10\% H$_{2}$O by mass; bottom panel: 9.6\% H$_{2}$O and 0.4\% H$_{2}$ by mass. Note that the vertical axis is logarithmic above the tilde and linear below it.}
    \label{fig:atm_escape}
\end{figure*}

We begin by estimating the atmospheric lifetime for TOI-1266c, assuming the observed mass and radius for the present day (\figureref{fig:atm_escape}). A pure-water atmosphere experiences substantial water loss over its lifetime, as can be seen in the top panel of Fig. \ref{fig:atm_escape}, but still retains abundant H$_{2}$O through $\sim$8 Gyr (vertical dashed line). That said, the \replaced{atmosphere}{volatile inventory} is roughly one-third oxygen by the present day\added{ (assuming the oxygen is not absorbed by the planet's interior)}, \replaced{and would have}{resulting in} spectroscopically-detectable oxygen (we will return to observations later). A second test, which includes a modest amount of hydrogen, can be seen in the bottom panel of Fig. \ref{fig:atm_escape}. This scenario demonstrates that a relatively minor amount of H$_{2}$ ($\sim$0.4\% of the planet's initial mass) can prevent significant loss of water and the commensurate accumulation of O$_{2}$, as the H$_{2}$ combines with any free oxygen to replenish H$_{2}$O\added{ (this is an explicit prescription in our model)}. This amount of H$_{2}$ is broadly consistent with what might remain following atmospheric boil-off \citep{owen2016atmospheres}. In both of these cases, the mass of the total volatile inventory does not substantially change throughout the planet's life, and in total the mass changes by $\sim$1\% and the radius by $\sim$0.3-0.5\% over this same interval. However, it is important to note that this is equivalent to losing $\sim$100 Earth oceans, substantially \replaced{over}{more than} the initial water reservoirs explored in other work \citep[e.g.,][]{luger2015extreme}. These evolutionary tracks are useful for illustrating the behavior of individual scenarios near the boundaries between regimes, but given the large uncertainties in some critical parameters, it is important to fully explore the impact of atmospheric escape on the present state of TOI-1266c.

\begin{figure*}[ht]
    \centering
    \includegraphics[width=\textwidth]{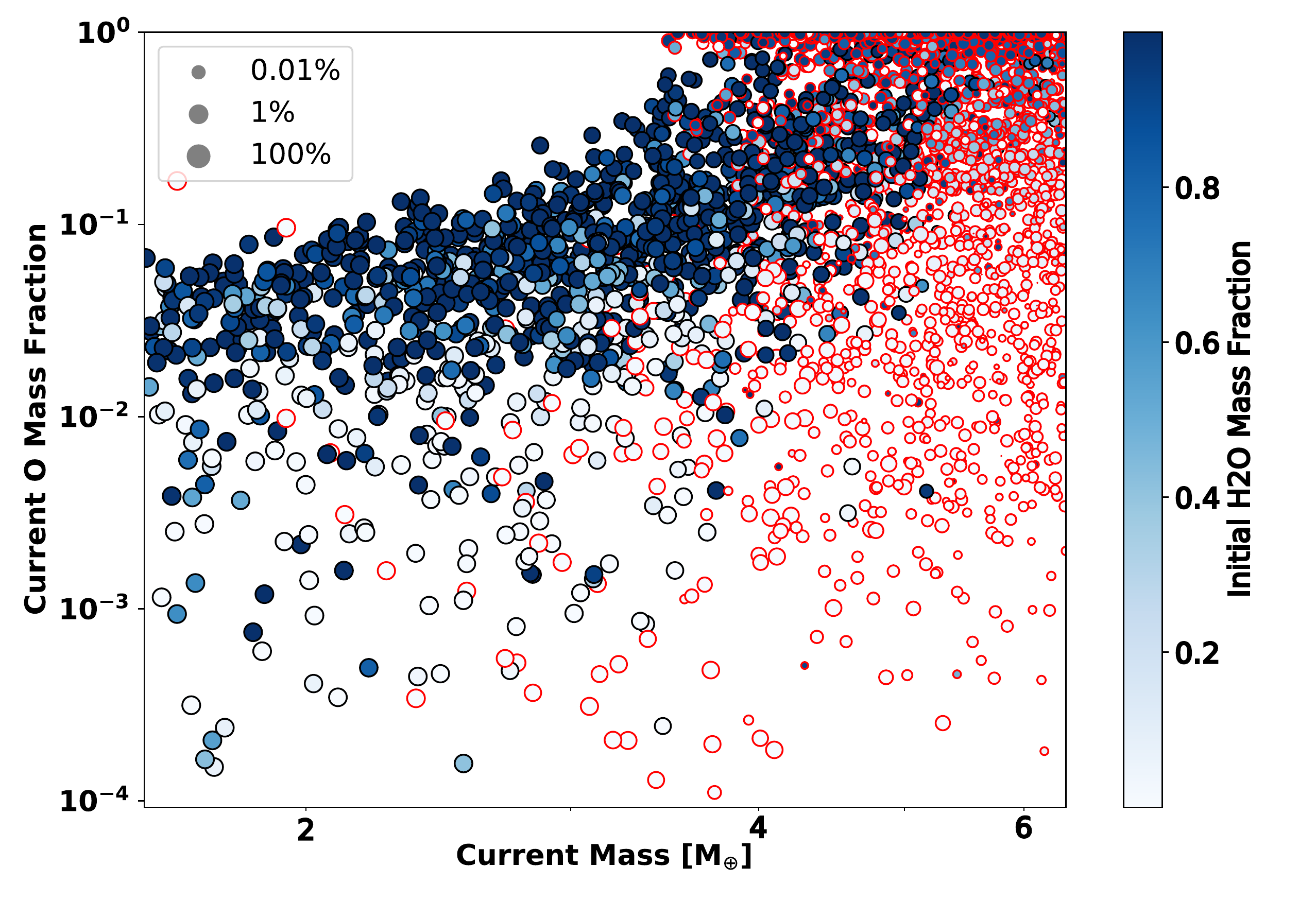}
    \caption{Water loss ensemble results calculated based on the range of planet parameter uncertainties from Table \ref{tab:mcranges}, plotted as a function of the current planet mass and the current envelope mass fraction of \added{free }oxygen. Symbols are colored by the initial envelope mass fraction of H$_{2}$O, and the size is scaled based on the fraction of the planet initially designated as envelope (taken to be the maximum water abundance permitted by the mass-radius relationship of \citet{noack2016water}). A red edge color for a given symbol denotes a scenario with less than 1 Earth ocean at present day, which would likely be short-lived \citep{kasting1983loss}.}
    \label{fig:atm_escapeMC}
\end{figure*}

As such, we \replaced{include}{ran} a suite of 10,000 atmospheric escape simulations covering the stated uncertainty ranges in Table \ref{tab:mcranges}. Several common-sense interpretations of this initial exploration can be gleaned from Fig. \ref{fig:atm_escapeMC}, namely 1) smaller initial planet masses for the same planetary radius correspond to larger potential water (volatile) inventories (denoted by the size of the points), which is a natural consequence of our experimental design; 2) larger volatile inventories are more difficult to lose completely, and suppress substantial accumulated oxygen mass fractions; and 3) it is unlikely that a planet \replaced{above}{more massive than} $\sim$3.5 M$_{\oplus}$ would have any remaining H$_{2}$O because of the small initial volatile inventories, and consequently, could have large oxygen mass fractions. The apparent gulf spanning intermediate oxygen mass fractions from 3.5-6.5 M$_{\oplus}$ reflects complete desiccation of initially hydrogen- and water-dominated states that are pulled up to the 100\% oxygen mass fraction state, barring a few simulations with escape efficiencies at the bottom of the tested range and/or young stellar ages. The remainder of the scenarios have small initial water fractions that do directly correspond to the oxygen mass fraction, but do not group up in the same way. There is also no significant trend with H$_{2}$ escape efficiency (\figureref{fig:H2_eff_test}) for the planet parameters and atmospheric compositions tested here, although this may not be the case for other regions of the parameter space.

\subsection{Initial Temperature/Pressure and Water Profiles\label{subsec:water_profiles}}

The initial water vapor profiles produced by the radiative-convective model were used to initialize the photochemical simulations. All of the following simulations assume the nominal radius and mass for TOI-1266c \citep[1.673 R$_{\oplus}$ and 1.9 M$_{\oplus}$;][]{stefansson2020}, as well as a water-dominated atmosphere. As an aside, the climatological and photochemical water vapor profiles for the same temperature/pressure conditions differ slightly. This is largely due to the combination of photolysis and vertical mixing (via both parameterized advection and molecular diffusion) in the photochemical model that modifies the water profiles in the upper atmosphere (above $\sim$10 mbar) by a factor of a few (\figureref{fig:water_profile_test}). For the atmospheric compositions explored here, this results in transmission spectra uniformly decreased by a few parts per million at all wavelengths between the climatological water profiles and the photochemical water profiles as a result of the change in mean molecular weight (not shown). This may not be the case for every scenario, however, particularly if water is more efficiently segregated to the lower atmosphere (for example, through weaker vertical mixing or efficient scavenging processes).

\begin{figure*}[ht]
    \centering
    \includegraphics[width=\textwidth]{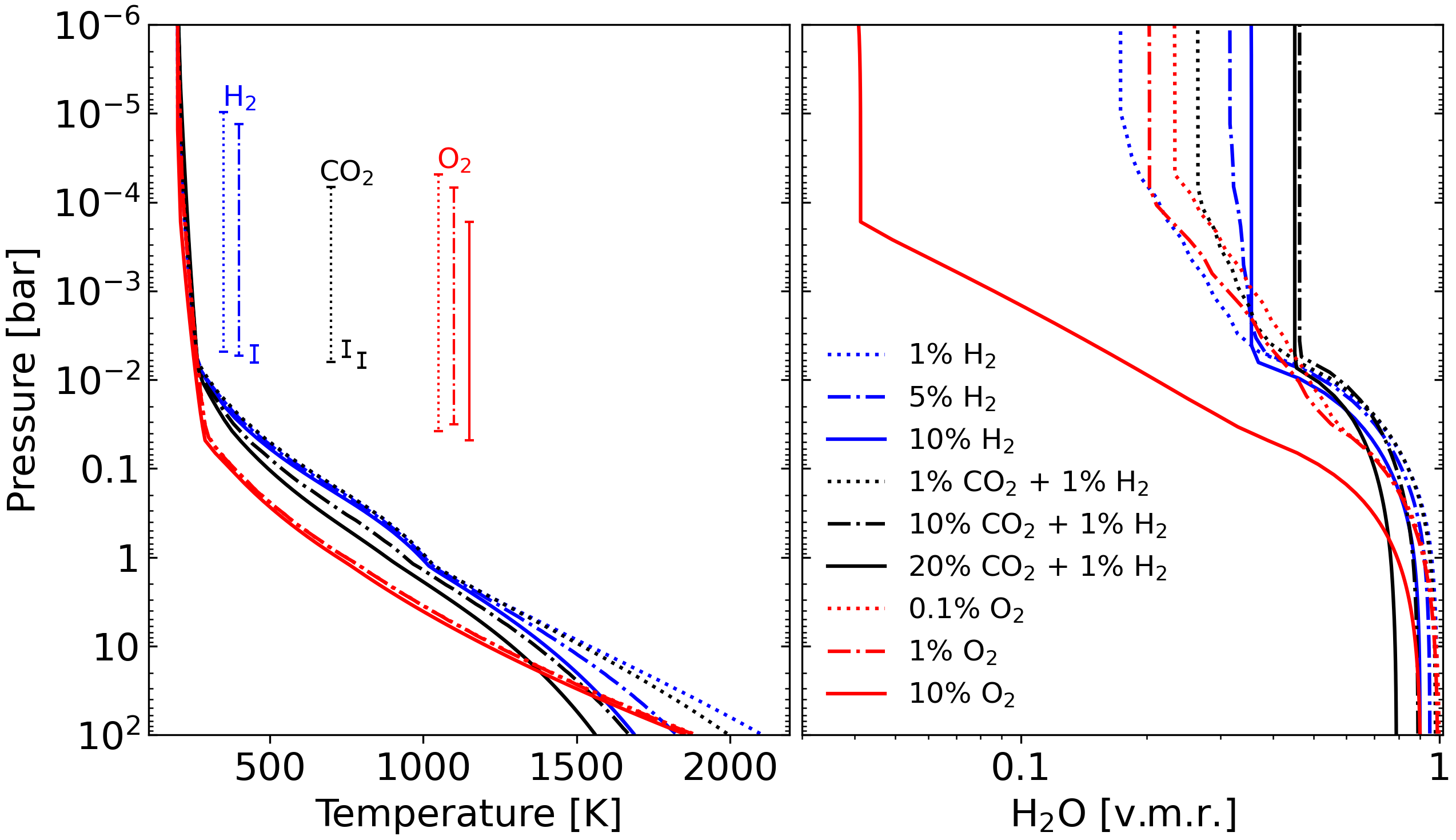}
    \caption{Temperature (left) and water vapor \replaced{(right)}{volume mixing ratio (v.m.r; right)} profiles for the cases outlined in the text. The onset of water vapor condensation (vertical range in the left panel) occurs just above the `knee' in the water profiles, consistently around $\sim$1-10 mbar.}
    \label{fig:temp_water}
\end{figure*}

The three families of atmospheres (H$_{2}$+H$_{2}$O; H$_{2}$+CO$_{2}$+H$_{2}$O; O$_{2}$+H$_{2}$O) have some shared attributes, including the same general pressure ranges for the condensation of water (\figureref{fig:temp_water}, left panel). Of the O$_{2}$-bearing scenarios, only the intermediate-concentration cases (0.1\% and 1\% O$_{2}$) have water vapor profiles with higher upper atmospheric concentrations than the case with the highest water fraction. This is in contrast to both the H$_{2}$ and CO$_{2}$+H$_{2}$ scenarios, which have more saturated upper atmospheres for higher mixing ratios of the diluting species. The CO$_{2}$ scenarios are warmer in the deep atmosphere because of CO$_{2}$'s efficacy as a greenhouse gas, while H$_{2}$ is more effective than O$_{2}$ as a collisional broadening partner, resulting in intermediate temperatures.

\subsection{Atmospheric Chemistry\label{subsec:chem}}

\begin{figure*}[ht]
    \centering
    \includegraphics[width=\textwidth]{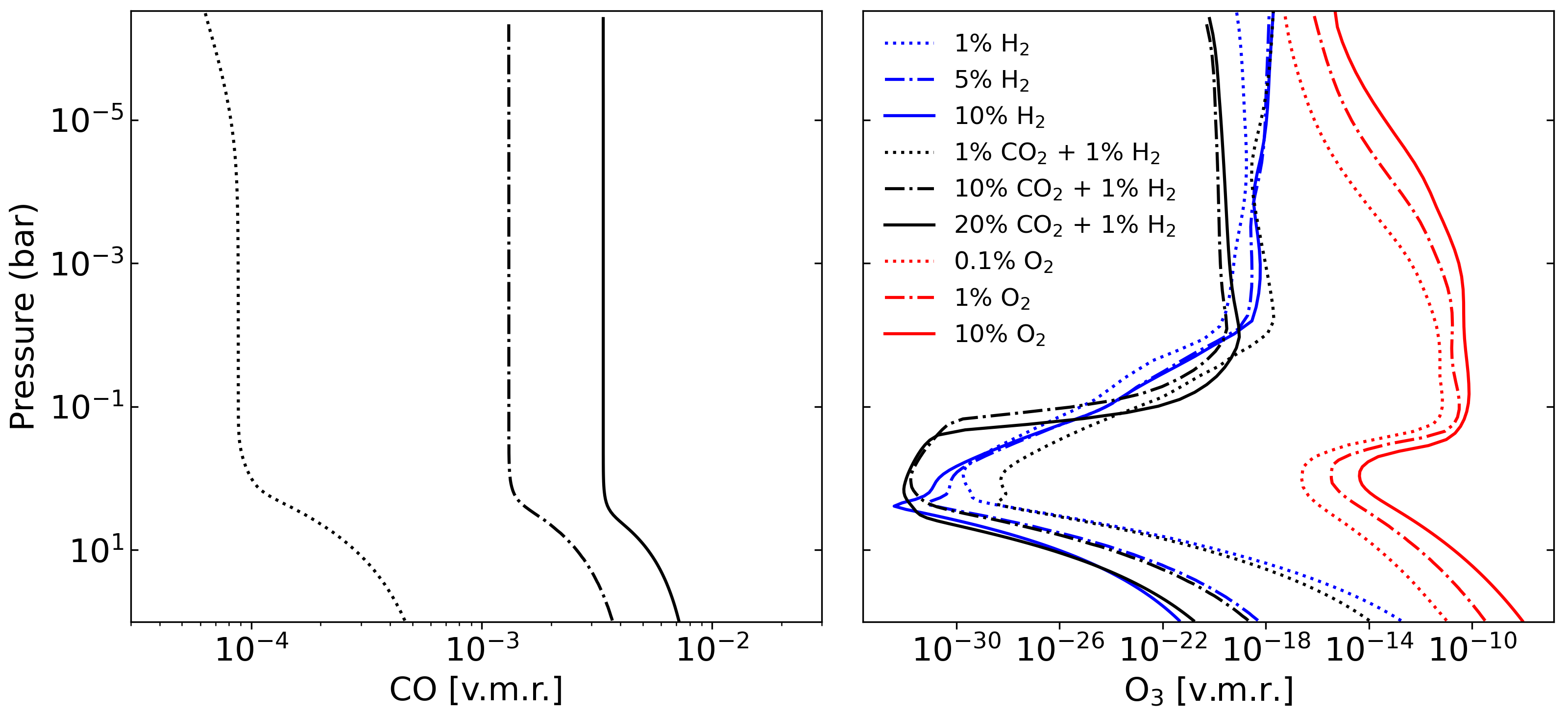}
    \caption{Left panel: CO mixing ratio profiles for the three atmospheres that contain CO$_{2}$ in significant abundance. Right panel: O$_{3}$ mixing ratio profiles for the nine principle scenarios included in this study.}
    \label{fig:comp_CO-O3}
\end{figure*}

Our photochemical modeling is informed by the atmospheric escape and radiative-convective simulations of TOI-1266c, focusing here on an initial exploration limited to H-C-O chemistry (future work will include other species). Each of the three families of water-dominated atmospheres have lesser amounts of H$_{2}$, H$_{2}$+CO$_{2}$, or O$_{2}$ which drive the chemistry of trace species. As we discuss later, many of these changes are not visible in the integrated planetary spectra, but they are integral to accurately capturing the major species. The Appendix has a collection of figures that highlight how each species changes with different major species' concentrations, but we will only focus on those that may be potentially observable (e.g., O$_{2}$, O$_{3}$, H$_{2}$O, CO$_{2}$, and CO). The water-dominated scenarios we focus on here are too oxidizing for substantial amounts of CH$_{4}$, C$_{2}$H$_{6}$, or other reduced carbon compounds, which likely precludes a hydrocarbon haze. Of the species that are likely to be observable, only O$_{3}$ and CO are essentially free to respond to instellation and compositional changes, while O$_{2}$, H$_{2}$O, and CO$_{2}$ are given fixed concentrations at the 100-bar pressure level that are then subject to dynamical and thermo- and photochemical processes. CO (\figureref{fig:comp_CO-O3}, left panel) is largely produced by photolysis of CO$_{2}$ in the upper atmosphere and then mixed downwards into the deeper atmosphere, where the background CO concentration is set by thermochemical reactions.

Ozone, much like CO, is dependent on the concentration of another species (O$_{2}$), and secondary trace species and photolysis reactions that rapidly convert atoms between these two reservoirs. In terrestrial photochemical studies \citep[e.g.][]{segura2003ozone}, the typical threshold to establish a robust O$_{3}$ layer is $\sim$1\% of Earth's present atmospheric level of O$_{2}$ (i.e., $\sim$2\% by volume O$_{2}$). On Earth, the ozone layer is maintained by photochemistry at roughly ppm concentrations between $\sim$0.5-50 mbar. This pressure range is comparable to the scenarios with more abundant O$_{3}$ in \figureref{fig:comp_CO-O3} (right panel), but the mixing ratios are lower by a factor of $\sim$10$^{3}$. The lower concentration of O$_{3}$ for these scenarios is driven by the higher abundance of OH radicals in the upper atmosphere derived from water vapor photolysis (see Appendix A), in line with earlier work that demonstrated a reduction in O$_{3}$ with warmer atmospheres and high OH abundances \citep{chen2019habitability}. Because all of the scenarios explored here have non-negligible H$_{2}$O abundances, increasing the O$_{2}$ abundance beyond 10\% by volume produces a roughly linear increase in the peak O$_{3}$ mixing ratio (not shown), still much less than the maximum ozone mixing ratio in Earth's atmosphere. We will return to remote detectability later.

\begin{figure*}[ht]
    \centering
    \includegraphics[width=0.9\textwidth]{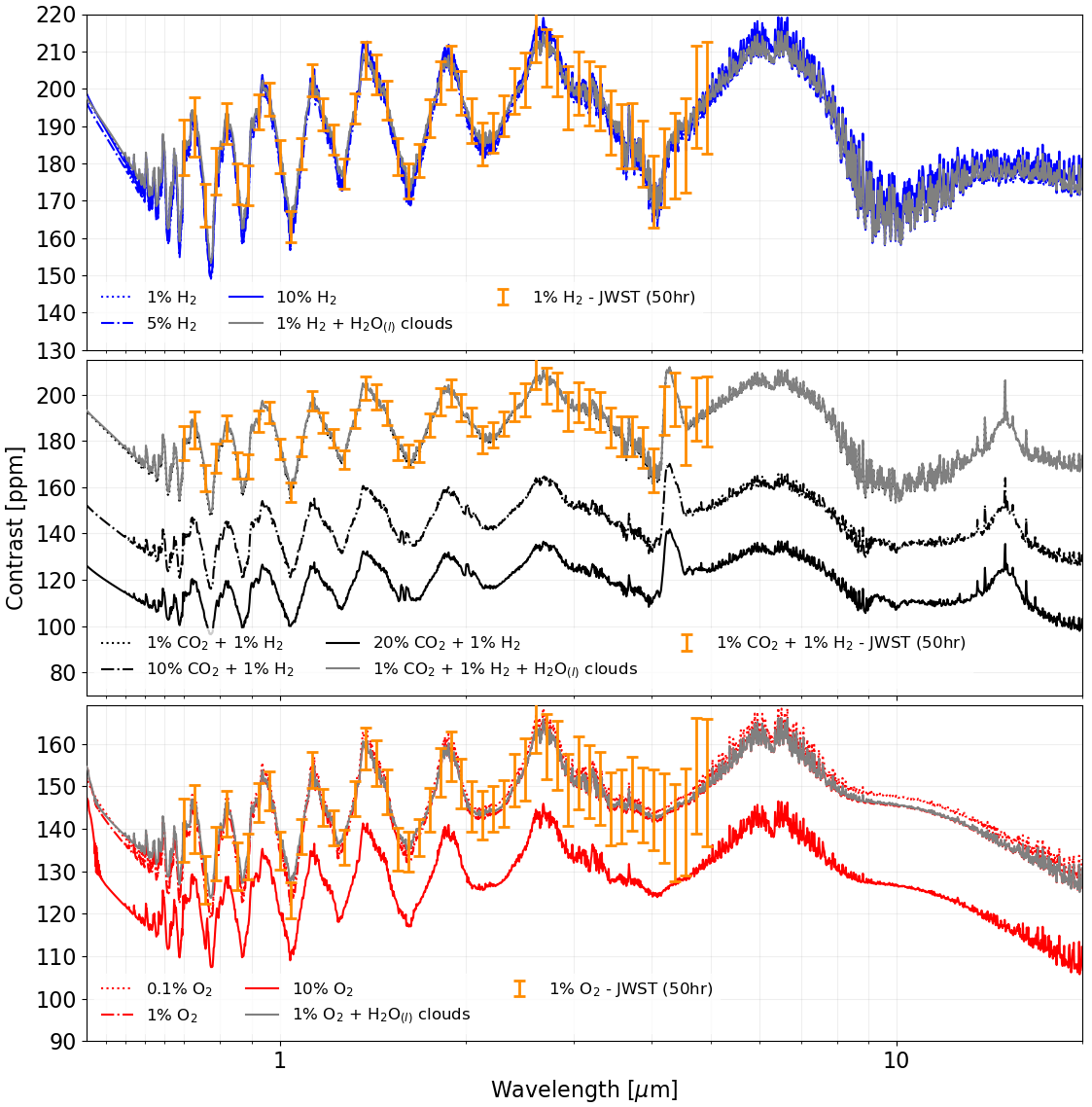}
    \caption{Synthetic spectra (R=500) of the nine cases highlighted previously. The panels are divided into families based on the gas mixture of interest: H$_{2}$+H$_{2}$O (top panel), CO$_{2}$+H$_{2}$+H$_{2}$O (middle panel), and O$_{2}$+H$_{2}$O (bottom panel). A zoomed-in view of the 4.3-$\mu$m CO$_{2}$ feature can be found in \figureref{fig:co2_spectra}, while \figureref{fig:o2_spectra} focuses on key O$_{2}$ and O$_{3}$ features. For each 1\% mixture, we overplot two other spectra. The orange error bars are for a simulated \replaced{observation with 50-hour \emph{JWST} retrieval with NIRSpec-Prism}{observation using JWST's NIRSpec-Prism} (R=100, but plotting only every fourth point). We also overplot a cloudy scenario, assuming 14-$\mu$m liquid water droplets with a volume mixing ratio of 0.1 ppm \added{\citep{kopparapu2021nitrogen} }are distributed throughout the pressure range identified in \figureref{fig:temp_water}.}
    \label{fig:all_spectra}
\end{figure*}

\begin{figure}[ht]
    \centering
    \includegraphics[width=\columnwidth]{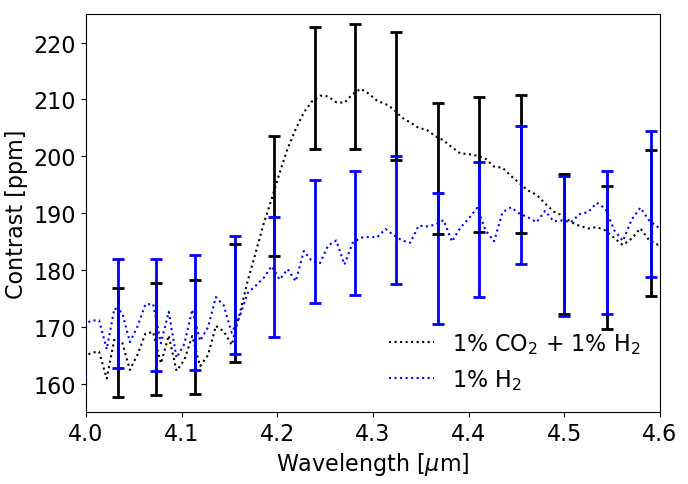}
    \caption{Synthetic spectra (R=500) for just the 1\% CO$_{2}$+H$_{2}$+H$_{2}$O and 1\% O$_{2}$+H$_{2}$O mixtures from Fig. \ref{fig:all_spectra}, focusing on the 4.3-$\mu$m CO$_{2}$ absorption feature. The corresponding error bars are for \replaced{the simulated 50-hour \emph{JWST} retrievals with NIRSpec-Prism}{the simulated 50-hour observation using JWST's NIRSpec-Prism} (R=100), and indicate that these two cases would be distinct from one another with \emph{JWST}.}
    \label{fig:co2_spectra}
\end{figure}

\begin{figure*}[ht]
    \centering
    \includegraphics[width=\textwidth]{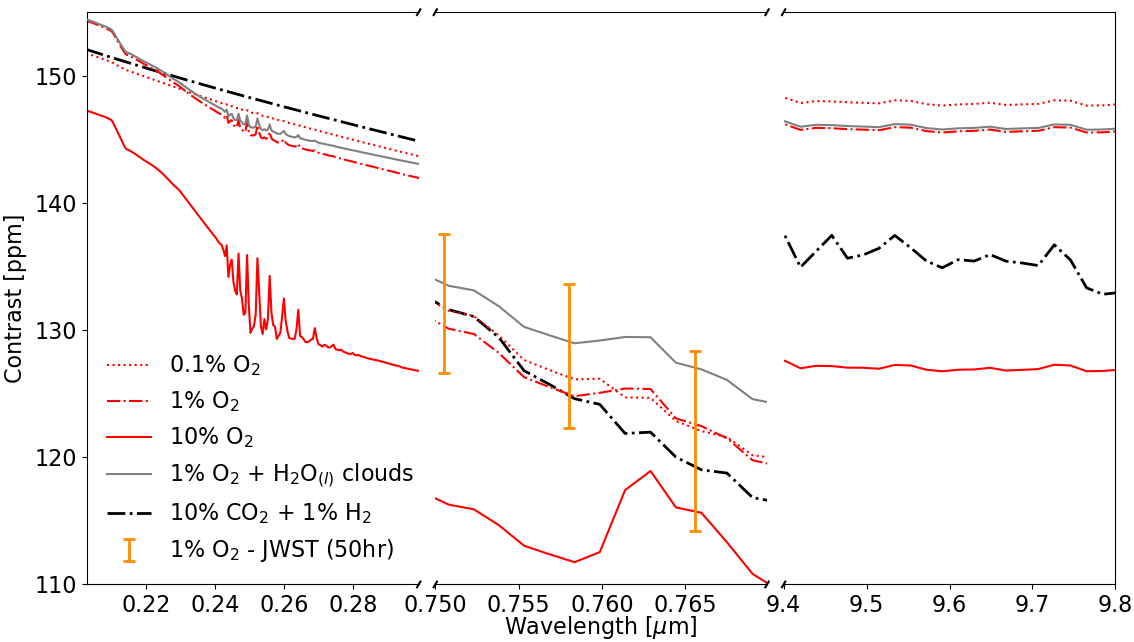}
    \caption{Synthetic spectra (R=500) for just the O$_{2}$+H$_{2}$O mixtures from Fig. \ref{fig:all_spectra}, compared with the 10\% CO$_{2}$ mixture. Note the O$_{2}$ A-band feature at 0.76 $\mu$m increases to $\sim$10 ppm at 10\% O$_{2}$. However, the 0.2-0.3 $\mu$m and 9.6 $\mu$m ozone features are absent due to the lack of a substantial O$_{3}$ column abundance (the 0.2-0.3 $\mu$m region has O$_{2}$ features from the Herzberg continuum, but no contributions from O$_{3}$). Variations between the spectra in this region are due to the increasing atmospheric mean molecular weight at higher O$_{2}$ abundances. The orange error bars are for \replaced{a simulated 50-hour \emph{JWST} retrieval with NIRSpec-Prism}{a simulated observation using JWST's NIRSpec-Prism} (R=100). We also overplot a cloudy scenario, assuming 14-$\mu$m liquid water droplets with a volume mixing ratio of 0.1 ppm \added{\citep{kopparapu2021nitrogen} }are distributed throughout the pressure range identified in \figureref{fig:temp_water}.}
    \label{fig:o2_spectra}
\end{figure*}

\begin{figure}[htb]
    \centering
    \includegraphics[width=\columnwidth]{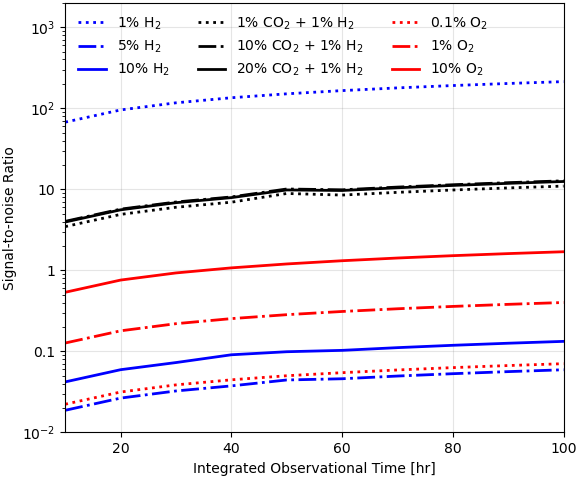}
    \caption{Signal-to-noise ratio for observations with \emph{JWST} at 0.7-5 $\mu$m. The composition is dominated by H$_{2}$O, and has secondary constituents denoted by the label.}
    \label{fig:SNR}
\end{figure}

\section{Discussion \label{sec:discussion}}

The atmospheric escape calculations showcase a number of evolutionary pathways that are in line with other estimates for super-Earths and sub-Neptunes \citep[e.g.][]{estrela2020evolutionary}, and TOI-1266c sits at the nexus of the potential states, although it would constitute a low-instellation terrestrial planet following \citeauthor{estrela2020evolutionary}. One possible outcome is that TOI-1266c was (and remains) a rocky planet composed of predominantly silicates and iron. TOI-1266c would then most resemble a super-Venus \citep[][]{barclay2013super, kane2013potential}, but even among Venus-like planets some variation is expected \citep[e.g.][]{schaefer2011atmospheric, kane2018climate}. Barring the potentially brief steam atmospheres immediately following formation and/or a later transition into the moist and runaway regimes \citep[e.g.][]{hamano2013emergence, driscoll2013divergent, way2016venus}, however, the lack of a substantial volatile inventory results in dry, rocky super-Venuses. On the other end of the compositional spectrum, hydrogen-dominated sub-Neptunes boast larger spectroscopic features requiring fewer transits to obtain sufficient signal-to-noise \citep[e.g.][see also \figureref{fig:SNR}]{chouqar2020properties}. For strongly irradiated objects, however, the impact of atmospheric escape should be considered when estimating atmospheric and bulk composition, much like how we have chosen to consider largely H$_{2}$O-dominated scenarios for TOI-1266c.

\subsection{Chemical considerations \label{subsec:disc-chemical}}

The carbon speciation is dependent on temperature \citep{lodders2002atmospheric}, so while we have used the conjectured water and methane ice fractions from the Uranus and Neptune as a starting point, the equilibrium speciation heavily favors CO$_{2}$ over CH$_{4}$ at the lower boundary. If the planet starts out as more reduced, CO$_{2}$ would shift towards CO and ultimately CH$_{4}$; however, even trace amounts of water vapor are able to rapidly convert photochemically-produced CO back into CO$_{2}$ such that the upper atmosphere would have a smaller abundance of CO than would be predicted solely from thermochemistry. The assumed `surface' pressure also affects the abundances of trace species \citep{yu2021identify}, but we have not tested this explicitly in our simulations. Beyond this, other factors, such as the choice of K$_{\text{zz}}$, can further modify the concentrations of trace species.

We have explored the sensitivity of atmospheric composition to changes in K$_{\text{zz}}$ by decreasing it from our default value of 10$^{10}$ cm$^{2}$/s down to 2$\times$10$^{7}$ cm$^{2}$/s. Below this value, our photochemical model has difficulty converging. This appears to be due to the descent of the homopause (also called the turbopause) into the warmer, denser parts of the atmosphere below the isothermal stratosphere (note the sharp decrease in concentration at the upper boundary in \figureref{fig:O2_edd_test}). Lower K$_{\text{zz}}$ values affect our chemical profiles in much the same way as they affect other models \citep[e.g.][]{visscher2011quenching, venot2014atmospheric, gao2018sedimentation}. Additionally, we find no significant deflection in the location of the water condensation region, which would have a much stronger effect on the \replaced{retrieval}{observations} (e.g., \figureref{fig:all_spectra}, top panel) than the variations in species' concentrations with K$_{\text{zz}}$. Uranus and Neptune have K$_{\text{zz}}$ values closer to $\sim10^{8}$ cm$^{2}$/s\replaced{, noting that s}{\citep{cavalie2017thermochemistry}. S}tronger mixing and/or different temperature profiles can give the appearance of lower metallicities \replaced{\citep{cavalie2017thermochemistry}}{(ibid.)}. \added{Mixing length theory \citep[e.g.,][]{visscher2010deep} suggests that K$_{\text{zz}}\sim 10^{6}$--10$^{8}$ cm$^{2}$/s throughout the model domain if the internal heat flux is 50-50,000 erg/cm$^{2}$/s, comparable to modern Earth and Jupiter, respectively \citep{gando2011partial, pearl1991albedo}.}

For this initial work, we have neglected species that could play an important role in modifying the atmospheric structure and evolution. For example, sulfur chemistry has been shown to significantly modify the thermal profile of hot Jupiters \citep{zahnle2009atmospheric}, while sulfuric acid aerosols have been suggested as an alternative way to form a cold trap \citep{walker1975evolution}, given their hygroscopic tendencies in Venus' modern atmosphere \citep[e.g.,][]{krasnopolsky1994h2o, yung2009evidence, tsang2010correlations}. Additionally, if NH$_{3}$ is a substantial component for ice giant cores \citep[e.g.,][]{nettelmann2016uranus}, then NH$_{3}$ could be present in the gas phase and contribute to the total reducing power available to the atmosphere. Particularly for the scenarios where TOI-1266c loses most of its hydrogen, nitrogen could oxidize into NO$_{x}$ compounds, analogous to NO$_{x}$ derived from persistent lightning storms \citep[e.g.,][]{ardaseva2017lightning}. However, the temperature profiles used here all lie above the N$_{2}$/NH$_{3}$ equal-abundance pressure-temperature curve \citep[e.g.][]{fortney2020hot}, suggesting that \deleted{any} ammonia \replaced{in the planet's ice component}{incorporated as ice} may \deleted{only} affect the total atmospheric pressure (as N$_{2}$) \replaced{while acting}{and act} as a source of reducing power by equilibrating to form H$_{2}$ at depth\added{ without NH$_{3}$ necessarily becoming a major constituent in the atmosphere}. Lastly, if TOI-1266c has a silicate core, then moderately volatile elements like Na and Cl could contribute to atmospheric composition either directly (that is, there may be a rock vapor atmosphere) or indirectly (e.g., through catalytic and secondary reactions with the major species).

Hazes, either driven by condensation or by photochemistry, represent significant hurdles for characterizing exoplanetary atmospheres. Here, we have only considered water and a few other potentially major species, but the presence of sulfuric acid clouds on Venus \citep[][]{kawabata1980cloud} or other sulfur-based aerosols \citep[e.g.,][]{zahnle2009atmospheric, zahnle2016photolytic, gao2017sulfur} are possible if sulfur is present in trace amounts. This could lead to observational degeneracies between a solid surface or a highly reflective cloud top \citep[e.g.,][]{lustig2019mirage}. Similarly, abundant carbon could lead to the formation of organic aerosols, but the relatively water-rich\added{ and lower temperature} scenarios tested here largely prevent carbon-carbon chemistry. Even with the uncertainties in TOI-1266c's mass allowing for a predominantly silicate composition, a modest amount of water loss would \added{produce enough free oxygen to }\deleted{also} effectively preclude organic aerosols \citep[e.g.,][]{horst2018exploring}, unless the \deleted{free }oxygen left over from water loss were absorbed by the solid planet \citep[e.g.][]{luger2015extreme}. However, if the planet started out relatively water-poor, or more diverse haze formation pathways are considered, hazes seem likely \citep[][]{moran2020chemistry, reed2020impact, vuitton2021h2so4}, and could be of various compositions with distinct optical properties \citep[e.g.][]{he2018photochemical, he2020haze, he2020sulfur}. Spectroscopic characterization, in combination with better mass constraints, would effectively narrow down the possibilities, much like it would for the TRAPPIST-1 system \citep{moran2018limits}. Other secondary condensate species could be present, such as potassium chloride (KCl) \citep[e.g.,][]{gao2018sedimentation}, which could enhance the effectiveness of (or serve in their own right as) cloud condensation nuclei (CCN) for water clouds.

\subsection{Redox considerations \label{subsec:disc-redox}}

As a super-Earth, TOI-1266c's size requires that we use caution with regards to the common assumptions about the atmospheric composition and evolution of warm Neptunes \citep[e.g.,][]{hu2014photochemistry, moses2020atmospheric}. As mentioned previously, TOI-1266c may be rocky, and if so, may have started out with a relatively H$_{2}$- and He-poor composition before subsequently losing the H$_{2}$ and He over its lifetime. In our Monte Carlo simulations, the average scenario lost $\lesssim$1\% of the planet's mass by the present day, but at the same time, the mean envelope fraction declined by $\sim$20\% of its initial value. This makes intuitive sense -- the largest impact of atmospheric escape is seen in those cases where the atmosphere is initially only a small fraction of the planet's mass. Studies suggest that more massive planets under higher instellation have comparable mass losses for higher H$_{2}$ mass fractions, which would produce larger variations in the planet's present-day radius \citep{estrela2020evolutionary}. 

Alternatively, if TOI-1266c is more massive, then water may be sequestered into and later outgassed from a magma ocean, preserving a relatively high water mass fraction \citep{kite2021water}. If TOI-1266c started with a water-dominated atmosphere without a sufficient buffer of H$_{2}$, then the persistent loss of H, derived from water vapor photolysis, would fundamentally alter the redox of the the planet through the accumulation of oxygen. Since we have hypothesized scenarios in which TOI-1266c has substantial amounts of water at present, this build-up of oxidants would still be happening today. These oxidants could react with a magma ocean and drive chemical alteration, or they could be sequestered through incorporation into high-pressure ice phases. Transport via convection through high-pressure ice layers has been studied for icy satellites \citep[e.g.,][]{deschamps2001thermal} and water-dominated super-Earths \citep{fu2009interior,noack2016water}, and would allow for both a supply of reducing gases from the interior and redox evolution of the interior driven by atmospherically-derived oxidants.

\subsection{Other factors affecting atmospheric loss \label{subsec:disc-other}}

Uranus and Neptune's water-dominated interiors\added{ have adiabats that} likely pass through the superionic portion of the high-pressure and high-temperature water phase diagram \citep{redmer2011phase, knudson2012probing, millot2018experimental}. This may explain why Uranus and Neptune are the only planets with multipolar rather than dipolar fields \citep[][]{schubert2011planetary}. If TOI-1266c is more water-dominated than the ice giants, then the pressure-temperature profile \replaced{is potentially steep enough to avoid}{does not cross through} the superionic regime, which could result in a weaker planetary magnetic field dominated by the dipolar component \citep[e.g.,][]{tian2013interior}. We note, however, that our temperature-pressure profiles for TOI-1266c are incompatible with those of \citet{tian2013interior} because we have assumed that the H$_{2}$ and H$_{2}$O are well-mixed. Additional components like ammonia or methane further complicate the conductivity of the high-pressure ice layers, but carbon and nitrogen may precipitate out together \citep[e.g.,][]{chau2011chemical}.

\replaced{O}{In terms of uncertainties related to the host star, o}ur assumed \added{stellar }luminosity evolution is based on the grid of \citet{hidalgo2018updated}, which includes luminosity evolution data for 0.4 and 0.45 M$_{\odot}$ stars. Given that TOI-1266 is $\sim$0.44 M$_{\odot}$, we could reasonably assume that it follows the 0.45-M$_{\odot}$ stellar evolution. However, the observed luminosity for TOI-1266 and 0.45-M$_{\odot}$ luminosity are different by $\sim$+3\% at 7.9 Gyr (TOI-1266's notional age). Taking the mass-weighted logarithmic mean of the \citeauthor{hidalgo2018updated} evolutionary tracks results in a $\sim$-6\% discrepancy between observed and estimated present-day luminosities. Normalizing the luminosity to match both the observed stellar mass and luminosity has the unintended side effect of producing higher fluxes than the 0.45-M$_{\odot}$ track early in the star's history (\figureref{fig:luminosity-uv-flux}). It is not immediately clear which approach is appropriate, but we find that using both the normalization and a weighted mean of the luminosities accurately reproduces the generic luminosity estimate derived from the stellar mass within -5\% \citep{cuntz2018mass}, as opposed to $\sim$-12\% when using the 0.45-M$_{\odot}$ evolution data (we use the \citeauthor{cuntz2018mass} generic mass-luminosity relationship because TOI-1266's mass estimate is on the cusp of where older formulations have a discontinuity \citet[e.g.,][]{kutner2003astronomy}. Uncertainties in mass and luminosity have knock-on effects for when the star enters the main sequence and on the estimated atmospheric loss. \added{Our methodology for the luminosity interpolation should be viewed with skepticism, and would improve with stronger constraints on stellar properties.} 

Additionally, the age uncertainties for TOI-1266 suggest that longer-term persistent atmospheric loss processes like interactions with the stellar wind \citep[e.g.,][]{cohen2015interaction, tilley2019modeling, gronoff2020atmospheric} could either play a major role in the current state of TOI-1266c if the star is older, or still represent a small fraction of the total loss when compared to the loss estimates from the pre-main sequence super-luminous phase.\added{ While no observations of flaring for TOI-1266 have been reported at this time, stellar flares can further contribute to atmospheric erosion. Losses due to flaring are typically smaller than the baseline XUV-driven escape rates \citep[e.g.,][]{atri2021stellar}, although this may not be universal, particularly for stars that flare more frequently \citep[e.g.,][]{france2020high}.} \deleted{Our methodology for the luminosity interpolation should be viewed with skepticism, and would improve with stronger constraints on stellar properties.} Magnetohydrodynamical models of H$_{2}$ \citep[e.g.,][]{johnstone2015evolution} and H$_{2}$O \citep[e.g.,][]{johnstone2020hydrodynamic} loss, as well as generically H-dominated super-Earth loss rates \citep{kislyakova2013xuv}, suggest that loss is a certainty, even if the magnitude\added{ and dominant mechanisms} remain\deleted{s an} open question\added{s}.

Another potential factor is the communication between the interior of the planet and its atmosphere. If, for example, TOI-1266c is water- or hydrogen-dominated, then the core component may effectively supply material to the escaping envelope \citep[e.g.,][]{wilson2011solubility}, especially if H$_{2}$ is effectively incorporated into water ices \citep{soubiran2015miscibility} or separates out slowly over the course of the planet's lifetime \citep{bailey2019thermodynamically}. However, if H$_{2}$ is ultimately immiscible \citep[e.g.,][and references therein]{bailey2019thermodynamically}, then the H$_{2}$ stranded in the atmosphere would be lost preferentially to H$_{2}$O, as discussed previously, leaving behind an ice-dominated core.

\subsection{Spectral signatures and observations \label{subsec:disc-spectral}}

In terms of differentiating the scenarios discussed here, CO$_{2}$ has a strong absorption feature at $\sim$4.3 $\mu$m ($\sim$40 ppm; \figureref{fig:co2_spectra}) even for 1\% CO$_{2}$ in a cloud-free atmosphere. This is broadly consistent with simulated retrievals of warm sub-Neptunes with \emph{JWST} \citep[e.g.][]{greene2016characterizing}. \added{The 4.3 $\mu$m feature would likely not be significantly impacted by the presence of haze \citep[e.g.,][]{kempton2017observational}.}The O$_{2}$ features in the ultraviolet are relatively small (\figureref{fig:o2_spectra}), while the 0.76-$\mu$m feature provides a relatively wide $\sim$15 ppm signal in comparison. Because of the abundant water vapor and the modest O$_{2}$ mixing ratios, O$_{2}$-O$_{2}$ dimer spectral features \citep{misra2014using, fauchez2020sensitive} are unlikely to be present or observable. From \figureref{fig:o2_spectra}, and more broadly \figureref{fig:SNR}, it is clear that \emph{JWST} will be unable to positively identify oxygen without a substantial investment of observational time, even for the relatively extended, warm atmospheres we consider here.

CO and O$_{3}$, derived from CO$_{2}$ and O$_{2}$, also have spectroscopic features that can help in distinguishing these scenarios. CO has \added{weak }features at 1.6, 2.3, and 4.7 $\mu$m \citep{wang2016detection, schwieterman2019rethinking}, but these are subsumed by strong H$_{2}$O and CO$_{2}$ features at those wavelengths. Interestingly, the appearance of the O$_{2}$ A-band at $\sim$0.76 $\mu$m (\figureref{fig:all_spectra}) does not result in the commensurate rise of an O$_{3}$ feature at 9.6 $\mu$m that is expected for temperate, O$_{2}$-rich atmospheres \citep[e.g.][]{segura2003ozone, segura2005biosignatures, rugheimer2013spectral, rugheimer2018spectra, meadows2018exoplanet}. This was noted by \citet{chen2019habitability} as a result of OH reducing the O$_{3}$ concentrations, but here is comparable to the results first shown by \citet{pidhorodetska2021L98}, where high temperatures force the rapid thermal decomposition of O$_{3}$ (the back reaction of Reaction \#309 in the Appendix). As a result, there is a 10$^{3}$-fold reduction in O$_{3}$, with O$_{3}$ being entirely absent from the integrated transmission spectra (\figureref{fig:o2_spectra}, 9.4-9.8 $\mu$m).

The possibility of clouds at temperate conditions (230-290 K; $\sim$0.5-5 mbar) acts in two ways to \replaced{obfuscate}{obscure} spectral features \citep{Fauchezclouds2019}: first, by limiting transmission through the deeper, warmer parts of the atmosphere, and by introducing strong intermediate-temperature water features. \replaced{These pressures are}{This pressure range for condensation is} in keeping with those reported in \deleted{some }other studies of steam atmospheres\replaced{ (}{. For example, }\citet[][]{nikolaou2019factors} report much lower pressures as an upper bound, \replaced{but}{although those experiments} have substantially more CO$_{2}$ that the cases described here\deleted{)}. Water clouds may also appear in the atmospheres of more temperate massive planets \citep{charnay2020formation}, again at around 10 mbar. However, some scenarios featuring clouds work to enhance spectroscopic features \citep[e.g.][]{kawashima2019theoretical}, which makes determining self-consistent cloud, climate, and photochemistry a critical next step. We have attempted to include the impact of clouds by assuming the clouds are composed of $\sim$14-$\mu$m droplets with a volume mixing ratio of 10$^{-7}$ or ice clouds composed of 25-$\mu$m crystals with a mixing ratio of 10$^{-9}$. In both cases, the clouds do not appear to substantially impact the water spectral features (\figureref{fig:all_spectra}, gray curves), with ice clouds having a smaller effect, largely due to their smaller abundance. The smaller reduction, as compared to more temperate atmospheres \citep{Fauchezclouds2019}, is likely due to the presence of water-vapor above the cloud-deck in a water-rich atmosphere that minimizes the impact of clouds on spectral features. A self-consistent cloud modeling effort is necessary to further this work.

TOI-1266c could be characterized by \emph{JWST} in the future, which could effectively distinguish between some of these cases. The signal-to-noise ratio of a series of observations of TOI-1266c in transit is shown in \figureref{fig:SNR}. We calculate signal-to-noise as the difference between two synthetic spectra, one of which ignores the spectral contributions from the chief secondary species (e.g., H$_{2}$, O$_{2}$, or CO$_{2}$), dividing by the simulated noise. The counter-intuitive reduction in signal-to-noise with increasing H$_{2}$ abundances stems from the decreasing apparent water column mass. The other compositions show the opposite trend, driven by those gases having distinct spectroscopic features of their own. Ultimately, CO$_{2}$ and H$_{2}$O in significant abundances could be identified in a few tens of hours, but oxygen to a sufficient signal-to-noise ratio may be beyond \emph{JWST}'s capabilities.\added{ As mentioned previously, the 4.3-$\mu$m CO$_{2}$ band would not be substantially affected by hazes, although further tests are necessary.}

Further investigations of the radius gap have the potential to provide key insights into the processes that shape planets over their lifetimes. As the community continues to find more transitional objects, it is becoming increasingly clear that some of the exoplanets that are likely to be characterized in the near future may not be precisely what we expect them to be. Volatile-depleted sub-Neptune cores masquerading as super-Earths could inadvertently skew our perspectives on habitability, such as if a water-dominated sub-Neptune is incorrectly classified as a terrestrial planet, since the large water fraction would suggest oceans so deep that they would suppress volatile exchange and plate tectonics \citep[][]{kite2009geodynamics}.

\section{Conclusions and Future Work\label{sec:conc}}

The potential to observe a Venus analogue, particularly if it remains in a steam-dominated runaway greenhouse at present, offers an unparalleled window into the history and evolution of a unique terrestrial planet as well as one of the first few steam atmospheres accessible with \emph{JWST}. Data about water-dominated atmospheres are also relevant to the bounds of habitability for terrestrial planets, particularly those that form around low-mass host stars \citep{luger2015extreme} and orbit older stars \citep[e.g.][]{rushby2013habitable, lehmer2020carbonate}. Lastly, observations of exoplanets that may have accumulated oxygen derived from water loss are important to establishing a baseline for larger planet sample size analyses, particularly in the context of biosignatures \citep[e.g.][]{bixel2020testing}. 

\added{Estimating the composition and any potential observables for TOI-1266c remains difficult. The degeneracies in relating bulk composition, atmosphere-to-solid planet fraction, and mean density \citep[e.g.][]{rogers2010framework, dorn2015can, welbanks2019degeneracies} pose significant challenges in predicting the atmospheres of exoplanets even if the mass and radius are well constrained, and even spectroscopic observations of the planet may not necessarily break this degeneracy \citep{batalha2017challenges}. Additional radial velocity observations--such as precision RVs with HPF \citep{mahadevan2012habitable, mahadevan2014habitable}, NEID or CARMENES--are required to constrain the planet's mass, which will narrow the range of possible compositions \citep[e.g.][]{valencia2013bulk}.
}

The modeling demonstrated here showcases the impact of dynamical, photochemical, and ancillary atmospheric processes on the disposition of some of the possible planetary states in the radius gap. These planets highlight the continuing need to understand the processes that shape highly-irradiated, volatile-rich planets in advance of observational campaigns with \emph{JWST} and other future instruments. Follow-up observations are planned to better constrain TOI-1266c's mass (and by extension its possible composition). Regardless of whether or not TOI-1266c is the first such target to be observed, this class of objects requires additional capabilities beyond thermochemical equilibrium models and assumptions about composition in terms of metallicity. In a future study, we plan to expand our photochemical scheme to include secondary species that were omitted in this work, as well as explore the feedbacks between climate, chemistry, and observability.

\begin{acknowledgements}
Harman and Kopparapu acknowledge support from the GSFC Sellers Exoplanet Environments Collaboration (SEEC), which is supported by NASA’s Planetary Science Division’s Research Program. This work was performed as part of NASA’s Virtual Planetary Laboratory, supported by the National Aeronautics and Space Administration through the NASA Astrobiology Institute under solicitation NNH12ZDA002C and Cooperative Agreement Number NNA13AA93A, and by the NASA Astrobiology Program under grant 80NSSC18K0829 as part of the Nexus for Exoplanet System Science (NExSS) research coordination network.

Harman would like to thank James Kasting for his invaluable comments on the manuscript; Kevin Zahnle for his help in hunting for technical issues during development, and both Mahmuda Afrin Badhan and Sandra Bastelberger for their work in improving and updating \texttt{Atmos}; Benjamin Drummond for his work in establishing an intercomparison of 1-D chemical kinetics codes, hosted as an ISSI International Team, which served as the nucleus of the code modifications used in this study, as well as Shami Tsai (also part of the Team) for his openness with VULCAN and insights on chemical kinetics. \added{The authors would also like to thank the two anonymous reviewers for their comments, which improved the quality of the manuscript.}
\end{acknowledgements}

\software{\texttt{Atmos} \citep{arney2017pale},
          Planetary Spectrum Generator (PSG) \citep{villanueva2018planetary}}

\appendix

\section{Atmospheric Escape Code Derivation and Description \label{app:escape}}
As described in the main text, our model for atmospheric escape is based on the stellar luminosity evolution models of \citet{hidalgo2018updated} in conjunction with the escape flux parameterizations of \citet{murray2009atmospheric, owen2016atmospheres, lopez2017born}. Here, we walk through the assumptions built into our model and how we implemented these processes. 

The \textsc{Snowball} repository has two branches -- the \texttt{main} branch, and the \texttt{monte\_carlo} branch. Both branches have largely the same code, save that the \texttt{monte\_carlo} branch features a refactored \texttt{main.py} such that it can be called by the Monte Carlo generator program and return the results. Because the Monte Carlo application has additional assumptions, we will use that as the basis for the remainder of this discussion. The code is laid out in an attempt to compartmentalize individual physical concepts and processes, but because some are interconnected it is not always possible to completely separate some of them. Below is a partial dependency tree for how we calculate the escape regimes and fluxes (omitting generic and optional functions):

\dirtree{%
.1 monte carlo branch.
.2 monte\_carlo.py.
.3 main.py. 
.4 run\_escape(age\_star, l\_star, m\_planet, r\_planet, envelope\_comp, efficiencies).
.5 modules.py.
.6 read\_hidalgo.
.6 read\_thermo.
.6 generic\_diffusion.
.6 crossover\_mass.
.6 bisect2.
.6 planet\_radius.
.6 f\_lum\_CW18.
.6 calc\_escape\_regime.
.5 constants.py.
.5 planet.py.
.6 envelope\_species.
.2 analyze\_MC.py.
}

The program \texttt{monte\_carlo.py} is responsible for calling the Latin Hypercube Sampling method \citep{bouhlel2019python, jin2003efficient} for the selected uncertainty ranges on the stellar age (\texttt{age\_star}, in Gyr), present-day stellar luminosity (\texttt{l\_star}, in terms of solar luminosity), initial planet mass (\texttt{m\_planet}, in Earth masses), initial planet radius (\texttt{r\_planet}, in Earth radii), initial volatile abundances for H$_{2}$, H$_{2}$O, He, and CO$_{2}$ (\texttt{envelope\_comp}, as mass fractions of the initial envelope), and the escape efficiencies for H$_{2}$, H$_{2}$O, and the other gases (\texttt{efficiencies}). There are two caveats with respect to the volatile abundances. First, we set the helium mass fraction as being less than or equal to the solar He/H$_{2}$ ratio (24.85\% He/73.8\% H$_{2}$ by mass $\sim$0.3367) by multiplying together the randomly-selected He `abundance' fraction, the solar He/H$_{2}$ ratio, and the randomly-selected H$_{2}$ abundance. Second, because the ranges for the abundances can result in the combined mass fraction exceeding unity, we normalize the abundances following parameter sampling. All of the selected variables (age, luminosity, planet mass and radius, volatile abundances and escape efficiencies) are then passed in as arguments to \texttt{run\_escape}.

The function \texttt{run\_escape} initializes a given simulation, runs the scenario to the specified end time, and returns a number of diagnostic values to the parent \texttt{monte\_carlo.py} program (\texttt{monte\_carlo.py} then archives these values to a saved file on disk). We break down each of these steps below, and include the specific descriptions of function calls as they arise.

\subsection*{\texttt{run\_escape} - setup step}

The lion's share of the setup deals with the stellar evolution track file read-in and interpolation. Initially, two evolution tracks for stars that bookend the given host star mass are identified, read in, and used to create two cubic interpolations for their luminosity evolution tracks as a function of stellar age. We then take the weighted logarithmic mean of the luminosities to represent the star in question -- for example:

\begin{equation*}
    \text{log}_{10}(\text{L}_{*}) =  \frac{(M_{*}-M_{i})}{(M_{k}-M_{i})}*\text{log}_{10}(\text{L}_{i}) + \frac{(M_{k}-M_{*})}{(M_{k}-M_{i})}*\text{log}_{10}(\text{L}_{k})
\end{equation*}
where M and L are the stellar mass and luminosity of each star, and the subscripts `*' and $i$ and $k$ stand for the host star and two bookend stellar masses, respectively, such that M$_{i}$ $<$ M$_{*}$ $<$ M$_{k}$. As a second step, we then normalize the synthetic luminosity at the estimated stellar age to the observed present-day luminosity. This results in luminosities that are higher than those of the upper bookend stellar evolution track at ages less than 0.1 Gyr because the observed luminosity \citep[0.02629 L$_{\odot}$;][]{stefansson2020} is higher than the interpolated luminosity ($\sim$0.0247 L$_{\odot}$). This assumption is necessary, given our focus on matching the observables as closely as possible. Because of the higher present-day \added{stellar }luminosity, the \replaced{mass loss is}{planet experiences} $\sim$3\% higher\added{ atmospheric mass loss over its lifetime}, but this does not qualitatively change any of our conclusions. We next produce several diagnostics, including identifying the transition onto the main sequence, na{\"i}vely assuming that this corresponds to the global minimum in stellar luminosity. Lastly, we calculate the X-ray and EUV flux ratio with respect to the star's evolving luminosity, with options available to use either the \citet{sanz2011estimation} or the \citet{peacock2020hazmat} relationships (as explained in the text, we have chosen the \citeauthor{peacock2020hazmat} parameterization). Once this is done, the other variables are initialized based on the values in the \texttt{planet.py} file. This file also has the options for specifying other choices for which luminosity evolution tracks are used (\citet{baraffe2015new} vs. \citet{hidalgo2018updated}), whether the luminosity is normalized, and which UV scalings are applied.

One additional assumption we make, as described in the main text, is that for each set of randomly-sampled mass and radius, we construct an estimated envelope fraction based on the maximum possible water inventory using the composition-mass-radius relationship from \citet{noack2016water}. We do this by iterating through the solution space starting with no iron core and checking to see if the planet radius falls between the 100\% silicate and 100\% water composition radii (using \texttt{bisect2} and \texttt{planet\_radius}). If it is still too large, we slowly increasing the fraction of iron to shift the planet from a large initial volatile inventory and planet radius towards a solution that satisfies both the mass and radius.

\subsection*{\texttt{run\_escape} - loop over time array}
At each time point, several parameters can be derived from the stellar luminosity, the planet's mass and radius, and the initial envelope composition. These include the:
\begin{itemize}
    \item `Surface' gravity $g = G M_{p}/R_{p}^{2}$ ~~~[m/s$^2$]
    \item From \citet{lopez2017born}, the \replaced{exobase pressure,}{pressure} where most of the XUV is absorbed $p_{XUV}\sim8.8\times 10^{-14} \times g$ ~~~[Pa] 
    \item The atmospheric scale height $H = k_{b} T_{eq} / (\overline{m}_{atm} g)$ ~~~[m]
    \item From \citet{lopez2017born}, the radius of the exobase $R_{exobase}\sim R_{p} + H \times p_{photo}/p_{XUV}$ ~~~[R$_{\oplus}$] 
\end{itemize}
where $G$ is the gravitational constant, $k_b$ is the Boltzmann constant, $M_p$ is the planet mass, $R_p$ is the planet radius, $\overline{m}_{atm}$ is the mean  molecular mass of the atmosphere, $T_{eq}$ is the equilibrium temperature (assuming the planetary albedo = 0.), and $p_{photo}$ is the transit radius pressure \citep[20 mbar;][]{lopez2017born}. \added{The approximation of the exobase as the region in which XUV is predominantly absorbed is sensitive to the spectral distribution of incoming stellar radiation, as well as to the composition of the atmosphere. The atmospheric scale height equation is valid only when the atmosphere is largely hydrostatic, because otherwise the vertical velocity of the escaping component `stretches' the scale height \citep[e.g.,][]{hunten1973escape}.}

Following the determination of these quantities, we estimate the escape regime following \citep{owen2016uv}. In brief, there are three surfaces in stellar flux--planet mass--planet radius phase space that correspond to three different escape regimes: energy-limited, photon-limited, and recombination-limited. \added{ Briefly, these regimes correspond to situations in which the escape rate is limited by a particular phenomena \citep[][]{owen2017evaporation, murray2009atmospheric}. The energy-limited escape rate is determined by the plausible upper bound of the amount of energy that could be absorbed and converted into kinetic energy that then drives atmospheric escape. The photon-limited escape rate is derived from the need for ionizing radiation to break up molecules and atoms into smaller and/or charged particles that can then escape. Since the radiation arrives at the top of the atmosphere as a flux of photons that cannot be divided further, there can only be a certain amount of material available for escape at any one time. Lastly, the radiation-recombination limit is moderated by a thin, fully-ionized gas layer in the escaping flow that prevents the further absorption of more energy until some of the dissociated material is allowed to recombine. }Generally, the larger the planet or the higher the incoming EUV flux, the more likely the planet is to be in the recombination-limited regime, while smaller planets or those experiencing low EUV fluxes fall into the photon-limited regime.

\added{ The individual escape rate parameterizations are taken from literature, and are most often defined in the context of uniform molecular or atomic hydrogen atmospheres, which may not be fully reflected by some of the scenarios described here. The photon-limited escape rate [in kg/s] is defined as:

\begin{equation} \label{app:photon}
    \Phi_{\text{photon}} = \frac{\pi R_{p}^{2} m_{H} \Phi_{XUV}}{h \bar{\nu}_{h}},
\end{equation}
where $R_{p}$ is the planet radius [m], $m_{H}$ is the mass of a hydrogen atom [kg], $\Phi_{XUV}$ is the XUV energy flux received by the planet [W/m$^{2}$], and $h \bar{\nu}_{h}$ is the mean photon energy for photons that heat the upper atmosphere [J/photon]. 

The radiation-recombination limit [in kg/s] is given by:
\begin{equation} \label{app:recombination}
    \Phi_{\text{recombination}} = 7.11\times 10^{4} \Phi_{XUV}^{0.5} R_{p}^{3/2},
\end{equation}  
which comes from \citet{luger2015habitable}, based on the 1-D photoevaporation modeling and resulting parameterization of hot Jupiter atmospheric escape by \citet{murray2009atmospheric} (see \citeauthor{luger2015habitable} for the derivation). The leading co-factor includes an additional unit conversion to MKS units from CGS units, but is otherwise identical. The equation relates the escape limit to $\Phi_{XUV}$, the XUV energy flux received by the planet [W/m$^{2}$], and includes an explicit dependence on the planet's radius ($R_{p}$, in meters).

The energy-limited escape rate [in kg/s] is:

\begin{equation} \label{app:energy}
    \Phi_{\text{energy}} = \frac{\eta \Phi_{XUV} R_{p} R_{exobase}^{2}}{G M_{p} K_{tide}},
\end{equation} 
using the formulation of \citet{luger2015extreme}. In this equation,  $\eta$ is the escape efficiency parameter (0$<\eta<$1; unit-less), $R_{p}$ is the radius of the planet [m], $R_{exobase}$ is the radius of the exobase \citep[in m; defined above, following][]{lopez2017born}, G is the gravitational constant [m$^{3}$/kg/s$^{2}$], $M_{p}$ is the mass of the planet [m], and $K_{tide}$ is the non-dimensional tidal enhancement factor \citep[taken from][]{erkaev2007roche}, which depends on the Roche lobe radius ($R_{Roche} = a (M_{p}/(3M_{*}))^{1/3}$, in meters) and the planet radius ($R_{p}$, in meters) such that

\begin{equation*}
    K_{tide} =  1 - \frac{3 R_{p}}{2 R_{Roche}} + \frac{R_{p}^{3}}{2 R_{Roche}^{3}}.
\end{equation*}
Note that this includes the assumption of \citet{luger2015extreme} where the location at which the XUV radiation is absorbed is approximately the planet radius.

We have also added the possibility of entering the diffusion-limited escape regime \citep[e.g.,][]{kasting1983loss} when the available hydrogen drops below 1\% (assuming that this is the smaller flux of both the energy and diffusion limits; otherwise, we continue to use the energy limit). The threshold value of $\sim$1\% roughly corresponds to the transition region between diffusion- and energy-limited escape regimes when the XUV flux is 5-10 times higher than what is received by the Earth \citep{kuramoto2013effective}. The diffusion limit can be a critical hurdle to the complete desiccation of a planet, limiting the supply of water and/or hydrogen to the upper atmosphere via the need to diffuse through the largely static, heavy background gases. The diffusion limit [molecules/cm$^{2}$/s] is defined by \citet{hunten1973escape} as:

\begin{equation} \label{app:difflimit}
    \Phi_{i, \text{diffusion}} = \frac{b_{i} \xi_{i}}{(1+ \xi_{i})} \Big(\frac{1}{H_{\text{heavy}}} - \frac{1}{H_{i}}\Big),
\end{equation} 
or by expanding the scale height terms and including the dependence on the size of the escaping surface, we see that

\begin{equation}\label{app:difflimit2}
    \Phi_{i, \text{diffusion}}' = 4 \pi R_{p}^{2} m_{H} \frac{b_{i} g \xi_{i} (m_{\text{heavy}}- m_{i})}{k_{b} T (1+ \xi_{i})}~~~~\text{[kg/s]}
\end{equation}
This limit exists only when the escaping flux is sufficiently small (in terms of the magnitude of the escape flux) such that the flow does not exceed the crossover mass for other constituents in the atmosphere. Above this limit, the assumption of a stationary heavy gas component is not valid, as the heavy gas is drug off with the escaping component. Determining what additional species are included in the flow is often estimated with the crossover mass \citep{hunten1987mass}:

\begin{equation} \label{app:eq4}
    m_{c} = m_{H} + \frac{k_{b} T \Phi_{\text{escape}}}{b g f_{H}}
\end{equation}
where $m_{H}$ is the mass of the hydrogen atom [kg], $k_{b}$ is the Boltzmann constant [m$^{2}$ kg/s$^{2}$/K], $T$ is the temperature [K] (taken to be the equilibrium temperature of the planet, a simplification used in lieu the homopause temperature), $\Phi_{escape}$ is the escape rate as calculated in the prior step [kg/s], $g$ is the planet's `surface' gravity [m/s$^{2}$], $f_{H}$ is the total hydrogen atom fraction ($=\xi_{H_{2}} + 2/3\times \xi_{H_{2}O}$, where $\xi$ is the volume mixing ratio of a given species). The total hydrogen fraction calculation combines the assumption that XUV radiation is effective at breaking molecules down into their constituent atoms, and that the new volume mixing ratios of the constituent atoms reflect the composition of the original molecules. This is analogous to the convention defined by  $b$ is the generic binary diffusion coefficient for two species from \citet{banks1973aeronomy}: 

\begin{equation}
    b=1.52\times 10^{20} \Big(\frac{m_{H}}{m_{\text{minor}}} + \frac{m_{H}}{m_{\text{major}}} \Big)^{0.5} T^{0.5} ~~~ \text{[/m/s]},
\end{equation}
where $m_{\text{minor}}$ and $m_{\text{major}}$ are the molecular masses of the minor (light) species and the major (heavy) species (noting that we assume the heavy species is the mean molecular mass of the species that are not escaping). The crossover mass and the diffusion limit are siblings, such that $\Phi_{\text{escape}} = \Phi_{\text{diffusion}}$ only when $m_{c} = m_{\text{heavy}}$). The crossover mass represents the upper limit on the `lifting' action of hydrogen as it escapes, meaning that anything with a smaller atomic weight than the crossover mass can become entrained in the escaping hydrogen and ultimately removed from the planet. 

Because the escaping component is not necessarily just hydrogen, the presence of heavier atoms and molecules changes the crossover mass and the flux of escaping material. The first step in our code assumes that the escaping material is atomic hydrogen, and the crossover mass in Eqn. \ref{app:eq4} is calculated using just the mass of hydrogen. We then extend the formalism adopted in \citet[][their Section 2.4.2]{luger2015extreme}, defining the flux of each species in relation to the flux of hydrogen and the crossover mass:

\begin{equation} \label{app:eq5}
    \Phi_{x} = \frac{\xi_{x}}{f_{H}} \Phi_{H} \Big(\frac{m_{c} - m_{x}}{m_{c} - m_{H}}\Big) \text{ if } m_{c} > m_{x}; \text{= 0 otherwise}.
\end{equation}
Note that an implicit assumption here is that there are sufficient photons to completely dissociate any hydrogen-bearing species, such that $\xi_{H} \approxeq f_{H}$. To calculate the flux of each species in the escaping flow, we can generalize Eqn. 6 of \citet{luger2015extreme}:
 
\begin{equation} \label{app:eq6}
    m_{H} \Phi_{H}^{\text{ref}} = \sum_{x}^{m_{x} < m_{c}} m_{x} \Phi_{x}.
\end{equation}
We can then follow the same derivation as \citet{luger2015extreme}, which will result in a crossover mass that accommodates more than one other gas. Substituting Eqn. \ref{app:eq5} into this relationship, we see that:

\begin{align}
    m_{H} \Phi_{H}^{\text{ref}} &= \sum_{x}^{m_{x} < m_{c}} m_{x} \frac{\xi_{x}}{f_{H}} \Phi_{H} \Big(\frac{m_{c} - m_{x}}{m_{c} - m_{H}}\Big) \\ 
    &= \frac{\Phi_{H}}{(m_{c} - m_{H})f_{H}} \sum_{x}^{m_{x} < m_{c}} m_{x} \xi_{x} \Big(m_{c} - m_{x}\Big) \\
    \frac{\Phi_{H}}{m_{H} f_{H}} &= \Phi_{H}^{\text{ref}}(m_{c} - m_{H}) \Bigg[\sum_{x}^{m_{x} < m_{c}} m_{x} \xi_{x} \Big(m_{c} - m_{x}\Big)\Bigg]^{-1}
\end{align}
If we substitute this expression back into Eqn. \ref{app:eq4}:

\begin{align}
    m_{c} =& m_{H} + m_{H}\frac{k_{b} T \Phi_{H}^{\text{ref}}}{b g} (m_{c} - m_{H}) \Bigg[\sum_{x}^{m_{x} < m_{c}} m_{x} \xi_{x} \Big(m_{c} - m_{x}\Big)\Bigg]^{-1}\\
    \frac{m_{c} - m_{H}}{m_{H}} =& \frac{k_{b} T \Phi_{H}^{\text{ref}}}{b g}(m_{c} - m_{H}) \Bigg[\sum_{x}^{m_{x} < m_{c}} m_{x} \xi_{x} \Big(m_{c} - m_{x}\Big)\Bigg]^{-1}\\
    \frac{1}{m_{H}} =& \frac{k_{b} T \Phi_{H}^{\text{ref}}}{b g} \Bigg[\sum_{x}^{m_{x} < m_{c}} m_{x} \xi_{x} \Big(m_{c} - m_{x}\Big)\Bigg]^{-1}\\
    \sum_{x}^{m_{x} < m_{c}} m_{x} \xi_{x} \Big(m_{c} - m_{x}\Big)  =& \frac{k_{b} T  \Phi_{H}^{\text{ref}} m_{H}}{b g}\\
    \sum_{x}^{m_{x} < m_{c}} m_{x} \xi_{x} m_{c} - \sum_{x}^{m_{x} < m_{c}} m_{x} \xi_{x} m_{x}  =& \frac{k_{b} T  \Phi_{H}^{\text{ref}} m_{H}}{b g}\\
    m_{c} \sum_{x}^{m_{x} < m_{c}} m_{x} \xi_{x}  =& \frac{k_{b} T  \Phi_{H}^{\text{ref}} m_{H}}{b g} + \sum_{x}^{m_{x} < m_{c}} m_{x} \xi_{x} m_{x}\\
    m_{c}  =& \Bigg[\sum_{x}^{m_{x} < m_{c}} m_{x} \xi_{x}\Bigg]^{-1} \bigg(\frac{k_{b} T  \Phi_{H}^{\text{ref}} m_{H}}{b g} + \sum_{x}^{m_{x} < m_{c}} m_{x} \xi_{x} m_{x}\bigg)
\end{align} 
As the sum of the mixing-ratio-weighted molecular masses gives the mean molecular mass, this expression simplifies to:
\begin{equation}
    m_{c}  = \frac{k_{b} T \Phi_{H}^{\text{ref}} m_{H}}{b g \bar{\mu}} + \frac{m_{H}}{\bar{\mu}}\sum_{x}^{m_{x} < m_{c}} m_{x} \xi_{x} \mu_{x}\label{eq:crossover}
\end{equation}
In our code, we describe $m_{c}$ in amu, not kg, which is the same as above once $m_{H}$ has been divided out of both sides of the equation. 

We estimate escape and the changing mean molecular weight of the flow by sequentially including each potential species in reverse order by mass (in our modeling, this would be helium, then atomic oxygen, then water, and so on). Each sequential addition forces us to check whether the new crossover mass (which has a slower relative flux but higher mean molecular weight) is still in excess of the molecular mass of the next-heaviest component, only exiting if this is not the case. Because the additional species is added before the crossover mass is calculated, the last loop exits without updating the crossover mass, ensuring that all the incorporated species are represented without adding a repeated term. Subsequent to this loop, the revised escape fluxes are computed (per Eqn. \ref{app:eq5}). Then the escaping mass is removed from the relevant inventories such that, if no hydrogen remains in the atmosphere, the hydrogen escape is subtracted from the water inventory (leaving behind the oxygen atoms), and any resulting negative inventories are instead set to zero and the remainder of the mass removed as hydrogen (using the same relationship in Eqn. \ref{app:eq5}).
}
\subsubsection*{Determining the escape regime}\label{app:regime}
As noted by \citet{owen2016uv}, there is no closed form of the equations governing which regime a planet would fall into, and so the boundaries of each regime must be solved for at each time step as a function of instellation and planet mass and radius. The three equations \citep[\#18, 19, and 20 from][]{owen2016uv} are:

\begin{align}
    M_{p} &= \eta \Big( \frac{h \bar{\nu}_{h}}{4 G m_{H}} \Big) R_{p} \label{app:eq18} \\
    W_{0} \Bigg[ - \Bigg(\frac{R_{p}}{R_{s}}\Bigg)^{-4} \text{exp} \Bigg( 3 - \frac{4R_{s}}{R_{p}}\Bigg) \Bigg] &= \frac{-J_{0}\alpha_{B}H}{4c_{s}^{2}} \label{app:eq19} \\
    W_{0} \Bigg[ - \Bigg(\frac{R_{p}}{R_{s}}\Bigg)^{-4} \text{exp} \Bigg( 3 - \frac{4R_{s}}{R_{p}}\Bigg) \Bigg] &= \frac{-\eta J_{0}\alpha_{B}H}{4c_{s}^{2}} \times \Big( \frac{R_{p} h \bar{\nu}_{h}}{4 G M_{p} m_{H}} \Big) \label{app:eq20}
\end{align}
where $M_p$ is the planet mass, $\eta$ is the escape efficiency parameter\added{ (0$<\eta<$1)}, $h \bar{\nu}_{h}$ is the mean photon energy for photons that heat the upper atmosphere, $G$ is the gravitational constant, $m_{H}$ is the mass of a hydrogen atom, $R_{p}$ is the planet radius, $W_{0}$ is the Lambert $W$ function (the Lambert $W$ function is the set of solutions to $W(x)$exp($W(x)) = x$, and $W_{0}$ is the principal branch such that $x$ and $W(x)$ are real numbers; we use the Python function \texttt{lambertw()} from the \texttt{scipy.special} library), $R_{s}$ is the sonic point (given by $G M_{p} / (2c_{s}^{2})$), $J_{0}$ is the ionizing photon flux, $\alpha_{B}$ is the case-B recombination coefficient ($=2.6\times 10^{-13}$ cm$^{3}$ s$^{-1} (T/10^{4}$)$^{-0.7}$, taking $T$ to be the exospheric temperature) acting as a stand-in for the recombination rate, $H$ is the atmospheric scale height, and $c_{s}$ is the sound speed ($c_{s}^{2}=k_{b} T / (\bar{m}_{atm})$, with $\bar{m}_{atm}$ defined as before to be the mean molecular mass of the atmosphere). Note that we include the additional parentheses to clarify exponentiation, following \citet{cranmer2004new}. The exospheric temperature, as described in the text, is set to either 10$^{4}$ K or 2000 K based on the incoming XUV flux, with a threshold of $\sim$180 erg cm$^{-2}$ s$^{-1}$ forcing a higher exospheric temperature\added{ when H$_{3}^{+}$ cooling becomes ineffective} \citep{koskinen2007stability, murray2009atmospheric}.

Equation \ref{app:eq18} defines the boundary between the photon- and energy-limited escape regimes, while Equation \ref{app:eq19} is for the recombination- and photon-limited regimes, and Equation \ref{app:eq20} is for the recombination- and energy-limited regimes. Both Eqns. \ref{app:eq19} and \ref{app:eq20} have either two solutions or none, if describing the system in mass-radius space \citep[as shown by][]{owen2016uv}, but since we have chosen a planet mass and radius as part of our initial conditions, this collapses the three equations down to one or no solution that is solely a function of the ionizing flux. We calculate $J_{0}$ by rearranging Eqns. \ref{app:eq19} and \ref{app:eq20} to solve for $J_{0}$, which we then compare against the stellar XUV flux calculated in the setup step. The boundary between energy- and photon-limited escape is in the form of planet mass, such that a planet more massive than the threshold mass (the left-hand side of Eqn. \ref{app:eq18}) will be in the energy- or recombination-limited regime. If the planet's mass is instead below the threshold and the XUV flux is below the recombination- and photon-limited threshold flux, the escape is in the photon-limited regime\replaced{, with a rate of}{.}
\deleted{
\begin{equation} \label{app:photon}
    \Phi_{\text{photon}} = \frac{\pi R_{p}^{2} m_{H} \Phi_{XUV}}{h \bar{\nu}_{h}},
\end{equation}
where $R_{p}$ is the planet radius, $m_{H}$ is the mass of a hydrogen atom, $\Phi_{XUV}$ is the XUV energy flux received by the planet, and $h \bar{\nu}_{h}$ is the mean photon energy for photons that heat the upper atmosphere. 

We then check to see if the stellar XUV flux is higher than the recombination- and energy-limited threshold flux, and if it is, the escape is recombination-limited, which is given by:

\begin{equation} \label{app:recombination}
    \Phi_{\text{recombination}} = 7.11\times 10^{4} \Phi_{XUV}^{0.5} R_{p}^{3/2},
\end{equation}  
which comes from \citet{luger2015habitable}, based on the 1-D photoevaporation modeling and resulting parameterization of hot Jupiter atmospheric escape by \citet{murray2009atmospheric} (see \citeauthor{luger2015habitable} for the derivation). The leading co-factor includes an additional unit conversion to MKS units from CGS units, but is otherwise identical. The equation relates the escape limit to $\Phi_{XUV}$, the XUV energy flux received by the planet, and includes an explicit dependence on the planet's radius ($R_{p}$).

Otherwise, the escape is in the energy-limited regime, with a rate given by:

\begin{equation} \label{app:energy}
    \Phi_{\text{energy}} = \frac{\eta \Phi_{XUV} R_{p} R_{exobase}^{2}}{G M_{p} K_{tide}},
\end{equation} 
using the formulation of \citet{luger2015extreme}. In this equation,  $\eta$ is the escape efficiency parameter, $R_{p}$ is the radius of the planet, $R_{exobase}$ is the radius of the exobase \citep[defined above, following][]{lopez2017born}, G is the gravitational constant, $M_{p}$ is the mass of the planet, and $K_{tide}$ is the tidal enhancement factor \citep[taken from][]{erkaev2007roche}, which depends on the Roche lobe radius ($R_{Roche} = a (M_{p}/(3M_{*}))^{1/3}$) and the planet radius ($R_{p}$) such that

\begin{equation*}
    K_{tide} =  1 + \frac{3 R_{p}}{2 R_{Roche}} + \frac{R_{p}^{3}}{2 R_{Roche}^{3}}.
\end{equation*}
Note that this includes the assumption of \citet{luger2015extreme} where the location at which the XUV radiation is absorbed is approximately the planet radius. 

We have also added the possibility of entering the diffusion-limited escape regime \citep[e.g.,][]{kasting1983loss} when the available hydrogen drops below 1\% (assuming that this is the smaller flux of both the energy and diffusion limits; otherwise, we continue to use the energy limit). This can be a critical hurdle to the complete desiccation of a planet, limiting the supply of water and/or hydrogen to the upper atmosphere via the need to diffuse through the largely static, heavy background gases. The diffusion limit is defined by \citet{hunten1973escape} as:

\begin{equation} \label{app:difflimit}
    \Phi_{i, \text{diffusion}} = \frac{b_{i} \xi_{i}}{(1+ \xi_{i})} \Big(\frac{1}{H_{i}} + \frac{1}{H_{\text{heavy}}}\Big),
\end{equation} 
or by expanding the scale height terms and including the dependence on the size of the escaping surface, we see that

\begin{equation}\label{app:difflimit2}
    \Phi_{i, \text{diffusion}} = 4 \pi R_{p}^{2} m_{H} \frac{b_{i} g \xi_{i} (m_{\text{heavy}}- m_{i})}{k_{b} T (1+ \xi_{i})}.
\end{equation}
This limit exists only when the escaping flux is sufficiently small (in terms of the magnitude of the escape flux) that the flow does not exceed the crossover mass for other constituents in the atmosphere (in fact, the crossover mass and the diffusion limit are siblings, such that $\Phi_{\text{escape}} = \Phi_{\text{diffusion}}$ only when $m_{c} = m_{\text{heavy}}$). 

Because the escape fluxes can be particularly high, there is the possibility that not only hydrogen is escaping. Determining what additional species are included in the flow is often estimated with the crossover mass (typically as $m_{c}$, but written here as $\mu_{c}$):

\begin{equation} \label{app:eq4}
    \mu_{c} = \mu_{H} + \frac{k_{b} T \Phi_{\text{escape}}}{b m_{H} g f_{H}}
\end{equation}
where $\mu_{H}$ is the mass of the hydrogen atom ($\approxeq$1 in atomic mass units, such that $\mu_{c}$ is also in amu), $k_{b}$ is the Boltzmann constant, $T$ is the temperature (taken to be the equilibrium temperature of the planet), $\Phi_{escape}$ is the escape rate as calculated in the prior step, $m_{H}$ is the mass of the hydrogen atom, $g$ is the planet's `surface' gravity, $f_{H}$ is the total hydrogen atom fraction ($=\xi_{H_{2}} + 2/3\times \xi_{H_{2}O}$, where $\xi$ is the volume mixing ratio of a given species), and $b$ is the generic binary diffusion coefficient for two species from \citet{banks1973aeronomy}: 

\begin{equation*}
    b=1.52\times 10^{20} \Big(\frac{1}{m_{\text{minor}}} + \frac{1}{m_{\text{major}}} \Big)^{0.5} T^{0.5} ~~~ \text{[/m/s]},
\end{equation*}
where $m_{\text{minor}}$ and $m_{\text{major}}$ are the molecular masses of the minor (light) species and the major (heavy) species (noting that we assume the heavy species is the mean molecular mass of the species that are not escaping). The crossover mass represents the upper limit on the `lifting' action of hydrogen as it escapes, meaning that anything with a smaller atomic weight than the crossover mass can become entrained in the escaping hydrogen and ultimately removed from the planet. 

Because the escaping component is not necessarily just hydrogen, the presence of heavier atoms and molecules changes the crossover mass and the flux of escaping material. Initially, the escaping material is assumed to be atomic hydrogen, and the crossover mass in Eqn. \ref{app:eq4} is calculated using just the mass of hydrogen ($m_{H}$ in the denominator). We then extend the formalism adopted in \citet[][their Section 2.4.2]{luger2015extreme}, defining the flux of each species in relation to the flux of hydrogen and the crossover mass:

\begin{equation} \label{app:eq5}
    \Phi_{x} = \frac{\xi_{x}}{f_{H}} \Phi_{H} \Big(\frac{m_{c} - m_{x}}{m_{c} - m_{H}}\Big).
\end{equation}
To calculate the flux of each species in the escaping flow, we can generalize Eqn. 6 of \citet{luger2015extreme}:
 
\begin{equation} \label{app:eq6}
    m_{H} \Phi_{H}^{\text{ref}} = \sum_{x}^{m_{x} < m_{c}} m_{x} \Phi_{x}.
\end{equation}
Substituting Eqn. \ref{app:eq5} into this relationship, we see that:

\begin{align}
    m_{H} \Phi_{H}^{\text{ref}} &= \sum_{x}^{m_{x} < m_{c}} m_{x} \frac{\xi_{x}}{f_{H}} \Phi_{H} \Big(\frac{m_{c} - m_{x}}{m_{c} - m_{H}}\Big) \\ 
    &= \frac{\Phi_{H}}{(m_{c} - m_{H})f_{H}} \sum_{x}^{m_{x} < m_{c}} m_{x} \xi_{x} \Big(\frac{m_{c} - m_{x}}{m_{c} - m_{H}}\Big) \\
    \frac{\Phi_{H}}{m_{H} f_{H}} &= \Phi_{H}^{\text{ref}}(m_{c} - m_{H}) \Bigg[\sum_{x}^{m_{x} < m_{c}} m_{x} \xi_{x} \Big(m_{c} - m_{x}\Big)\Bigg]^{-1}
\end{align}
If we substitute this expression back into Eqn. \ref{app:eq4}:

\begin{align}
    \mu_{c} =& \mu_{H} + \frac{k_{b} T \Phi_{H}^{\text{ref}}}{b g} (m_{c} - m_{H}) \Bigg[\sum_{x}^{m_{x} < m_{c}} m_{x} \xi_{x} \Big(m_{c} - m_{x}\Big)\Bigg]^{-1}\\
    \frac{m_{c} - m_{H}}{m_{H}} =& \frac{k_{b} T \Phi_{H}^{\text{ref}}}{b g}(m_{c} - m_{H}) \Bigg[\sum_{x}^{m_{x} < m_{c}} m_{x} \xi_{x} \Big(m_{c} - m_{x}\Big)\Bigg]^{-1}\\
    \frac{1}{m_{H}} =& \frac{k_{b} T \Phi_{H}^{\text{ref}}}{b g} \Bigg[\sum_{x}^{m_{x} < m_{c}} m_{x} \xi_{x} \Big(m_{c} - m_{x}\Big)\Bigg]^{-1}\\
    \sum_{x}^{m_{x} < m_{c}} m_{x} \xi_{x} \Big(m_{c} - m_{x}\Big)  =& \frac{k_{b} T  \Phi_{H}^{\text{ref}} m_{H}}{b g}\\
    \sum_{x}^{m_{x} < m_{c}} m_{x} \xi_{x} m_{c} - \sum_{x}^{m_{x} < m_{c}} m_{x} \xi_{x} m_{x}  =& \frac{k_{b} T  \Phi_{H}^{\text{ref}} m_{H}}{b g}\\
    m_{c} \sum_{x}^{m_{x} < m_{c}} m_{x} \xi_{x}  =& \frac{k_{b} T  \Phi_{H}^{\text{ref}} m_{H}}{b g} + \sum_{x}^{m_{x} < m_{c}} m_{x} \xi_{x} m_{x}\\
    m_{c}  =& \Bigg[\sum_{x}^{m_{x} < m_{c}} m_{x} \xi_{x}\Bigg]^{-1} \bigg(\frac{k_{b} T  \Phi_{H}^{\text{ref}} m_{H}}{b g} + \sum_{x}^{m_{x} < m_{c}} m_{x} \xi_{x} m_{x}\bigg)\\
    \mu_{c}  =& \Bigg[\sum_{x}^{m_{x} < m_{c}} m_{x} \xi_{x}\Bigg]^{-1} \bigg(\frac{k_{b} T  \Phi_{H}^{\text{ref}}}{b g} + \sum_{x}^{m_{x} < m_{c}} m_{x} \xi_{x} \mu_{x}\bigg)
\end{align} 
As the sum of the mixing-ratio-weighted molecular masses gives the mean molecular mass, this expression simplifies to:
\begin{equation}
    \mu_{c}  = \frac{k_{b} T  \Phi_{H}^{\text{ref}}}{b g \bar{\mu}} + \frac{1}{\bar{\mu}}\sum_{x}^{m_{x} < m_{c}} m_{x} \xi_{x} \mu_{x}
\end{equation}
We estimate escape and the changing mean molecular weight of the flow by sequentially including each potential species in reverse order by mass (in our modeling, this would be helium, then atomic oxygen, then water, and so on). Each sequential addition forces us to check whether the new crossover mass (which has a slower relative flux but higher mean molecular weight) is still in excess of the molecular mass of the next-heaviest component, only exiting if this is not the case. Because the additional species is added before the crossover mass is calculated, the last loop exits without updating the crossover mass, ensuring that all the incorporated species are represented without adding a repeated term. Subsequent to this loop, the revised escape fluxes are computed (per Eqn. \ref{app:eq5}). Then the escaping mass is removed from the relevant inventories such that, if no hydrogen remains in the atmosphere, the hydrogen escape is subtracted from the water inventory (leaving behind the oxygen atoms), and any resulting negative inventories are instead set to zero and the remainder of the mass removed as hydrogen (using the same relationship in Eqn. \ref{app:eq5}). 
}
\subsection*{\texttt{run\_escape} - returning values to \texttt{monte\_carlo.py}}

The code currently returns summary statistics for individual runs to \texttt{monte\_carlo.py}, including the planet's mass, radius, core mass fraction, envelope fraction and composition, and total mass change over the duration of the simulation (this is always longer than the estimated age of the host star in this parameter sweep). Individual time evolution data is not returned, in an effort to provide manageable data volumes and run times; by default, the simulations have 3,000 points across the whole of the available stellar evolution data, and only returning 3 of each of the reported parameters is a thousand-fold reduction in data.

\section{Photochemical reaction list and thermodynamic data \label{app:reactions}}

\begin{longtable}{l|c|c|c}
\caption{Reactions and rates.}\\
 Rxn \# & Reaction & Rate & Notes \\ \hline \endfirsthead 
 Rxn \# & Reaction & Rate & Notes \\ \hline \endhead 
1 & H + H$_{2}$O $\rightarrow$ OH + H$_{2}$ &  $ 7.500\times 10^{-16} $ $\cdot T^{1.60}$$\cdot$ exp(-9720/T)& 1 \\
3 & O + H$_{2}$ $\rightarrow$ OH + H &  $ 8.520\times 10^{-20} $ $\cdot T^{2.67}$$\cdot$ exp(-3160/T)& 1 \\
5 & O + H$_{2}$O $\rightarrow$ OH + OH &  $ 8.200\times 10^{-14} $ $\cdot T^{0.95}$$\cdot$ exp(-8570/T)& 1 \\ \hline 
7 & H + CH $\rightarrow$ H$_{2}$ + C &  $ 1.310\times 10^{-10} $ $\cdot$ exp(-80/T)& 1 \\
9 & H + CH$_{2}$ $\rightarrow$ CH + H$_{2}$ &  $ 1.000\times 10^{-11} $ $\cdot$ exp(900/T)& 1 \\
11 & CH$_{2}$ + H$_{2}$ $\rightarrow$ H + CH$_{3}$ &  $ 7.320\times 10^{-19} $ $\cdot T^{2.30}$$\cdot$ exp(-3699/T)& 1 \\
13 & H + CH$_{4}$ $\rightarrow$ CH$_{3}$ + H$_{2}$ &  $ 2.200\times 10^{-20} $ $\cdot T^{3.00}$$\cdot$ exp(-4040/T)& 1 \\
15 & C + CH $\rightarrow$ C$_{2}$ + H &  $ 1.050\times 10^{-12} $ $\cdot T^{0.50}$& 1 \\ \hline 
17 & H$_{2}$ + C$_{2}$H $\rightarrow$ H + C$_{2}$H$_{2}$ &  $ 9.200\times 10^{-18} $ $\cdot T^{2.17}$$\cdot$ exp(-478/T)& 1 \\
19 & CH + CH$_{2}$ $\rightarrow$ H + C$_{2}$H$_{2}$ &  $ 6.640\times 10^{-11} $ & 1 \\
21 & H + C$_{2}$H$_{3}$ $\rightarrow$ C$_{2}$H$_{2}$ + H$_{2}$ &  $ 2.010\times 10^{-11} $ & 1 \\
23 & H$_{2}$ + C$_{2}$H$_{3}$ $\rightarrow$ H + C$_{2}$H$_{4}$ &  $ 5.000\times 10^{-20} $ $\cdot T^{2.63}$$\cdot$ exp(-4300/T)& 1 \\
25 & CH + CH$_{4}$ $\rightarrow$ H + C$_{2}$H$_{4}$ &  $ 5.000\times 10^{-11} $ $\cdot$ exp(200/T)& 1 \\ \hline 
27 & CH$_{2}$ + CH$_{3}$ $\rightarrow$ H + C$_{2}$H$_{4}$ &  $ 7.010\times 10^{-11} $ & 1 \\
29 & H + C$_{2}$H$_{5}$ $\rightarrow$ CH$_{3}$ + CH$_{3}$ &  $ 5.990\times 10^{-11} $ & 1 \\
31 & H + C$_{2}$H$_{5}$ $\rightarrow$ C$_{2}$H$_{4}$ + H$_{2}$ &  $ 3.010\times 10^{-12} $ & 1 \\
33 & H + C$_{2}$H$_{6}$ $\rightarrow$ C$_{2}$H$_{5}$ + H$_{2}$ &  $ 9.190\times 10^{-22} $ $\cdot T^{3.50}$$\cdot$ exp(-2600/T)& 1 \\
35 & OH + CO $\rightarrow$ H + CO$_{2}$ &  $ 1.050\times 10^{-17} $ $\cdot T^{1.50}$$\cdot$ exp(259/T)& 1 \\ \hline 
37 & CH + CH$_{3}$ $\rightarrow$ H$_{2}$ + C$_{2}$H$_{2}$ &  $ 1.000\times 10^{-11} $ & 1 \\
39 & C$_{2}$ + O $\rightarrow$ C + CO &  $ 1.050\times 10^{-12} $ & 1 \\
41 & CH$_{2}$ + CH$_{2}$ $\rightarrow$ C$_{2}$H$_{2}$ + H + H &  $ 1.800\times 10^{-10} $ $\cdot$ exp(-400/T)& 1 \\
43 & CH$_{2}$ + CH$_{2}$ $\rightarrow$ CH + CH$_{3}$ &  $ 3.980\times 10^{-10} $ $\cdot$ exp(-5000/T)& 1 \\
45 & CH$_{2}$ + CH$_{2}$ $\rightarrow$ H + C$_{2}$H$_{3}$ &  $ 3.320\times 10^{-11} $ & 1 \\ \hline 
47 & CH$_{2}$ + CH$_{4}$ $\rightarrow$ CH$_{3}$ + CH$_{3}$ &  $ 4.090\times 10^{-18} $ $\cdot T^{2.00}$$\cdot$ exp(-4162/T)& 1 \\
49 & CH$_{2}$ + C$_{2}$H$_{5}$ $\rightarrow$ CH$_{3}$ + C$_{2}$H$_{4}$ &  $ 3.010\times 10^{-11} $ & 1 \\
51 & CH$_{3}$ + OH $\rightarrow$ CH$_{2}$ + H$_{2}$O &  $ 1.850\times 10^{-21} $ $\cdot T^{3.00}$$\cdot$ exp(-1400/T)& 1 \\
53 & C$_{2}$H + OH $\rightarrow$ CH$_{2}$ + CO &  $ 3.010\times 10^{-11} $ & 1 \\
55 & C$_{2}$H$_{2}$ + O $\rightarrow$ CH$_{2}$ + CO &  $ 6.780\times 10^{-16} $ $\cdot T^{1.50}$$\cdot$ exp(-854/T)& 1 \\ \hline 
57 & CH$_{3}$ + C$_{2}$H $\rightarrow$ C$_{2}$H$_{2}$ + CH$_{2}$ &  $ 1.000\times 10^{-11} $ & 1 \\
59 & CH$_{4}$ + C$_{2}$H $\rightarrow$ CH$_{3}$ + C$_{2}$H$_{2}$ &  $ 3.010\times 10^{-12} $ $\cdot$ exp(-250/T)& 1 \\
61 & CH$_{3}$ + C$_{2}$H$_{3}$ $\rightarrow$ CH$_{4}$ + C$_{2}$H$_{2}$ &  $ 6.510\times 10^{-13} $ & 1 \\
63 & CH$_{4}$ + C$_{2}$H$_{3}$ $\rightarrow$ CH$_{3}$ + C$_{2}$H$_{4}$ &  $ 2.130\times 10^{-24} $ $\cdot T^{4.02}$$\cdot$ exp(-2754/T)& 1 \\
65 & CH$_{3}$ + C$_{2}$H$_{5}$ $\rightarrow$ CH$_{4}$ + C$_{2}$H$_{4}$ &  $ 3.250\times 10^{-11} $ $\cdot T^{-0.50}$& 1 \\ \hline 
67 & CH$_{3}$ + C$_{2}$H$_{6}$ $\rightarrow$ CH$_{4}$ + C$_{2}$H$_{5}$ &  $ 9.120\times 10^{-25} $ $\cdot T^{4.00}$$\cdot$ exp(-4170/T)& 1 \\
69 & C$_{2}$H$_{2}$ + OH $\rightarrow$ CH$_{3}$ + CO &  $ 8.040\times 10^{-28} $ $\cdot T^{4.00}$$\cdot$ exp(1010/T)& 1 \\
71 & C$_{2}$ + H$_{2}$ $\rightarrow$ H + C$_{2}$H &  $ 1.100\times 10^{-10} $ $\cdot$ exp(-4000/T)& 1 \\
73 & C$_{2}$ + CH$_{4}$ $\rightarrow$ CH$_{3}$ + C$_{2}$H &  $ 5.050\times 10^{-11} $ $\cdot$ exp(-297/T)& 1 \\
75 & C$_{2}$H + CH$_{2}$ $\rightarrow$ CH + C$_{2}$H$_{2}$ &  $ 3.010\times 10^{-11} $ & 1 \\ \hline 
77 & C$_{2}$H + C$_{2}$H$_{6}$ $\rightarrow$ C$_{2}$H$_{2}$ + C$_{2}$H$_{5}$ &  $ 5.990\times 10^{-12} $ & 1 \\
79 & C$_{2}$H + O $\rightarrow$ CH + CO &  $ 1.690\times 10^{-11} $ & 1 \\
81 & C$_{2}$H + OH $\rightarrow$ C$_{2}$H$_{2}$ + O &  $ 3.010\times 10^{-11} $ & 1 \\
83 & C$_{2}$H + H$_{2}$O $\rightarrow$ C$_{2}$H$_{2}$ + OH &  $ 2.200\times 10^{-21} $ $\cdot T^{3.05}$$\cdot$ exp(-376/T)& 1 \\
85 & C$_{2}$H$_{3}$ + C$_{2}$H$_{3}$ $\rightarrow$ C$_{2}$H$_{4}$ + C$_{2}$H$_{2}$ &  $ 1.600\times 10^{-12} $ & 1 \\ \hline 
87 & C$_{2}$H$_{3}$ + C$_{2}$H$_{5}$ $\rightarrow$ C$_{2}$H$_{4}$ + C$_{2}$H$_{4}$ &  $ 8.000\times 10^{-13} $ & 1 \\
89 & C$_{2}$H$_{3}$ + C$_{2}$H$_{5}$ $\rightarrow$ C$_{2}$H$_{6}$ + C$_{2}$H$_{2}$ &  $ 8.000\times 10^{-13} $ & 1 \\
91 & C$_{2}$H$_{4}$ + OH $\rightarrow$ C$_{2}$H$_{3}$ + H$_{2}$O &  $ 2.600\times 10^{-20} $ $\cdot T^{2.75}$$\cdot$ exp(-2100/T)& 1 \\
93 & C$_{2}$H$_{5}$ + C$_{2}$H$_{5}$ $\rightarrow$ C$_{2}$H$_{4}$ + C$_{2}$H$_{6}$ &  $ 2.310\times 10^{-12} $ & 1 \\
95 & C$_{2}$H$_{5}$ + C$_{2}$H$_{4}$ $\rightarrow$ C$_{2}$H$_{3}$ + C$_{2}$H$_{6}$ &  $ 1.050\times 10^{-21} $ $\cdot T^{3.13}$$\cdot$ exp(-9060/T)& 1 \\ \hline 
97 & C$_{2}$H$_{6}$ + OH $\rightarrow$ C$_{2}$H$_{5}$ + H$_{2}$O &  $ 1.470\times 10^{-14} $ $\cdot T^{1.04}$$\cdot$ exp(-913/T)& 1 \\
99 & CH + O $\rightarrow$ OH + C &  $ 2.520\times 10^{-11} $ $\cdot$ exp(-2381/T)& 1 \\
101 & O + CH $\rightarrow$ H + CO &  $ 6.590\times 10^{-11} $ & 1 \\
103 & O + CH$_{3}$ $\rightarrow$ CH$_{2}$ + OH &  $ 1.000\times 10^{-11} $ $\cdot$ exp(-3970/T)& 1 \\
105 & CH$_{3}$ + OH $\rightarrow$ O + CH$_{4}$ &  $ 1.160\times 10^{-19} $ $\cdot T^{2.20}$$\cdot$ exp(-2240/T)& 1 \\ \hline 
107 & O + C$_{2}$H$_{6}$ $\rightarrow$ OH + C$_{2}$H$_{5}$ &  $ 2.000\times 10^{-12} $ $\cdot T^{0.60}$$\cdot$ exp(-3680/T)& 1 \\
109 & OH + C $\rightarrow$ CO + H &  $ 1.050\times 10^{-12} $ $\cdot T^{0.50}$& 1 \\
111 & OH + CH$_{2}$ $\rightarrow$ H$_{2}$O + CH &  $ 1.430\times 10^{-18} $ $\cdot T^{2.02}$$\cdot$ exp(-3410/T)& 1 \\
113 & OH + CH$_{4}$ $\rightarrow$ H$_{2}$O + CH$_{3}$ &  $ 3.190\times 10^{-19} $ $\cdot T^{2.40}$$\cdot$ exp(-1060/T)& 1 \\
115 & OH + C$_{2}$H$_{3}$ $\rightarrow$ H$_{2}$O + C$_{2}$H$_{2}$ &  $ 5.000\times 10^{-11} $ & 1 \\ \hline 
117 & OH + C$_{2}$H$_{5}$ $\rightarrow$ H$_{2}$O + C$_{2}$H$_{4}$ &  $ 4.000\times 10^{-11} $ & 1 \\
119 & CH$_{2}$OH + H $\rightarrow$ OH + CH$_{3}$ &  $ 1.600\times 10^{-10} $ & 1 \\
121 & H$_{2}$CO + H $\rightarrow$ HCO + H$_{2}$ &  $ 3.640\times 10^{-16} $ $\cdot T^{1.77}$$\cdot$ exp(-1510/T)& 1 \\
123 & O + C$_{2}$H$_{4}$ $\rightarrow$ HCO + CH$_{3}$ &  $ 2.190\times 10^{-16} $ $\cdot T^{1.55}$$\cdot$ exp(-215/T)& 1 \\
125 & H$_{2}$CO + CH$_{3}$ $\rightarrow$ CH$_{4}$ + HCO &  $ 9.200\times 10^{-21} $ $\cdot T^{2.81}$$\cdot$ exp(-2950/T)& 1 \\ \hline 
127 & CH$_{3}$ + CH$_{2}$OH $\rightarrow$ H$_{2}$CO + CH$_{4}$ &  $ 4.000\times 10^{-12} $ & 1 \\
129 & HCO + H $\rightarrow$ CO + H$_{2}$ &  $ 1.500\times 10^{-10} $ & 1 \\
131 & HCO + OH $\rightarrow$ CO + H$_{2}$O &  $ 1.690\times 10^{-10} $ & 1 \\
133 & CO$_{2}$ + CH $\rightarrow$ HCO + CO &  $ 5.710\times 10^{-12} $ $\cdot$ exp(-345/T)& 1 \\
135 & CH$_{3}$ + O $\rightarrow$ H$_{2}$CO + H &  $ 1.400\times 10^{-10} $ & 1 \\ \hline 
137 & CH$_{3}$O + O $\rightarrow$ H$_{2}$CO + OH &  $ 1.000\times 10^{-11} $ & 1 \\
139 & CH$_{3}$O + OH $\rightarrow$ H$_{2}$CO + H$_{2}$O &  $ 3.010\times 10^{-11} $ & 1 \\
141 & CH$_{3}$OH + H $\rightarrow$ CH$_{3}$O + H$_{2}$ &  $ 6.820\times 10^{-20} $ $\cdot T^{2.69}$$\cdot$ exp(-4643/T)& 1 \\
143 & CH$_{3}$OH + H $\rightarrow$ CH$_{3}$ + H$_{2}$O &  $ 4.910\times 10^{-19} $ $\cdot T^{2.49}$$\cdot$ exp(-10380/T)& 1 \\
145 & CH$_{2}$ + O $\rightarrow$ CO + H + H &  $ 1.330\times 10^{-10} $ & 1 \\ \hline 
147 & CH$_{2}$ + OH $\rightarrow$ H$_{2}$CO + H &  $ 3.010\times 10^{-11} $ & 1 \\
149 & CO$_{2}$ + CH$_{2}$ $\rightarrow$ H$_{2}$CO + CO &  $ 3.900\times 10^{-14} $ & 1 \\
151 & CH$_{3}$O + CO $\rightarrow$ CH$_{3}$ + CO$_{2}$ &  $ 2.610\times 10^{-11} $ $\cdot$ exp(-5940/T)& 1 \\
153 & CH$_{3}$OH + H $\rightarrow$ CH$_{2}$OH + H$_{2}$ &  $ 1.090\times 10^{-19} $ $\cdot T^{2.73}$$\cdot$ exp(-2240/T)& 1 \\
155 & HCO + C$_{2}$H $\rightarrow$ C$_{2}$H$_{2}$ + CO &  $ 1.000\times 10^{-10} $ & 1 \\ \hline 
157 & CH$_{2}$OH + C$_{2}$H $\rightarrow$ H$_{2}$CO + C$_{2}$H$_{2}$ &  $ 5.990\times 10^{-11} $ & 1 \\
159 & CH$_{3}$O + C$_{2}$H $\rightarrow$ H$_{2}$CO + C$_{2}$H$_{2}$ &  $ 4.000\times 10^{-11} $ & 1 \\
161 & CH$_{3}$OH + C$_{2}$H $\rightarrow$ CH$_{2}$OH + C$_{2}$H$_{2}$ &  $ 1.000\times 10^{-11} $ & 1 \\
163 & CH$_{3}$OH + C$_{2}$H $\rightarrow$ CH$_{3}$O + C$_{2}$H$_{2}$ &  $ 2.010\times 10^{-12} $ & 1 \\
165 & O + C$_{2}$H$_{3}$ $\rightarrow$ C$_{2}$H$_{2}$ + OH &  $ 1.760\times 10^{-12} $ $\cdot T^{0.20}$$\cdot$ exp(215/T)& 1 \\ \hline 
167 & CH$_{2}$ + C$_{2}$H$_{3}$ $\rightarrow$ C$_{2}$H$_{2}$ + CH$_{3}$ &  $ 3.000\times 10^{-11} $ & 1 \\
169 & O + CH$_{2}$ $\rightarrow$ CO + H$_{2}$ &  $ 9.960\times 10^{-11} $ & 1 \\
171 & O + C$_{2}$H$_{3}$ $\rightarrow$ HCO + CH$_{2}$ &  $ 2.000\times 10^{-11} $ & 1 \\
173 & HCO + CH$_{2}$ $\rightarrow$ CO + CH$_{3}$ &  $ 3.010\times 10^{-11} $ & 1 \\
175 & O + C$_{2}$H$_{4}$ $\rightarrow$ H$_{2}$CO + CH$_{2}$ &  $ 1.350\times 10^{-17} $ $\cdot T^{1.80}$$\cdot$ exp(-90/T)& 1 \\ \hline 
177 & CH$_{2}$OH + CH$_{2}$ $\rightarrow$ OH + C$_{2}$H$_{4}$ &  $ 4.000\times 10^{-11} $ & 1 \\
179 & CH$_{2}$OH + CH$_{2}$ $\rightarrow$ H$_{2}$CO + CH$_{3}$ &  $ 2.010\times 10^{-12} $ & 1 \\
181 & CH$_{3}$O + CH$_{2}$ $\rightarrow$ H$_{2}$CO + CH$_{3}$ &  $ 3.000\times 10^{-11} $ & 1 \\
183 & CH$_{3}$OH + CH$_{2}$ $\rightarrow$ CH$_{3}$O + CH$_{3}$ &  $ 2.390\times 10^{-23} $ $\cdot T^{3.10}$$\cdot$ exp(-3490/T)& 1 \\
185 & CH$_{3}$OH + CH$_{2}$ $\rightarrow$ CH$_{2}$OH + CH$_{3}$ &  $ 5.290\times 10^{-23} $ $\cdot T^{3.20}$$\cdot$ exp(-3609/T)& 1 \\ \hline 
187 & HCO + CH$_{3}$ $\rightarrow$ CO + CH$_{4}$ &  $ 2.010\times 10^{-10} $ & 1 \\
189 & CH$_{3}$O + CH$_{3}$ $\rightarrow$ H$_{2}$CO + CH$_{4}$ &  $ 4.000\times 10^{-11} $ & 1 \\
191 & H$_{2}$CO + CH $\rightarrow$ CO + CH$_{3}$ &  $ 8.000\times 10^{-11} $ $\cdot$ exp(260/T)& 1 \\
193 & CH$_{3}$OH + CH$_{3}$ $\rightarrow$ CH$_{3}$O + CH$_{4}$ &  $ 2.390\times 10^{-23} $ $\cdot T^{3.10}$$\cdot$ exp(-3490/T)& 1 \\
195 & CH$_{3}$CO + H $\rightarrow$ HCO + CH$_{3}$ &  $ 3.320\times 10^{-11} $ & 1 \\ \hline 
197 & CH$_{3}$CO + CH$_{3}$ $\rightarrow$ CO + C$_{2}$H$_{6}$ &  $ 4.900\times 10^{-11} $ & 1 \\
199 & O + OH $\rightarrow$ O$_{2}$ + H &  $ 7.470\times 10^{-10} $ $\cdot T^{-0.50}$$\cdot$ exp(-30/T)& 1 \\
201 & H + CH$_{3}$O $\rightarrow$ H$_{2}$CO + H$_{2}$ &  $ 3.010\times 10^{-11} $ & 1 \\
203 & H + CH$_{2}$CO $\rightarrow$ CO + CH$_{3}$ &  $ 1.290\times 10^{-15} $ $\cdot T^{1.45}$$\cdot$ exp(-1399/T)& 1 \\
205 & O + C$_{2}$H$_{3}$ $\rightarrow$ CH$_{2}$CO + H &  $ 1.600\times 10^{-10} $ & 1 \\ \hline 
207 & C$_{2}$H$_{2}$ + O $\rightarrow$ HCCO + H &  $ 1.500\times 10^{-11} $ $\cdot$ exp(-2280/T)& 1 \\
209 & HCCO + H $\rightarrow$ CO + CH$_{2}$ &  $ 2.490\times 10^{-10} $ & 1 \\
211 & O + H$_{2}$CO $\rightarrow$ HCO + OH &  $ 6.850\times 10^{-13} $ $\cdot T^{0.57}$$\cdot$ exp(-1390/T)& 1 \\
213 & HCO + HCO $\rightarrow$ H$_{2}$CO + CO &  $ 3.010\times 10^{-11} $ & 1 \\
215 & CH$_{2}$OH + CH$_{3}$O $\rightarrow$ H$_{2}$CO + CH$_{3}$OH &  $ 4.000\times 10^{-11} $ & 1 \\ \hline 
217 & CH$_{3}$O + CH$_{3}$O $\rightarrow$ H$_{2}$CO + CH$_{3}$OH &  $ 1.000\times 10^{-10} $ & 1 \\
219 & H + H $\rightarrow$ H$_{2}$ + M & $\Bigg \{$ \parbox{8cm}{\centering k$_{0}$ =  $ 2.700\times 10^{-31} $ $\cdot T^{-0.60}$ \\  k$_{\infty}$ =  $ 3.310\times 10^{-06} $ $\cdot T^{-1.00}$ } & 2 \\
221 & H + O $\rightarrow$ OH + M & $\Bigg \{$ \parbox{8cm}{\centering k$_{0}$ =  $ 1.300\times 10^{-29} $ $\cdot T^{-1.00}$ \\  k$_{\infty}$ =  $ 1.000\times 10^{-11} $  } & 2 \\
223 & OH + H $\rightarrow$ H$_{2}$O + M & $\Bigg \{$ \parbox{8cm}{\centering k$_{0}$ =  $ 3.890\times 10^{-25} $ $\cdot T^{-2.00}$ \\  k$_{\infty}$ =  $ 4.260\times 10^{-11} $ $\cdot T^{0.23}$ } & 2 \\
225 & H + CH $\rightarrow$ CH$_{2}$ + M & $\Bigg \{$ \parbox{8cm}{\centering k$_{0}$ =  $ 2.760\times 10^{-29} $ $\cdot T^{-1.00}$ \\  k$_{\infty}$ =  $ 1.000\times 10^{-12} $  } & 2 \\ \hline 
227 & H + CH$_{3}$ $\rightarrow$ CH$_{4}$ + M & $\Bigg \{$ \parbox{8cm}{\centering k$_{0}$ =  $ 1.760\times 10^{-24} $ $\cdot T^{-1.80}$ \\  k$_{\infty}$ =  $ 2.060\times 10^{-10} $ $\cdot T^{-0.40}$ } & 2 \\
229 & H + C$_{2}$H$_{2}$ $\rightarrow$ C$_{2}$H$_{3}$ + M & $\Bigg \{$ \parbox{8cm}{\centering k$_{0}$ =  $ 1.050\times 10^{-07} $ $\cdot T^{-7.27}$$\cdot$ exp(-3630/T) \\  k$_{\infty}$ =  $ 9.130\times 10^{-12} $ $\cdot$ exp(-3630/T) } & 2 \\
231 & H + C$_{2}$H$_{3}$ $\rightarrow$ C$_{2}$H$_{4}$ + M & $\Bigg \{$ \parbox{8cm}{\centering k$_{0}$ =  $ 1.500\times 10^{-27} $  \\  k$_{\infty}$ =  $ 6.400\times 10^{-11} $ $\cdot T^{0.20}$ } & 2 \\
233 & H + C$_{2}$H$_{4}$ $\rightarrow$ C$_{2}$H$_{5}$ + M & $\Bigg \{$ \parbox{8cm}{\centering k$_{0}$ =  $ 7.690\times 10^{-30} $ $\cdot$ exp(-380/T) \\  k$_{\infty}$ =  $ 1.270\times 10^{-15} $ $\cdot T^{1.49}$$\cdot$ exp(-380/T) } & 2 \\
235 & H + C$_{2}$H$_{5}$ $\rightarrow$ C$_{2}$H$_{6}$ + M & $\Bigg \{$ \parbox{8cm}{\centering k$_{0}$ =  $ 4.000\times 10^{-19} $ $\cdot T^{-3.00}$$\cdot$ exp(-600/T) \\  k$_{\infty}$ =  $ 9.040\times 10^{-11} $ $\cdot T^{0.16}$$\cdot$ exp(-600/T) } & 2 \\ \hline 
237 & H$_{2}$ + C $\rightarrow$ CH$_{2}$ + M & $\Bigg \{$ \parbox{8cm}{\centering k$_{0}$ =  $ 6.890\times 10^{-32} $  \\  k$_{\infty}$ =  $ 2.060\times 10^{-11} $  } & 2 \\
239 & CH + M $\rightarrow$ C + H + M & $\Bigg \{$ \parbox{8cm}{\centering k$_{0}$ =  $ 3.160\times 10^{-10} $ $\cdot$ exp(-33700/T) \\  k$_{\infty}$ =  $ 1.000\times 10^{-12} $ $\cdot$ exp(-33700/T) } & 2 \\
241 & CH$_{2}$ + H $\rightarrow$ CH$_{3}$ + M & $\Bigg \{$ \parbox{8cm}{\centering k$_{0}$ =  $ 9.000\times 10^{-32} $ $\cdot$ exp(550/T) \\  k$_{\infty}$ =  $ 8.550\times 10^{-12} $ $\cdot T^{0.15}$$\cdot$ exp(550/T) } & 2 \\
243 & CH + H$_{2}$ $\rightarrow$ CH$_{3}$ + M & $\Bigg \{$ \parbox{8cm}{\centering k$_{0}$ =  $ 3.400\times 10^{-31} $ $\cdot$ exp(736/T) \\  k$_{\infty}$ =  $ 7.300\times 10^{-11} $ $\cdot$ exp(736/T) } & 2 \\
245 & CH$_{3}$ + CH$_{3}$ $\rightarrow$ C$_{2}$H$_{6}$ + M & $\Bigg \{$ \parbox{8cm}{\centering k$_{0}$ =  $ 3.500\times 10^{-07} $ $\cdot T^{-7.00}$$\cdot$ exp(-1390/T) \\  k$_{\infty}$ =  $ 1.580\times 10^{-09} $ $\cdot T^{-0.54}$$\cdot$ exp(-1390/T) } & 2 \\ \hline 
247 & C$_{2}$H + H $\rightarrow$ C$_{2}$H$_{2}$ + M & $\Bigg \{$ \parbox{8cm}{\centering k$_{0}$ =  $ 1.260\times 10^{-18} $ $\cdot T^{-3.10}$$\cdot$ exp(-721/T) \\  k$_{\infty}$ =  $ 3.000\times 10^{-10} $ $\cdot$ exp(-721/T) } & 2 \\
249 & C$_{2}$H$_{4}$ + M $\rightarrow$ C$_{2}$H$_{2}$ + H$_{2}$ + M & $\Bigg \{$ \parbox{8cm}{\centering k$_{0}$ =  $ 5.800\times 10^{-08} $ $\cdot$ exp(-36000/T) \\  k$_{\infty}$ =  $ 7.950\times 10^{+12} $ $\cdot T^{0.44}$$\cdot$ exp(-36000/T) } & 2 \\
251 & C$_{2}$H$_{6}$ + M $\rightarrow$ C$_{2}$H$_{4}$ + H$_{2}$ + M & $\Bigg \{$ \parbox{8cm}{\centering k$_{0}$ =  $ 3.800\times 10^{-07} $ $\cdot$ exp(-34000/T) \\  k$_{\infty}$ =  $ 1.320\times 10^{+15} $ $\cdot$ exp(-34000/T) } & 2 \\
253 & CO + O $\rightarrow$ CO$_{2}$ + M & $\Bigg \{$ \parbox{8cm}{\centering k$_{0}$ =  $ 1.700\times 10^{-33} $ $\cdot$ exp(-1510/T) \\  k$_{\infty}$ =  $ 2.660\times 10^{-14} $ $\cdot$ exp(-1510/T) } & 2 \\
255 & CH$_{2}$OH + M $\rightarrow$ H + H$_{2}$CO + M & $\Bigg \{$ \parbox{8cm}{\centering k$_{0}$ =  $ 1.660\times 10^{-10} $ $\cdot$ exp(-12630/T) \\  k$_{\infty}$ =  $ 3.000\times 10^{+09} $ $\cdot$ exp(-12630/T) } & 2 \\ \hline 
257 & H + CO $\rightarrow$ HCO + M & $\Bigg \{$ \parbox{8cm}{\centering k$_{0}$ =  $ 5.290\times 10^{-34} $ $\cdot$ exp(-370/T) \\  k$_{\infty}$ =  $ 1.960\times 10^{-13} $ $\cdot$ exp(-370/T) } & 2 \\
259 & H$_{2}$O + CH $\rightarrow$ CH$_{2}$OH + M & $\Bigg \{$ \parbox{8cm}{\centering k$_{0}$ =  $ 1.000\times 10^{-31} $  \\  k$_{\infty}$ =  $ 9.480\times 10^{-12} $  } & 2 \\
261 & CH$_{3}$O + M $\rightarrow$ H + H$_{2}$CO + M & $\Bigg \{$ \parbox{8cm}{\centering k$_{0}$ =  $ 9.000\times 10^{-11} $ $\cdot$ exp(-6790/T) \\  k$_{\infty}$ =  $ 1.560\times 10^{+15} $ $\cdot T^{-0.39}$$\cdot$ exp(-6790/T) } & 2 \\
263 & CH$_{2}$OH + H $\rightarrow$ CH$_{3}$OH + M & $\Bigg \{$ \parbox{8cm}{\centering k$_{0}$ =  $ 1.200\times 10^{-16} $ $\cdot T^{-4.65}$$\cdot$ exp(-2557/T) \\  k$_{\infty}$ =  $ 2.300\times 10^{-10} $ $\cdot T^{0.04}$$\cdot$ exp(-2557/T) } & 2 \\
265 & OH + C$_{2}$H$_{2}$ $\rightarrow$ CH$_{3}$CO + M & $\Bigg \{$ \parbox{8cm}{\centering k$_{0}$ =  $ 4.990\times 10^{-25} $ $\cdot T^{-2.00}$ \\  k$_{\infty}$ =  $ 1.060\times 10^{-07} $ $\cdot T^{-1.90}$ } & 2 \\ \hline 
267 & CO + CH$_{3}$ $\rightarrow$ CH$_{3}$CO + M & $\Bigg \{$ \parbox{8cm}{\centering k$_{0}$ =  $ 3.950\times 10^{-10} $ $\cdot T^{-7.50}$$\cdot$ exp(-5490/T) \\  k$_{\infty}$ =  $ 5.140\times 10^{-19} $ $\cdot T^{2.20}$$\cdot$ exp(-5490/T) } & 2 \\
269 & HCO + H $\rightarrow$ H$_{2}$CO + M & $\Bigg \{$ \parbox{8cm}{\centering k$_{0}$ =  $ 7.330\times 10^{-24} $ $\cdot T^{-2.57}$$\cdot$ exp(-215/T) \\  k$_{\infty}$ =  $ 7.770\times 10^{-14} $ $\cdot$ exp(-215/T) } & 2 \\
271 & CO + H$_{2}$ $\rightarrow$ H$_{2}$CO + M & $\Bigg \{$ \parbox{8cm}{\centering k$_{0}$ =  $ 2.800\times 10^{-20} $ $\cdot T^{-3.42}$$\cdot$ exp(-42450/T) \\  k$_{\infty}$ =  $ 7.140\times 10^{-17} $ $\cdot T^{1.50}$$\cdot$ exp(-42450/T) } & 2 \\
273 & OH + CH$_{3}$ $\rightarrow$ CH$_{3}$OH + M & $\Bigg \{$ \parbox{8cm}{\centering k$_{0}$ =  $ 4.370\times 10^{-04} $ $\cdot T^{-8.20}$ \\  k$_{\infty}$ =  $ 1.000\times 10^{-10} $  } & 2 \\
275 & C + C $\rightarrow$ C$_{2}$ + M &  $ 4.970\times 10^{-27} $ $\cdot T^{-1.60}$& 1 \\ \hline 
277 & C$_{2}$H + M $\rightarrow$ C$_{2}$ + H + M &  $ 2.920\times 10^{+11} $ $\cdot T^{-5.16}$$\cdot$ exp(-57400/T)& 1 \\
279 & O + C $\rightarrow$ CO + M &  $ 9.100\times 10^{-22} $ $\cdot T^{-3.10}$$\cdot$ exp(-2114/T)& 1 \\
281 & H + O$_{2}$ $\rightarrow$ HO$_{2}$ + M & $\Bigg \{$ \parbox{8cm}{\centering k$_{0}$ =  $ 5.240\times 10^{-28} $ $\cdot T^{-1.60}$ \\  k$_{\infty}$ =  $ 7.500\times 10^{-11} $  } & 2 \\
283 & H + HO$_{2}$ $\rightarrow$ H$_{2}$ + O$_{2}$ &  $ 7.200\times 10^{-12} $ & 1 \\
285 & H + HO$_{2}$ $\rightarrow$ H$_{2}$O + O &  $ 1.600\times 10^{-12} $ & 1 \\ \hline 
287 & H + HO$_{2}$ $\rightarrow$ OH + OH &  $ 7.120\times 10^{-11} $ & 1* \\
289 & OH + HO$_{2}$ $\rightarrow$ H$_{2}$O + O$_{2}$ &  $ 4.800\times 10^{-11} $ $\cdot$ exp(-250/T)& 1* \\
291 & OH + O$_{3}$ $\rightarrow$ HO$_{2}$ + O$_{2}$ &  $ 1.600\times 10^{-12} $ $\cdot$ exp(940/T)& 1* \\
293 & HO$_{2}$ + O $\rightarrow$ OH + O$_{2}$ &  $ 3.000\times 10^{-11} $ $\cdot$ exp(-200/T)& 1* \\
295 & H$_{2}$O$_{2}$ + OH $\rightarrow$ HO$_{2}$ + H$_{2}$O &  $ 2.900\times 10^{-12} $ $\cdot$ exp(160/T)& 1* \\ \hline 
297 & HCO + O$_{2}$ $\rightarrow$ HO$_{2}$ + CO &  $ 5.200\times 10^{-12} $ & 1* \\
299 & H$_{2}$O$_{2}$ + O $\rightarrow$ OH + HO$_{2}$ &  $ 1.400\times 10^{-12} $ $\cdot$ exp(2000/T)& 1* \\
301 & CH$_{3}$ + O$_{3}$ $\rightarrow$ H$_{2}$CO + HO$_{2}$ &  $ 5.400\times 10^{-12} $ $\cdot$ exp(2200/T)& 1* \\
303 & CH$_{3}$O + O$_{2}$ $\rightarrow$ H$_{2}$CO + HO$_{2}$ &  $ 7.200\times 10^{-14} $ $\cdot$ exp(1080/T)& 1* \\
305 & OH + OH $\rightarrow$ H$_{2}$O$_{2}$ + M & $\Bigg \{$ \parbox{8cm}{\centering k$_{0}$ =  $ 2.070\times 10^{-28} $ $\cdot T^{-1.0}$ \\  k$_{\infty}$ =  $ 2.600\times 10^{-11} $  } & 2* \\ \hline 
307 & H + O$_{3}$ $\rightarrow$ OH + O$_{2}$ &  $ 1.400\times 10^{-10} $ $\cdot$ exp(470/T)& 1* \\
309 & O + O$_{2}$ $\rightarrow$ O$_{3}$ + M & $\Bigg \{$ \parbox{8cm}{\centering k$_{0}$ =  $ 5.290\times 10^{-28} $ $\cdot T^{-2.4}$ \\  k$_{\infty}$ =  $ 3.000\times 10^{-11} $  } & 2* \\
311 & O + O$_{3}$ $\rightarrow$ O$_{2}$ + O$_{2}$ &  $ 8.000\times 10^{-12} $ $\cdot$ exp(2060/T)& 1* \\
313 & CH$_{3}$ + O$_{3}$ $\rightarrow$ CH$_{3}$O + O$_{2}$ &  $ 5.400\times 10^{-12} $ $\cdot$ exp(220/T)& 1* \\
315 & $^{1}$CH$_{2}$ + CH$_{4}$ $\rightarrow$ CH$_{3}$ + CH$_{3}$ &  $ 3.600\times 10^{-11} $ & 1* \\ \hline 
316 & $^{1}$CH$_{2}$ + O$_{2}$ $\rightarrow$ HCO + OH &  $ 3.000\times 10^{-11} $ & 1* \\
317 & $^{1}$CH$_{2}$ + M $\rightarrow$ $^{3}$CH$_{2}$ + M &  $ 8.800\times 10^{-12} $ & 1* \\
318 & $^{1}$CH$_{2}$ + H$_{2}$ $\rightarrow$ CH$_{3}$ + H &  $ 5.000\times 10^{-15} $ & 1* \\
319 & $^{1}$CH$_{2}$ + CO$_{2}$ $\rightarrow$ H$_{2}$CO + CO &  $ 1.000\times 10^{-12} $ & 1* \\
320 & CH + H$_{2}$ $\rightarrow$ $^{3}$CH$_{2}$ + H &  $ 2.380\times 10^{-10} $ $\cdot$ exp(1760/T)& 1* \\ \hline 
321 & C$_{2}$H$_{2}$ + O $\rightarrow$ $^{3}$CH$_{2}$ + CO &  $ 2.900\times 10^{-11} $ $\cdot$ exp(1600/T)& 1* \\
322 & $^{3}$CH$_{2}$ + H$_{2}$ $\rightarrow$ CH$_{3}$ + H &  $ 5.000\times 10^{-14} $ & 1* \\
323 & $^{3}$CH$_{2}$ + CH$_{4}$ $\rightarrow$ CH$_{3}$ + CH$_{3}$ &  $ 7.100\times 10^{-12} $ $\cdot$ exp(5051/T)& 1* \\
324 & $^{3}$CH$_{2}$ + O$_{2}$ $\rightarrow$ HCO + OH &  $ 4.100\times 10^{-11} $ $\cdot$ exp(750/T)& 1* \\
325 & $^{3}$CH$_{2}$ + O $\rightarrow$ HCO + H &  $ 1.000\times 10^{-11} $ & 1* \\ \hline 
326 & $^{3}$CH$_{2}$ + O $\rightarrow$ CH + OH &  $ 8.000\times 10^{-12} $ & 1* \\
327 & $^{3}$CH$_{2}$ + O $\rightarrow$ CO + H + H &  $ 8.300\times 10^{-11} $ & 1* \\
328 & $^{3}$CH$_{2}$ + CO$_{2}$ $\rightarrow$ H$_{2}$CO + CO &  $ 1.000\times 10^{-14} $ & 1* \\
329 & $^{3}$CH$_{2}$ + H $\rightarrow$ CH + H$_{2}$ &  $ 4.700\times 10^{-10} $ $\cdot$ exp(370/T)& 1* \\
330 & $^{3}$CH$_{2}$ + $^{3}$CH$_{2}$ $\rightarrow$ C$_{2}$H$_{2}$ + H$_{2}$ &  $ 5.300\times 10^{-11} $ & 1* \\ \hline 
331 & $^{3}$CH$_{2}$ + CH$_{3}$ $\rightarrow$ C$_{2}$H$_{4}$ + H &  $ 7.000\times 10^{-11} $ & 1* \\
332 & $^{3}$CH$_{2}$ + C$_{2}$H$_{3}$ $\rightarrow$ CH$_{3}$ + C$_{2}$H$_{2}$ &  $ 3.000\times 10^{-11} $ & 1* \\
333 & $^{3}$CH$_{2}$ + C$_{2}$H$_{5}$ $\rightarrow$ CH$_{3}$ + C$_{2}$H$_{4}$ &  $ 3.000\times 10^{-11} $ & 1* \\
334 & H$_{2}$O + O($^{1}$D) $\rightarrow$ OH + OH &  $ 2.200\times 10^{-10} $ & 1* \\
335 & H$_{2}$ + O($^{1}$D) $\rightarrow$ OH + H &  $ 1.100\times 10^{-10} $ & 1* \\ \hline 
336 & O($^{1}$D) + M $\rightarrow$ O + M &  $ 1.800\times 10^{-11} $ $\cdot$ exp(-110/T)& 1* \\
337 & O($^{1}$D) + O$_{2}$ $\rightarrow$ O + O$_{2}$ &  $ 3.200\times 10^{-11} $ $\cdot$ exp(-70/T)& 1* \\
338 & CH$_{4}$ + O($^{1}$D) $\rightarrow$ CH$_{3}$ + OH &  $ 1.125\times 10^{-10} $ & 1* \\
339 & CH$_{4}$ + O($^{1}$D) $\rightarrow$ H$_{2}$CO + H$_{2}$ &  $ 7.500\times 10^{-12} $ & 1* \\
340 & CH$_{4}$ + O($^{1}$D) $\rightarrow$ CH$_{3}$O + H &  $ 3.000\times 10^{-11} $ & 1* \\ \hline 
341 & C$_{2}$H$_{6}$ + O($^{1}$D) $\rightarrow$ C$_{2}$H$_{5}$ + OH &  $ 6.200\times 10^{-10} $ & 1* \\
342 & CO + O($^{1}$D) $\rightarrow$ CO + O &  $ 7.000\times 10^{-11} $ & 1* \\
343 & O$_{2}$ + h$\nu$ $\rightarrow$ O + O($^{1}$D) & $2.18 \times 10^{-6}$ & 3* \\
344 & O$_{2}$ + h$\nu$ $\rightarrow$ O + O & $5 \times 10^{-8}$ & 3* \\
345 & H$_{2}$O + h$\nu$ $\rightarrow$ H + OH & $8.16 \times 10^{-6}$ & 3 \\ \hline 
346 & OH + h$\nu$ $\rightarrow$ O($^{1}$D) + H & $4.9 \times 10^{-6}$ & 3 \\
347 & O$_{3}$ + h$\nu$ $\rightarrow$ O$_{2}$ + O($^{1}$D) & $9.98 \times 10^{-3}$ & 3* \\
348 & O$_{3}$ + h$\nu$ $\rightarrow$ O$_{2}$ + O & $1.92 \times 10^{-3}$ & 3* \\
349 & H$_{2}$O$_{2}$ + h$\nu$ $\rightarrow$ OH + OH & $1.27 \times 10^{-4}$ & 3 \\
350 & CO$_{2}$ + h$\nu$ $\rightarrow$ CO + O & $2.31 \times 10^{-9}$ & 3 \\ \hline 
351 & CO$_{2}$ + h$\nu$ $\rightarrow$ CO + O($^{1}$D) & $2.51 \times 10^{-7}$ & 3 \\
352 & CO + h$\nu$ $\rightarrow$ C + O & $1.54 \times 10^{-6}$ & 3 \\
353 & H$_{2}$CO + h$\nu$ $\rightarrow$ H$_{2}$ + CO & $1.15 \times 10^{-4}$ & 3 \\
354 & H$_{2}$CO + h$\nu$ $\rightarrow$ HCO + H & $1.32 \times 10^{-4}$ & 3 \\
355 & HO$_{2}$ + h$\nu$ $\rightarrow$ OH + O & $6.13 \times 10^{-4}$ & 3 \\ \hline 
356 & CH + h$\nu$ $\rightarrow$ C + H & $4.25 \times 10^{-5}$ & 3 \\
357 & CH$_{3}$ + h$\nu$ $\rightarrow$ $^{1}$CH$_{2}$ + H & 0. & 3 \\
358 & CH$_{4}$ + h$\nu$ $\rightarrow$ $^{1}$CH$_{2}$ + H$_{2}$ & $1.78 \times 10^{-6}$ & 3 \\
359 & CH$_{4}$ + h$\nu$ $\rightarrow$ CH$_{3}$ + H & $3.41 \times 10^{-6}$ & 3 \\
360 & CH$_{4}$ + h$\nu$ $\rightarrow$ $^{3}$CH$_{2}$ + H + H & $1.67 \times 10^{-6}$ & 3 \\ \hline 
361 & C$_{2}$H$_{2}$ + h$\nu$ $\rightarrow$ C$_{2}$H + H & $1.14 \times 10^{-6}$ & 3 \\
362 & C$_{2}$H$_{2}$ + h$\nu$ $\rightarrow$ C$_{2}$ + H$_{2}$ & $6.55 \times 10^{-7}$ & 3 \\
363 & C$_{2}$H$_{3}$ + h$\nu$ $\rightarrow$ C$_{2}$H$_{2}$ + H & $2.94 \times 10^{-4}$ & 3 \\
364 & C$_{2}$H$_{4}$ + h$\nu$ $\rightarrow$ C$_{2}$H$_{2}$ + H$_{2}$ & $2.08 \times 10^{-5}$ & 3 \\
365 & C$_{2}$H$_{4}$ + h$\nu$ $\rightarrow$ C$_{2}$H$_{2}$ + H + H & $2.17 \times 10^{-5}$ & 3 \\ \hline 
366 & C$_{2}$H$_{6}$ + h$\nu$ $\rightarrow$ $^{3}$CH$_{2}$ + $^{3}$CH$_{2}$ + H$_{2}$ & $4. \times 10^{-6}$ & 3 \\
367 & C$_{2}$H$_{6}$ + h$\nu$ $\rightarrow$ CH$_{4}$ + $^{1}$CH$_{2}$ & $1. \times 10^{-6}$ & 3 \\
368 & CH$_{2}$CO + h$\nu$ $\rightarrow$ $^{3}$CH$_{2}$ + CO & $7.03 \times 10^{-4}$ & 3 \\
369 & $^{1}$CH$_{2}$ + H$_{2}$ $\rightarrow$ $^{3}$CH$_{2}$ + H$_{2}$ &  $ 1.260\times 10^{-11} $ & 1 \\
 \end{longtable} 
 
The reaction list is largely composed of forward reactions that are then used to calculate the reverse reactions, based on the thermodynamic properties of the species involved in the reaction. However, some reactions do not have reversed reactions, including a) the ones where a molecule in an excited state relaxes into the ground state (e.g., reaction 369), and b) photolysis reactions (such as reaction 368). The majority of these reactions are detailed in \citet{tsai2017vulcan} as the reduced C-H-O system, to which we have added reactions for oxidizing species (denoted by the asterisk*: \#287--342; \#343--344; \#347--348).

 1: cm$^{3}$ molecules$^{-2}$ s$^{-1}$

2: These reaction rates take the form: k(M,T) =  $k_{0}(T)[M]/[1+k_{0}(T)[m]/k_{\infty}(T)]\cdot$ 0.6\^{}$[1+[log_{10}[k_{0}(T)[M]/k_{\infty    }(T)]]^{2}]^{-1}$, where $k_{0}(T)$ has units of cm$^{6}$ molecules$^{-2}$ s$^{-1}$ and k$_{\infty}(T)$ has units of cm$^{3}$ molecules$^{-2}$ s$^{-1}$.

3: The photolysis rates (in /s) presented here are taken from the uppermost layer in the model. Caution: the rates in the upper atmosphere are not good indicators for rates in the lower atmosphere.




\section{Composition \label{app:comp}}

\begin{figure*}[ht]
    \centering
    \includegraphics[width=\textwidth]{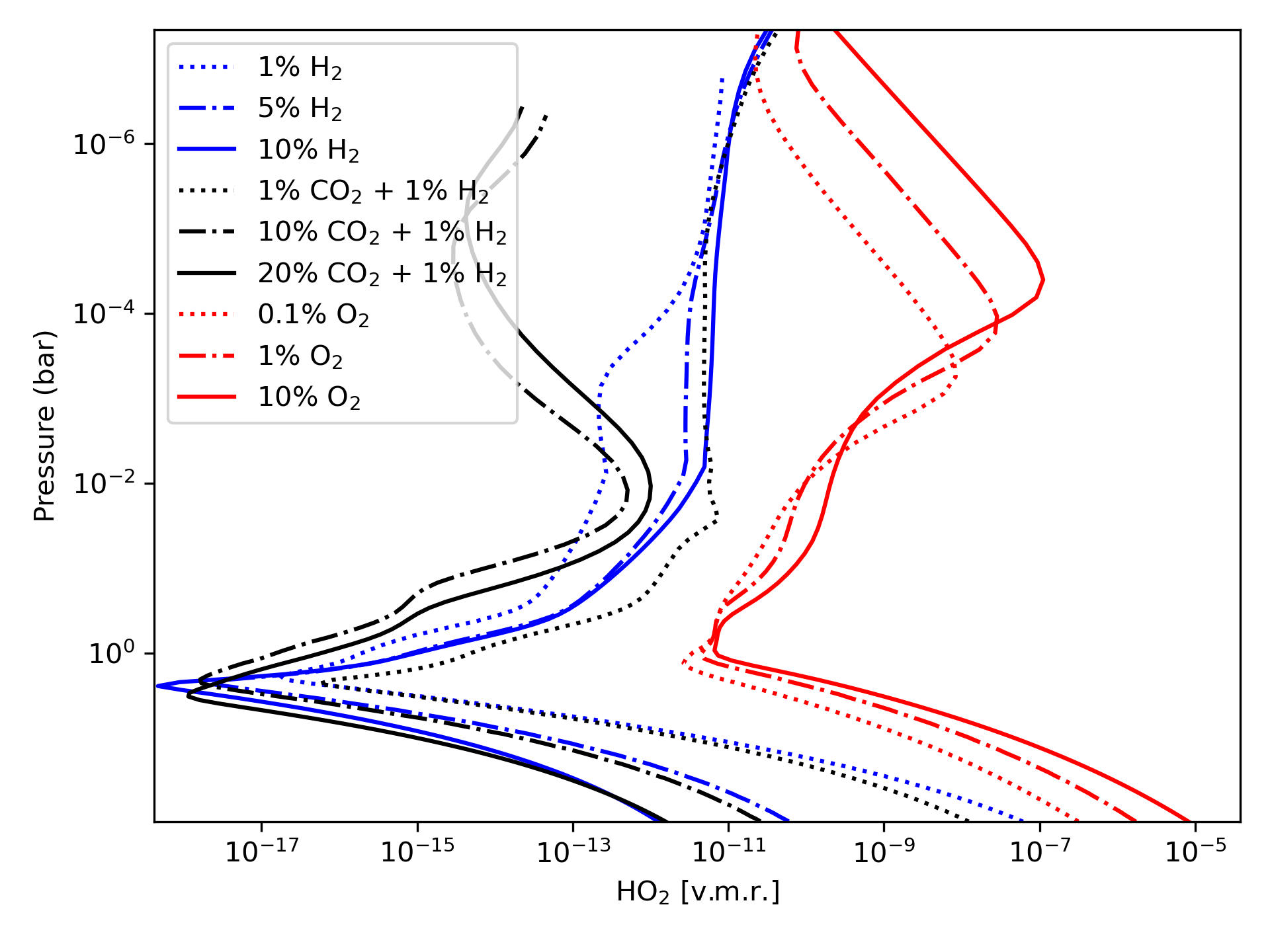}
    \caption{HO$_{2}$ profiles for the cases seen in Fig. \ref{fig:temp_water}, and sharing the same color palette.}
    \label{fig:HO2_profiles}
\end{figure*}

\begin{figure*}[ht]
    \centering
    \includegraphics[width=\textwidth]{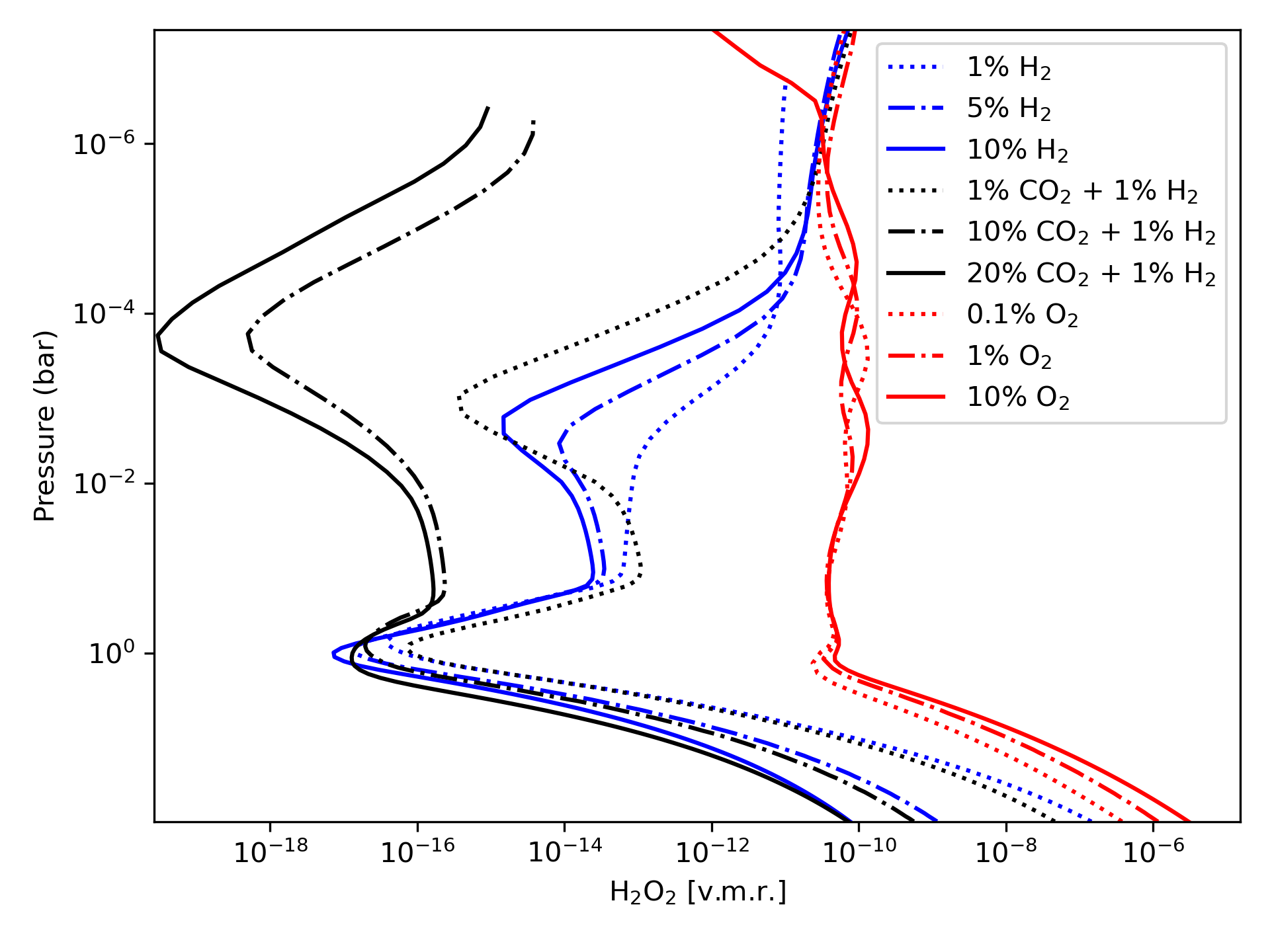}
    \caption{H$_{2}$O$_{2}$ profiles for the cases seen in Fig. \ref{fig:temp_water}, and sharing the same color palette.}
    \label{fig:H2O2_profiles}
\end{figure*}

\clearpage
\section{Secondary analyses \label{app:secondary}}

\begin{figure*}[ht]
    \centering
    \includegraphics[width=\textwidth]{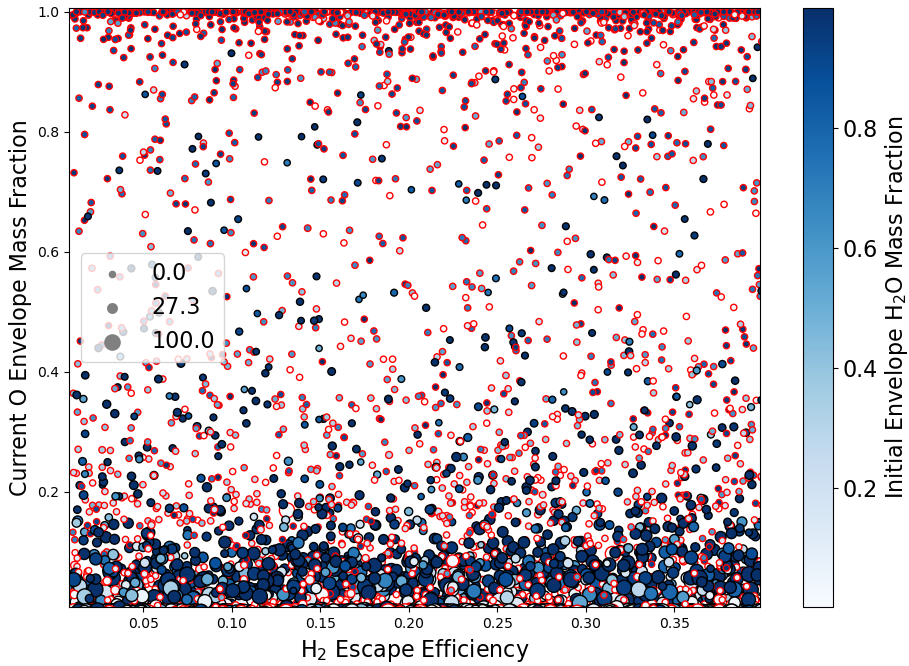}
    \caption{The water inventory for planets in our Monte Carlo atmospheric escape ensemble as a function of the assumed H$_{2}$ escape efficiency. The circle size indicates the initial envelope mass as a fraction of the total planet mass. A red edge color for a given symbol denotes a scenario with less than 1 Earth ocean at present day, which would likely be short-lived \citep{kasting1983loss}.}
    \label{fig:H2_eff_test}
\end{figure*}

\begin{figure*}[ht]
    \centering
    \includegraphics[width=\textwidth]{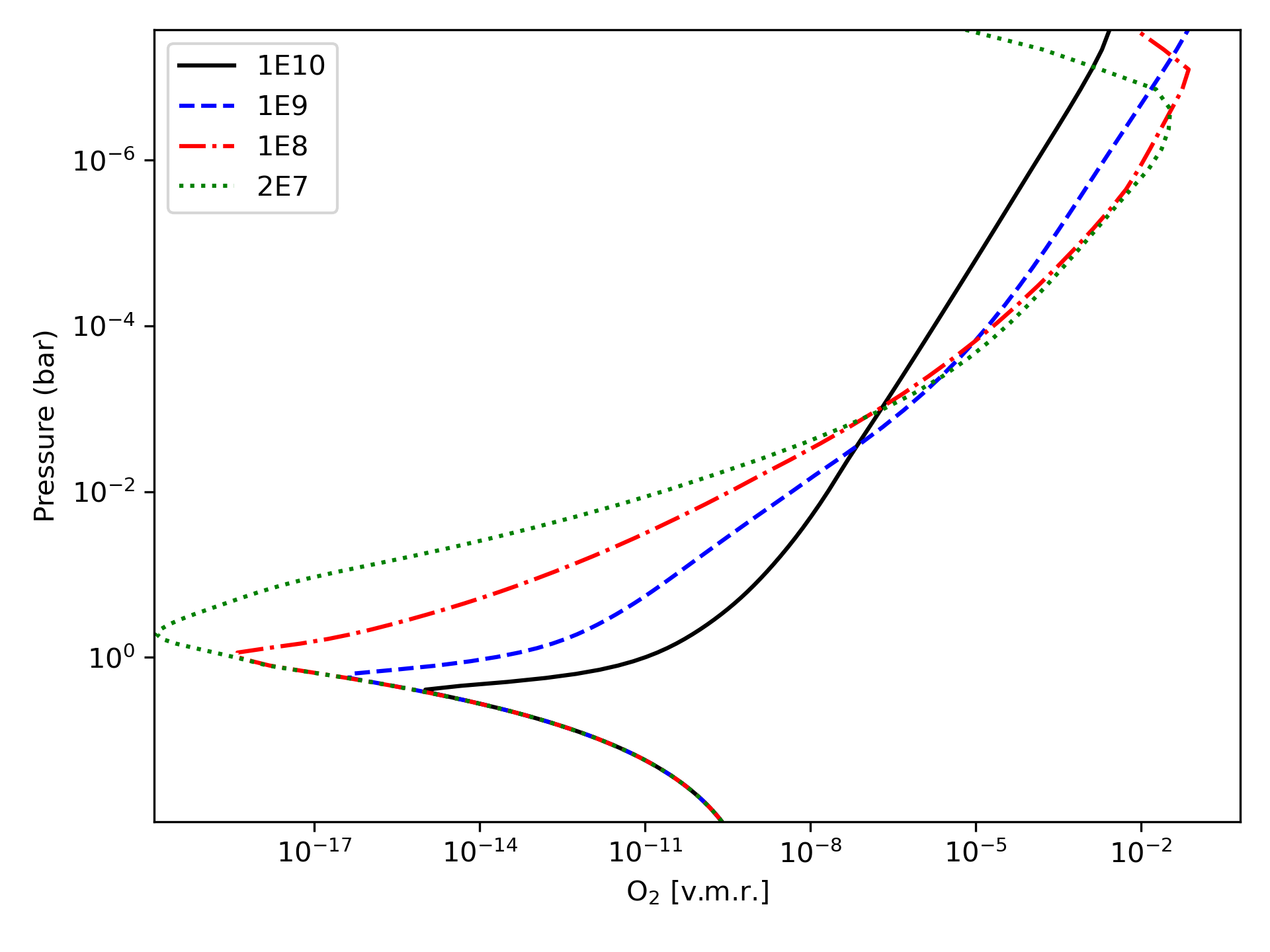}
    \caption{Molecular oxygen response to changing eddy diffusion profile strengths (values indicated in the key, with units of cm$^{2}$/s) for the 10\% H$_{2}$ scenario. Smaller values of K$_{\text{zz}}$ prevent photochemically-produced O$_{2}$ from being mixed down into the middle atmosphere, resulting in larger upper atmospheric concentrations at the expense of lower atmospheric concentrations.}
    \label{fig:O2_edd_test}
\end{figure*}

\begin{figure*}[ht]
    \centering
    \includegraphics[width=\textwidth]{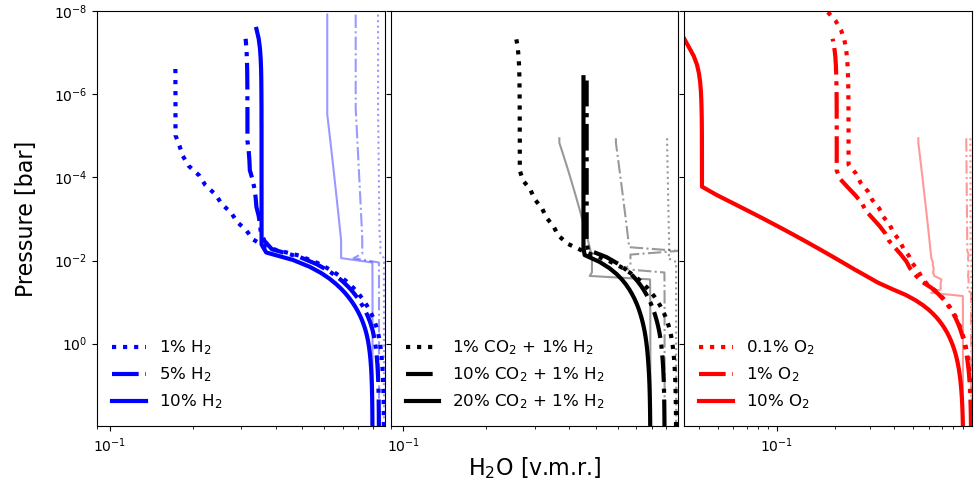}
    \caption{The water vapor profiles for the scenarios outlined in the main text from the photochemical model (heavy lines) compared with the water profiles produced by the radiative-convective model (light, pale lines). Differences between the two profiles are confined to the upper atmosphere (above $\sim$10 mbar) and do not produce more than a $\sim$5 ppm change in the resulting transmission spectra.}
    \label{fig:water_profile_test}
\end{figure*}

\clearpage

\bibliography{toi1266}{}
\bibliographystyle{aasjournal}

\end{document}